\documentclass[aps,prx,twocolumn,nofootinbib,groupedaddress,superscriptaddress
,longbibliography]{revtex4-2}

\usepackage{amsmath,amssymb}
\usepackage{graphicx}
\usepackage{multirow}
\usepackage{bm}
\usepackage{mathtools}
\usepackage{dsfont}
\usepackage{amsfonts}
\usepackage{xcolor}
\usepackage[normalem]{ulem}
\usepackage{url}
\usepackage{wasysym}
\usepackage{bbm}
\usepackage{hyperref}
\usepackage{braket}
\usepackage{booktabs}

\newcommand{\ourtitle}{
Intertwining Josephson and Vortex Topologies in Conventional Superconductors}

\allowdisplaybreaks

\begin{document}
\title{\textbf{\ourtitle}}

\author{Zhuo Chen} 
\altaffiliation{These authors contributed equally to this work}
\affiliation{Department of Physics and Astronomy, The University of Tennessee, Knoxville, TN 37996, USA}

\author{Jiangxu Li} 
\altaffiliation{These authors contributed equally to this work}
\affiliation{Department of Physics and Astronomy, The University of Tennessee, Knoxville, TN 37996, USA}

\author{Lun-Hui Hu} 
\affiliation{Department of Physics and Astronomy, The University of Tennessee, Knoxville, TN 37996, USA}
\affiliation{Center for Correlated Matter and School of Physics, Zhejiang University, Hangzhou, China}

\author{Zhen Bi} 
\affiliation{Department of Physics, The Pennsylvania State University, University Park, Pennsylvania 16802, USA}

\author{Rui-Xing Zhang}
\email{ruixing@utk.edu}
\affiliation{Department of Physics and Astronomy, The University of Tennessee, Knoxville, TN 37996, USA}
\affiliation{Department of Materials Science and Engineering, The University of Tennessee, Knoxville, TN 37996, USA}

\begin{abstract}
Recent experimental advances have unveiled promising evidence of vortex-bound Majorana quasiparticles in multiple superconducting compounds. However, theoretical progress in understanding these phenomena, especially from ab initio approaches, has been limited by the computational complexity of simulating vortex structures. To bridge this gap, we introduce the Josephson–vortex correspondence (JVC), a theoretical framework that systematically maps the bound-state topological properties of vortices to those of $\pi$-phase Josephson junctions in the same superconductor. This correspondence allows vortex phase diagrams to be constructed directly from Josephson junction calculations, thereby eliminating the need for large-scale vortex calculations. We demonstrate the validity and predictive power of JVC across a variety of effective models, and further extend the framework to the first-principles level. Applying our approach to 2M-WS$_2$ and Sr$_3$SnO, we identify them as realistic, doping-tunable platforms for realizing vortex Majorana zero modes. Our theory will pave the way for ab initio Majorana material discovery and design.
\end{abstract}

\maketitle

\let\oldaddcontentsline\addcontentsline
\renewcommand{\addcontentsline}[3]{}

\section{Introduction}

Topological principles in quantum condensed matters can enable exotic quasiparticle excitations that are otherwise unattainable~\cite{wilczek1982anyon,kitaev2006anyons,hasan2010RMP, qi2011RMP,barry2016science}. For instance, Majorana zero modes (MZMs), a class of non-Abelian quasiparticles, are predicted to exist in topological superconductors (TSCs) with unconventional pairing symmetries~\cite{read2000paired,kitaev2001unpaired,sau2010generic,lutchyn2010majorana,oreg2010helical,alicea2012new}. These MZMs hold transformative potential for quantum technologies~\cite{nayak2008RMP}. While definitive proof of TSCs remains elusive~\cite{dassarma2023search}, zero-bias peaks (ZBP) in the tunneling spectroscopy have been observed inside Abrikosov vortices of multiple 3D $s$-wave superconductors with a trivial ground-state topology, including FeTe$_{1-x}$Se$_{x}$~\cite{wang2018evidence,machida2019zero,kong2019half,zhu2020nearly}, (Li$_{0.84}$Fe$_{0.16}$)OHFeSe~\cite{liu2018LiFeOH}, 2M-WS$_2$~\cite{yuan2019evidence,fan2024stripe}, and LiFeAs~\cite{kong2021majorana,liu2022tunable,li2022ordered}. These ZBPs are often interpreted as signatures of vortex-bound MZMs, but the definitive existence of such modes remains debated.

The identification of the above compounds follows a key insight: the presence of a normal-state Dirac surface state is considered crucial for the emergence of vortex MZMs, as first highlighted by Fu and Kane in the context of superconducting topological insulators (TIs)~\cite{fu2008vortex}. While the Fu-Kane paradigm has been instrumental in shaping our understanding of vortex MZMs, real-world superconductors, such as those discussed here, invariably feature bulk Fermi pockets that facilitate Cooper pairing instabilities~\cite{zhang2019multiple}. These low-energy metallic bands can obscure the surface states and potentially influence the fate of vortex MZMs, complicating the direct application of the Fu-Kane theory~\cite{chiu2012stabilization,xu2016topological,hu2022competing,hu2023topological}. Moreover, while the observation of a ZBP is necessary, it is not sufficient to definitively identify MZMs experimentally. For instance, near-surface impurities bind low-energy vortex states that effectively mimic the phenomenology of MZMs, leading to potential false positives~\cite{pan2020physical}. These theoretical and experimental challenges underscore the difficulties in conclusively interpreting the nature of ZBP signals in these systems.

Recent theoretical advances have introduced an alternative bulk perspective for understanding the origin of vortex MZMs, viewing vortex lines in 3D type-II superconductors as effective 1D class-D nanowires. Along these ``nanowires", Caroli–de Gennes–Matricon (CdGM) bound states necessarily form and disperse~\cite{caroli1964bound}. If these 1D CdGM states acquire topologically nontrivial characteristics, MZMs will emerge as boundary modes at the ends of the vortex line~\cite{hosur2011vortex}. This vortex-line topology (VLT) framework directly quantifies the condition for realizing vortex MZMs, without requiring the disentanglement of surface and bulk contributions. However, evaluating VLT through vortex simulations presents significant computational challenges, as the vortex geometry breaks all in-plane translational symmetries, substantially increasing computational complexity. Consequently, VLT studies to date have been largely limited to effective ${\bf k\cdot p}$ or tight-binding models with idealized, often unrealistic, parameters, leaving their applicability to real-world materials in question.

The overarching goal of this work is to {\it enable VLT characterization at the first-principles level}, while overcoming the numerical challenges discussed earlier. Our approach is driven by a key geometric insight: a vortex line can be smoothly deformed into a pair of perpendicular $\pi$-phase Josephson junctions. By focusing on a single $\pi$-junction, or a “halved” vortex, we observe the emergence of 2D dispersing Andreev bound states within the bulk pairing gap. Similar to CdGM modes, these Andreev states can, in principle, develop emergent topological features through sub-gap band inversions. We refer to this phenomenon as {\it Josephson topology} (JT) and show that specific lattice symmetries compatible with the junction can protect a variety of novel JT phases, including those with higher-order topology.

\begin{figure*}[t] 
    \centering
    \includegraphics[width=0.8\textwidth]{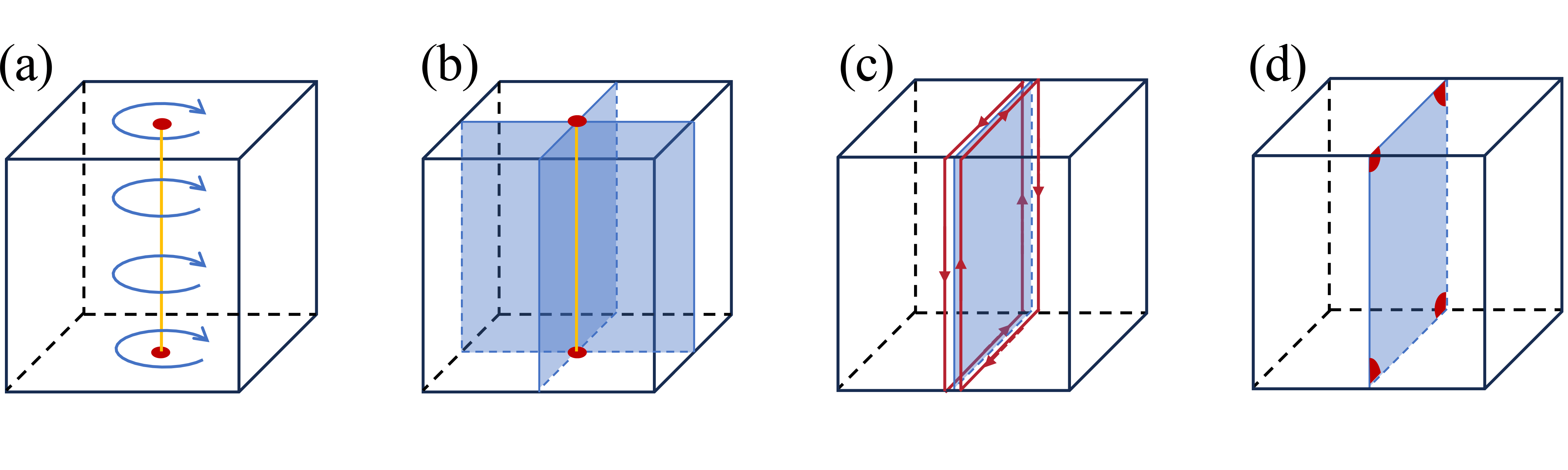}
    \caption{A conventional $s$-wave superconductor can feature emergent ``subsystem" topological phenomena due to various spatial structures of the pairing phase: (a) A continuous $U(1)$ Abrikosov vortex with a $\mathbb{Z}_2$ nontrivial vortex-line topology. 0D MZMs will be trapped at the ends of the vortex line (marked in red). (b) Deforming an $U(1)$ vortex to a discrete $\mathbb{Z}_4$ vortex described by Eq.~(\ref{eq:vortex discretization}), which further manifests as a superposition of two perpendicular $\pi$-phase Josephson junctions. (c) A single $\pi$-junction of a conventional superconductor can feature a $\mathbb{Z}_2$ nontrivial Josephson topology with helical Majorana edge states. (d) Schematic of a second-order Josephson topology with corner-localized Majorana Kramers pairs. }
    \label{fig:Schematic diagram}
\end{figure*}

The central conceptual advance of this work is the discovery of a topological mapping between 2D $\pi$-junctions and 1D vortex lines, a framework we term the {\bf Josephson-Vortex Correspondence} (JVC). Using a dimensional reduction strategy~\cite{qi2008TQFT,huang2024class,zhang2022bulk}, we derive these JVC relations by coupling a target $\pi$-junction with its complementary counterpart, thereby completing the $2\pi$ phase winding required of a vortex. Remarkably, we find that the mapping between Josephson and vortex-line topological phases is often one-to-one, with exceptions only arising in systems where high-fold rotational symmetries of the vortex tube are geometrically incompatible with planar $\pi$-junctions. In both situations, the topological phase can be tuned by varying the chemical potential $\mu$, with transitions marked by the gap closing of CdGM states in the vortex, and similarly, by Andreev states in the junction. We analytically demonstrate this $\mu$-tunable behavior of Josephson topology using a combination of effective theory and dimensional reduction. These mapping relations between JT and VLT provide a powerful and scalable alternative to direct vortex simulations, enabling systematic exploration of vortex phase diagrams through low-cost $\pi$-junction modeling.

As a proof of concept, we numerically evaluate and benchmark the JVC across several effective model systems, chosen such that both $\pi$-junction and vortex geometries remain computationally tractable. Our first example is a four-band minimal model of a 3D $\mathbb{Z}_2$ topological insulator~\cite{fu20073dTI,moore20073dTI,roy20093dTI}, a paradigmatic Fu-Kane system with $s$-wave superconducting pairing. This system is known to exhibit a $\mathbb{Z}_2$-nontrivial VLT at low chemical potential $\mu$~\cite{fu2008vortex,hosur2011vortex}, resembling a class-D Kitaev chain with localized Majorana zero modes~\cite{kitaev2001unpaired}. Through $\pi$-junction modeling, we demonstrate that the same system hosts a 2D $\mathbb{Z}_2$-nontrivial JT over an identical $\mu$ range, thereby providing direct numerical evidence for the JVC. We further apply our analysis to a superconducting Dirac semimetal~\cite{armitage2018RMP}, which features a gapless, rotation-protected nodal vortex phase~\cite{konig2019nodal,qin2019nodal}. In this case, the corresponding JT phase is found to be quasi-nodal and $\mathbb{Z}_2$ topological, again in agreement with the JVC framework.

Expanding beyond minimal models, we investigate the JVC in multi-band systems that more faithfully capture the complexity of real materials, including a six-band model of iron-based superconductors and an eight-band ``Dirac octet” model. The former has been successful in reproducing vortex-bound state phenomenology observed in LiFeAs~\cite{hu2022competing}, while the latter was originally proposed to describe the low-energy topological bands of certain superconducting anti-perovskites~\cite{hsieh2014topological,kariyado2011three,kariyado2012low,kawakami2018topological,fang2020SSO}. Using comprehensive $\pi$-junction simulations, we uncover rich JT phase diagrams in both models, featuring multiple phases distinguished by their Bogoliubov–de Gennes (BdG) mirror Chern numbers and inversion indicators. Applying the JVC framework, we derive the corresponding VLT phase diagrams, which show strong agreement with those obtained from direct vortex simulations. 

Finally, we turn to the investigation of vortex topological physics in two realistic material platforms: (i) 2M-WS$_2$, a widely regarded Fu-Kane superconductor with experimentally observed vortex ZBP signals~\cite{yuan2019evidence}, and (ii) Sr$_3$SnO, an anti-perovskite superconductor that has been synthesized but remains unexplored from a topological perspective~\cite{oudah2016superconductivity}. For both compounds, we carry out density functional theory (DFT) calculations to determine the normal-state electronic structure and construct symmetry-preserving Wannier models of the low-energy bands. Accurately capturing the complex band features necessitates models with approximately 40 Wannier orbitals per lattice site. Vortex simulations using these models pose a significant computational challenge: to avoid finite-size artifacts for a pairing gap on the meV scale, one would require system sizes on the order of $10^6$ lattice sites, corresponding to a Hilbert space dimension of ${\cal O}(10^8)$. In sharp contrast, the $\pi$-junction geometry involves only ${\cal O}(10^4)$ degrees of freedom. This dramatic reduction in complexity highlights the practical advantage of the JVC framework in enabling ab initio characterization of vortex-line topology.

Notably, our $\pi$-junction simulations for 2M-WS$_2$ reveal a $\mathbb{Z}_2$ nontrivial JT phase at $\mu = 0$. Following JVC, this numerical result supports a Majorana interpretation of the vortex ZBPs reported in Refs.~\cite{yuan2019evidence,fan2024stripe}. By continuously varying the chemical potential $\mu$, we construct a comprehensive JT phase diagram and identify a sequence of topological critical points. In particular, there exists a JT critical point approximately 25 meV below the Fermi level, below which the JT phase becomes $\mathbb{Z}_2$ trivial. Remarkably, 
this criticality emerges exactly when the electron pocket at $L$ of the Brillouin zone undergoes a Lifshitz transition. As such, it originates from physics beyond the conventional Fu-Kane paradigm, which in the case of 2M-WS$_2$ is primarily governed by the $\mathbb{Z}_2$ topology associated with the band inversion at the $\Gamma$ point. 

The ab initio JT phase diagram for Sr$_3$SnO similarly reveals multiple critical points at distinct doping levels. Our $\pi$-junction modeling indicates that the undoped system ($\mu = 0$) hosts a nontrivial $\mathbb{Z}_2$ JT phase, predicting a single MZM at the surface vortex core. Upon lowering the chemical potential, we uncover a JT phase transition into a state characterized by a BdG mirror Chern number of 2. This distinguishes Sr$_3$SnO from 2M-WS$_2$ and points to richer vortex topology. According to the JVC, this mirror-Chern JT phase corresponds to a novel vortex topological phase with a pair of spatially overlapping, yet symmetry-protected, decoupled MZMs. 

This paper is organized as follows. In Sec.~\ref{sec:JVC}, we introduce the notion of Josephson topologies for $s$-wave superconductors, classify them under different symmetries, and discuss how they correspond to various vortex-line topologies through the dimensional reduction approach. Sec.~\ref{sec:Criti Corresp} focuses on classifying and understanding JT transitions, based on which we analytically derive a set of criticality mapping relations between $\pi$-junctions and vortices. A series of effective model studies aiming at benchmarking JVC are provided and discussed in Sec.~\ref{sec:effective_models}, which include superconducting TI, Dirac semimetal, iron-based system, and the Dirac octet model. We present both ab-initio junction modeling and JVC analysis for 2M-WS$_2$ and Sr$_3$SnO in Sec.~\ref{sec:ab-initio}, where we predict concrete experimental consequences. We conclude our results in Sec.~\ref{sec:conclusion}, as well as an outlook of possible future directions.

\section{Josephson-vortex correspondence} 
\label{sec:JVC}

\subsection{From vortex to $\pi$-junction}

Our system of interest is a 1D field-induced vortex line along $\hat{z}$ in a 3D superconductor with an isotropic $s$-wave spin-singlet pairing. We use a Bogoliubov-de Gennes (BdG) Hamiltonian to describe the vortex-line system, 
\begin{equation}
    H({\bf r}, k_z) = \begin{pmatrix}
        h_0 ({\bf r}, k_z)-\mu & -i\Delta({\bf r}) s_y \\
        i \Delta({\bf r}) s_y & \mu -h_0^T({\bf r}, -k_z)
    \end{pmatrix}.
\label{eq:BdG_form}
\end{equation}
Here, $h_0({\bf r})$ describes the normal state for electrons, where ${\bf r}=(r,\theta)$ denotes the in-plane polar coordinates. $\mu$ is the chemical potential and $s_{x,y,z}$ are the Pauli matrices for electron spins. The vortex geometry is encoded in the spatially varying pairing function $\Delta({\bf r}) \equiv \Delta_0 \tanh{(r/\xi)} e^{i\theta}$, where $\xi$ controls the size of the vortex core. We set $\xi=0$ in this work for simplicity. While the bulk SC ground state is topology-free~\cite{haim2016no-go}, the vortex line can be topologically or trivially gapped~\cite{hosur2011vortex}, or gapless~\cite{konig2019nodal,qin2019nodal}, as determined by both symmetry and topological properties of the in-gap CdGM modes. 

The key inspiration behind our work is the Euler formula, allows us to express the vortex-dressed pairing function as real and imaginary parts, $\Delta_R= \Delta_0\cos\theta$ and $\Delta_I = \Delta_0\sin\theta$. Giving the sign structures of $\cos\theta$ and $\sin\theta$, we consider a geometric approximation of $\Delta({\bf r})$
\begin{equation}
    \Delta({\bf r})\approx \Delta_0 [\text{sgn}(x) +i \text{sgn}(y)],
    \label{eq:vortex discretization}
\end{equation}
which, as shown in Figs.~\ref{fig:Schematic diagram} (a) and (b), discretizes the original $U(1)$ vortex into a $\mathbb{Z}_4$ vortex while respecting the topological phase winding structure. We show in the Appendix {\bf B} that the bound-state spectrum of a $\mathbb{Z}_4$ vortex quantitatively matches that of a $U(1)$ vortex, which numerically proves the efficiency of this simplification. Physically, Eq.~\ref{eq:vortex discretization} implies that a vortex is topologically equivalent to the superposition of two perpendicular pairing domain walls, with each domain forming a $\pi$-phase Josephson junction. This raises the question: does a $\pi$-junction in the same superconductor exhibit similar topological physics? 

Without loss of generality, we consider a time-reversal-invariant $\pi$-junction along $\hat{x}$ direction by setting $\Delta({\bf r})=\Delta_0\text{sgn}(x)$, which consists of three parts: left domain, right domain, and the domain wall. Notably, electrons living around the domain wall will experience an effective odd-parity pairing order since $\Delta(-{\bf r})=-\Delta({\bf r})$, which locally goes around the no-go theorem of $s$-wave systems in Ref.~\cite{haim2016no-go}. Although each domain itself is topologically trivial, Andreev-bound states trapped by the $\pi$-junction can, in principle, develop 2D emergent interfacial topological physics. We dub this phenomenon {\it Josephson topology} (JT). 

This new notion naturally raises a series of interesting conceptual questions: (1) What are possible types of JT? (2) Is JT related to VLT? In the following, we will address the above key questions by classifying JT phases in different symmetry classes. Further exploiting a dimensional reduction strategy inspired by Eq.~\ref{eq:vortex discretization}, we will establish mapping relations between VLT and its ``parent" JT, which are generally dubbed {\it Josephson-vortex correspondence} (JVC).   

\subsection{$\mathbb{Z}_2$ Josephson topology}  
\label{sec:Z2_JT}

We start with a 3D superconductor with a space-group symmetry of $P1$ (No. 1), where the only lattice symmetry is the trivial identity operation $E$. The JT of a $\hat{x}$-oriented $\pi$-junction is characterized by three $\mathbb{Z}_2$ topological invariants $(\nu_0;\nu_y,\nu_z)$ due to the inherent time-reversal symmetry (TRS) $\Theta$~\cite{qi2009TSC,zhang2013TSC}. Specifically, the nontrivial strong index $\nu_0=1$ will enforce a pair of 1D helical Majorana modes circulating all ``edges" of the junction, as schematically shown in Fig.~\ref{fig:Schematic diagram} (c). When $\nu_0=0$, the value of weak index $\nu_{y}$ ($\nu_z$) further informs the existence of a Kramers pair of Majorana flat bands along the edge normal to $\hat{y}$ ($\hat{z}$) direction. 

Starting with a JT phase with $\nu_0=1$, we can recover a $\mathbb{Z}_4$ vortex by including a second $\hat{y}$-directional $\pi$-junction that explicitly spoils the TRS. The helical Majorana edge modes around the original $\pi$-junction will then experience a TRS-breaking mass domain, proportional to $\Delta_0 \text{sgn}(y)$. As a result, a single MZM will appear as a domain-wall zero mode at the crossing between two orthogonal $\pi$-junctions, i.e., the surface vortex center [as shown in Fig.~\ref{fig:Schematic diagram} (b)]. On the other hand, a $\pi$-junction with $\nu_0=0$ and a nontrivial weak index (e.g., $\nu_y=1$) will correspond to a trivial vortex line. This is because the degenerate edge flat bands enforced by $\nu_y=1$ are smoothly deformable into two pairs of helical Majorana modes at $k_y=0$ and $k_y=\pi$, respectively. After dimensional reduction, we now have two overlapping vortex MZMs, which, without additional protection, can always get hybridized and lose their Majorana nature. 

The above boundary matching process is generalizable to a vortex line along a general direction $\hat{n}$. Notably, the dimensional reduction will bridge a $\pi$-junction and a vortex line that are {\it parallel} to each other. Namely, the pairing-flipping domain plane $\Sigma$ of the junction must fulfill $\hat{n}\in \Sigma$, where $\hat{n}$ is the direction of the vortex tube. For example, a $\pi$-junction along $\hat{x}$ features a domain plane with a Miller index of $(100)$, which is parallel to any vortex line perpendicular to $\hat{x}$, e.g., one along $\hat{z}$. 

Knowing the $\mathbb{Z}_2$ index $\nu_0$ of the $\pi$-junction, we immediately arrive at
\begin{equation}
    \zeta = \nu_0,
    \label{eq:JVC_z2}
\end{equation}
where $\zeta\in\mathbb{Z}_2$ is the 1D class-D topological invariant for the parallel vortex line~\cite{kitaev2001unpaired}. When $\zeta=1$, the vortex line is topologically equivalent to a Kitaev Majorana chain with one exponentially localized MZM at each line end. We dub this fully gapped VLT phase a {\it Kitaev vortex}.

\subsection{Inversion symmetry and second-order Josephson topology}

Lattice symmetries compatible with junction geometry can enrich the topological structure of $\pi$ junctions. For example, the odd-parity nature of the $\pi$-junction requires the electrons and holes to transform oppositely under a spatial inversion operation ${\cal P}$. Specifically, we have ${\cal P} = \tau_z \otimes {\cal P}^{(e)}$, where ${\cal P}^{(e)}$ is the normal-state inversion operation and $\tau_{0,x,y,z}$ denote the Pauli matrices for the particle-hole degrees of freedom. Clearly, ${\cal P}$ anticommutes with the particle-hole symmetry $\Xi=\tau_x {\cal K}$, where ${\cal K}$ is the complex conjugation. This property allows us to define a $\mathbb{Z}_4$ inversion symmetry indicator $\kappa$~\cite{skurativska2020atomic,huang2021faithful} for the target $\pi$-junction Hamiltonian $H_{\pi J}$,
\begin{equation}
    \kappa \equiv  \sum_{\mathbf{K}_i\in \text{TRIM}} \frac{1}{2} [n^+_{{\bf K}_i}(H_{\pi J}) - n^+_{{\bf K}_i}(H_0)]\ \ (\text{mod } 4),
    \label{eq:inversion_indicator}
\end{equation}
where $n^+_{{\bf K}_i}(h)$ is the number of even-parity occupied BdG states of $h$ at a time-reversal-invariant momentum (TRIM) ${\bf K}_i$. Notably, $H_0({\bf k}) \equiv \tau_z \otimes \mathbb{I}_N$ is a constant reference Hamiltonian that shares the same inversion operator and matrix rank as $H_{\pi J}({\bf k})$. Since $\kappa\equiv \nu_0$ modulo 2, $\kappa=1$ or $3$ indicates a $\mathbb{Z}_2$ topological JT and hence a nontrivial Kitaev VLT as well. 

On the other hand, $\kappa=2$ implies a novel {\it second-order} JT phase with two inversion-related corner-localized Majorana Kramers pairs that are protected by both ${\cal P}$ and $\Theta$, as schematically shown in Fig.~\ref{fig:Schematic diagram} (d). This phenomenon resembles the boundary physics of bulk second-order topological superconductors with odd-parity Cooper pairings~\cite{khalaf2018higher,hsu2020inversion}. Similar to the case of a weak $\mathbb{Z}_2$ topological junction, the corner Majorana Kramer pairs here can be smoothly deformed into two pairs of helical Majorana edge modes through a fine-tuning of the edge physics. As a result, we expect the second-order JT phase to inform a trivial VLT after the dimensional reduction. 

\subsection{Mirror-Chern Josephson topology}
\label{sec:mirror-Chern JT}

Another crystalline symmetry of interest is the domain-flipping mirror reflection $M_x$, with $\{M_x, \Xi\}=0$ due to the $\pi$-junction geometry. We find $M_x = \tau_0 \otimes M_x^{(e)}$, where the normal-state mirror operation $(M_x^{(e)})^2=-1$ for spinful fermions. Notably, $M_x$ manifests as an on-site $\mathbb{Z}_2$ symmetry for junction-bound states. As a result, the $\pi$-junction Hamiltonian always admits a block-diagonal form with $H_{\pi J}(k_y,k_z) = H_+ \oplus H_-$, where $H_{\pm}$ carries an mirror index $m_x=\pm i$. One can thus define a mirror Chern number ${\cal C}_M = ({\cal C}_+ - {\cal C}_-)/2$, where ${\cal C}_\pm$ is the Chern number defined for each mirror sector~\cite{teo2008surface,hsieh2012topological}. The value of ${\cal C}_M \in \mathbb{Z}$ informs the number of $M_x$-protected helical edge modes around the $\pi$-junction. Notably, these helical modes are generally {\it not} Majorana modes since $\{M_x, \Xi\}=0$, and they may cross zero energy at some generic momentum $k_0$. Clearly, a junction with an odd mirror Chern number is $\mathbb{Z}_2$ topological, following $\nu_0 \equiv {\cal C}_M$ (mod 2).  

Updating a $\pi$-junction to a vortex breaks both $M_x$ and $\Theta$ simultaneously. A magnetic mirror symmetry $\Theta_M=M_x \Theta$, however, is preserved, which transforms the vortex Hamiltonian as $\Theta_M H_v(k_z) \Theta_M^{-1} = H_v(-k_z)$~\cite{chen2014magnetic}. Since $\Theta_M^2=1$, this magnetic mirror behaves as a 1D spinless TRS, thus promoting the symmetry of $H_v(k_z)$ to class BDI. Hence, the product of $\Theta_M$ and the particle-hole symmetry $\Xi$ generates a chiral symmetry ${\cal S}\equiv \Theta_M \Xi$ for the vortex line, with which a $\mathbb{Z}$-valued chiral winding number ${\cal W_S}$ can be defined to describe the VLT. In particular, there will be $|{\cal W_S}|$ decoupled MZMs, with the same ${\cal S}$ index, appearing at the surface vortex core. 

Remarkably, we find a simple JVC relation, with
\begin{equation}
     {\cal W_S} = {\cal C}_M.
    \label{eq:mirror_dimensional relation}
\end{equation}
Namely, the number of ${\cal S}$-stabilized vortex MZMs exactly equals that of $M_x$-protected helical edge modes of a $\pi$-junction in the same superconductor. When ${\cal C}_M=\pm 1$, we have $\nu_0 = 1$ and a $\mathbb{Z}_2$ JT. Then the system must host a single vortex MZM $({\cal W_S}=\pm 1)$ as concluded in Sec.~\ref{sec:Z2_JT}. The situation for ${\cal C}_M = \pm 2$ is, however, not as intuitive. The $\pi$-junction now features two pairs of helical edge modes, which, upon the dimensional reduction, generate two vortex zero modes that normally would hybridize with each other. However, as discussed in Appendix {\bf A}, an effective boundary theory analysis reveals that these zero modes must carry the same chiral symmetry index and remain decoupled. From the bulk-boundary correspondence, we thus arrive at ${\cal W_S} = {\cal C}_M=\pm 2$. Generalization to JT with a larger mirror Chern number can be proved similarly and is also provided in Appendix {\bf A}. 

\subsection{Rotation symmetries and symmetry mismatch}
\label{sec:rotation sym mismatch}

Finally, let us focus on the effect of rotation symmetries. We consider a superconductor candidate with a space group $P4$ (No. 75) generated by a 4-fold rotation symmetry $C_{4z}$ around $\hat{z}$. If the vortex line orientation $\hat{n}\neq \hat{z}$, its symmetry group ${\cal G}_v$ contains only the PHS, with which the Kitaev vortex phase is the only possible VLT following the tenfold-way classification~\cite{kitaev2009periodic}. In this case, we can unambiguously conclude the existence of Kitaev vortex by checking the $\mathbb{Z}_2$ JT of any $\pi$-junction parallel to the vortex tube. In other words, the JVC relation is expected to be one-to-one.      

When $\hat{n}=\hat{z}$, the CdGM modes inside the vortex are categorized into four distinct 1D irreducible representations (irreps) of $C_{4z}$, characterized by the $\hat{z}$-component angular momentum $J_z \in \{0,1,2,3\}$. Specifically, Kitaev vortex topology can arise from a band inversion of CdGM states labeled by either $J_{z}=0$ or $2$. Each scenario will contribute to a $\mathbb{Z}_2$ invariant, denoted as $\zeta_{0,2}$ respectively. Meanwhile, $J_z=1,3$ sectors must form a pair of PHS-related bands in the spectrum. They can cross to form a pair of $C_{4z}$-protected point nodes, which is dubbed a {\it nodal vortex} phase and is characterized by a $\mathbb{Z}$ topological charge ${\cal Q}_1$. Therefore, $C_{4z}$ enriches the VLT to a $(\mathbb{Z}_2)^2\times \mathbb{Z}$ topological class~\cite{hu2023topological}.  

However, a $\pi$-junction parallel to the above vortex line only respects a two-fold $C_{2z}$ due to its geometric structure. The JT is captured by the strong $\mathbb{Z}_2$ index $\nu_0$. Therefore, the junction-vortex symmetry mismatch here results in a JVC mapping that is not one-to-one. For instance, either $\zeta_0=1$ or $\zeta_2=1$ can correspond to a non-trivial $\nu_0=1$, while $\zeta_0=\zeta_2=1$ implies a trivial JT. In addition, $\nu_0=1$ can also arise in a system with an odd ${\cal Q}_1$. As a result, we find that
\begin{equation}
    \nu_0 \equiv \zeta_0 + \zeta_2 + {\cal Q}_1 \ \text{(mod 2)}.
    \label{eq:JVC_C4}
\end{equation}
Similar JVC relations can be concluded for systems with a two-fold, three-fold, or six-fold rotation symmetry. In general, when the vortex line is rotational invariant, the JVC does not definitively inform the explicit type of VLT. Notably, when we tilt the vortex line to misalign with the rotation axis, both $\zeta_2$ and ${\cal Q}_1$ become ill-defined, with which JVC in Eq.~\ref{eq:JVC_C4} can always be reduced to the exact mapping relation in Eq.~\ref{eq:JVC_z2}.   

\begin{table}[t]
\label{table:JVC relations}
\begin{tabular}{c|c|c}
\hline
Symmetry ${\cal G}$ & Josephson Topology & Vortex Topology                   \\ \hline
$E$           & $\mathbb{Z}_2$ & $\mathbb{Z}_2$       \\
$\cal{P}$ & $\mathbb{Z}_4$ & $\mathbb{Z}_2$       \\
$M$  & $\mathbb{Z}$   & $\mathbb{Z}$          \\
$C_2$         & $\mathbb{Z}_2$ & $\mathbb{Z}_2\times\mathbb{Z}_2$       \\
$C_3$         & $\mathbb{Z}_2$ & $\mathbb{Z}_2\times\mathbb{Z}$     \\
$C_4$         & $\mathbb{Z}_2$ & $(\mathbb{Z}_2)^2\times\mathbb{Z}$  \\
$C_6$         & $\mathbb{Z}_2$ & $(\mathbb{Z}_2)^2\times(\mathbb{Z})^2$
\end{tabular}
\caption{Classification of Josephson topology and vortex-line topology when the bulk superconductor features a crystalline symmetry ${\cal G}$. Note that ${\cal G}$ may be reduced in the presence of a $\pi$-junction or vortex geometry. }
\end{table}

\section{Topological critical points of $\pi$ junctions}
\label{sec:Criti Corresp}

We now turn to a key implication of the Josephson–vortex correspondence (JVC): the evolution of their topological character with chemical potential $\mu$. As pointed out in Ref.~\cite{hosur2011vortex}, vortex lines can undergo topological phase transitions as $\mu$ is tuned, marking the onset of Majorana zero modes at critical values $\mu_c$. If JVC generally holds, we naturally expect the topological physics of the $\pi$-junction to exhibit similar $\mu$-dependent behavior. In particular, the $\pi$-junction and the vortex line should undergo {\it simultaneous} topological transitions at matching values of $\mu_c$. The focus of this section is to analytically derive the relations between JT and VLT critical points.

We now consider 3D phase space $({\bf k}_\parallel, \mu)$, in which the $\pi$-junction Hamiltonian is updated to ${\cal H}({\bf k}_\parallel, \mu)$ and ${\bf k}_\parallel=(k_y,k_z)$ for a $\pi$-junction along [100]. Unlike physical crystal momentum, $\mu$ does not respond to the operation of either TRS $\Theta$ or PHS $\Xi$, with $\Theta {\cal H}({\bf k}_\parallel, \mu) \Theta^{-1} = {\cal H}(-{\bf k}_\parallel, \mu)$ and $\Xi {\cal H}({\bf k}_\parallel, \mu) \Xi^{-1} = -{\cal H}(-{\bf k}_\parallel, \mu)$. Therefore, if a JT gap closes at $({\bf k}_c, \mu_c^{(J)})$ with ${\bf k}_c\notin$TRIM, there must exist a second JT critical point at $(-{\bf k}_c, \mu_c^{(J)})$. On the other hand, such a ``fermion doubling" of critical points can be avoided if and only if ${\bf k}_c$ is a TRIM. Hence, JT transitions generally fall into two categories: (a) a single on-TRIM JT transition and (b) a pair of off-TRIM JT transitions at the same $\mu_c^{(J)}$.

Meanwhile, in the low-energy sector, a JT critical point manifests as a 4-fold-degenerate massless Dirac fermion living in the 3D phase space spanned by ${\bf k}_\parallel$ and $\mu$. Similarly, a VLT transition at some $\mu=\mu_c^{(v)}$ can be viewed as a 2D doubly degenerate Dirac fermion in the $(k_z,\mu)$ space. Motivated by these observations, our strategy to bridge between JT and VLT criticalities again exploits the idea of dimensional reduction, further guided by an intuitive physical picture:
\begin{itemize}
    \item A VLT transition is the domain-wall fermion of a JT critical point.   
\end{itemize}
In the following, we will establish effective theories for both on-TRIM and off-TRIM transitions and further explore their connection to VLT criticalities. 

\begin{figure}[tb!]
\includegraphics[width=1.0\columnwidth]{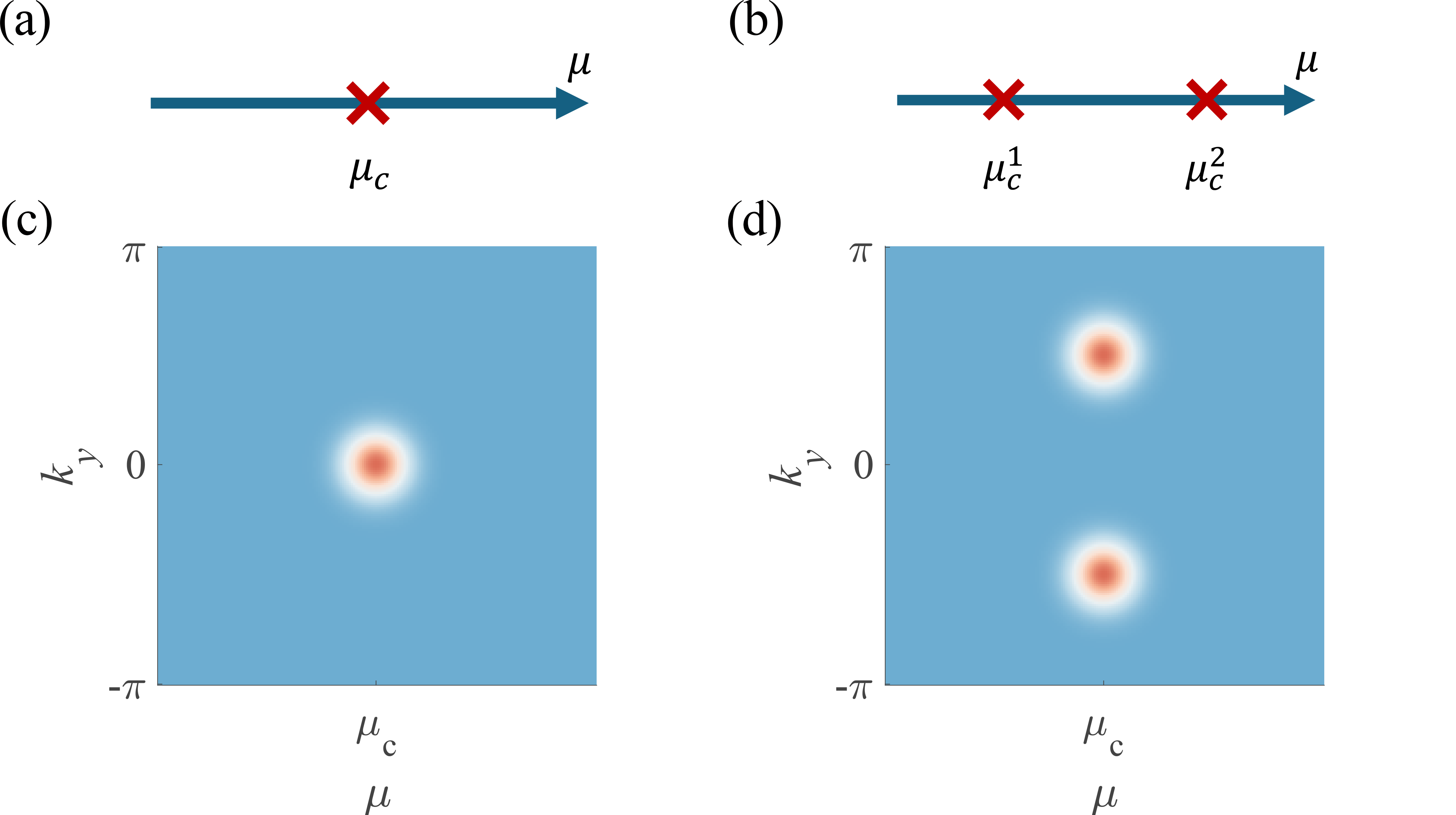}
    \caption{Schematics of mapping relations between VLT and JT transition points. 
    The crossings in panels (a) and (b) denote the 1D VLT transition points as a function of the chemical potential $\mu$, whereas the red spots in panels (c) and (d) denote the 2D JT transitions in the ($\mu$, $k_y$) parameter space. The on-TRIM JT transition shown in (c) will correspond to a VLT transition around the same $\mu_c$, as shown in (a). Meanwhile, off-TRIM JT transitions must come in pairs, as shown in (d). If appearing at opposite $k_y$, they will correspond to two VLT transitions that split along $\mu$ axis, as shown in (b).} 
    \label{fig:effective_model_TI}
\end{figure}

\subsection{On-TRIM JT transition}
\label{sec:on-TRIM JT} 

In the low-energy sector, a single on-TRIM JT criticality at $({\bf k}_c,\mu_c^{(J)})$ is a Dirac fermion that respects both TRS and PHS. We thus consider an effective Hamiltonian, 
\begin{equation}
    h_{s,0} ({\bf k}_\parallel, \mu) = v_1 k_z \gamma_1 + v_2 k_y \gamma_2  + v_3 \mu \gamma_3, 
\end{equation}
where $k_{y,z}$ and $\mu$ are defined relative to $({\bf k}_c,\mu_c^{(J)})$. Without loss of generality, we choose the convention of Dirac $\gamma$ matrices as $\gamma_1 = \tau_x \otimes s_z,\ \gamma_2 = \tau_y \otimes s_0,\ \gamma_3 = \tau_z \otimes \sigma_0, \gamma_4 = \tau_x \otimes s_x,\ \gamma_5 = \tau_x \otimes s_y$, where $\{ \gamma_i, \gamma_j\} = 2\delta_{ij}$ and other $\gamma$ matrices can be generated via $\gamma_{ij}=[\gamma_i,\gamma_j]/(2i)$. Under this Dirac basis, the corresponding PHS and TRS are given by $\Xi = \tau_x \otimes s_0 {\cal K}$ and $\Theta = i \tau_0 \otimes s_y {\cal K}$, respectively. 

We now add the $\hat{y}$ pairing domain to arrive at the vortex geometry, completing the dimensional reduction. As a mass term to $h_s$, such a pairing term will need to anti-commute with $\gamma_{1,2,3}$, as well as both $\Xi$ (to respect PHS) and $\Theta$ (to break TRS). It is straightforward to show that the only compatible term is $h_{s,1} = \tilde{\Delta} \text{sgn}(y) \gamma_5$, where $\tilde{\Delta}$ represents the projected pairing amplitude onto the low-energy basis. Taking $v_1k_z\gamma_1+v_3\mu\gamma_3$ as a perturbation, a pair of domain-wall bound states are obtained by solving the zero-mode equation,
\begin{eqnarray}
    [\partial_y - \frac{\tilde{\Delta} }{v_2} \text{sgn}(y) \gamma_{25}] \psi(y) = 0.
\end{eqnarray}
We consider an ansatz wavefunction $\psi(y)={\cal N}f(y)\xi_\alpha$ with $\alpha=\pm$, where the spinor part satisfies $\gamma_{25} \xi_\pm =\pm \xi_\pm$. The spatial part $f(y)$ shall be localized around $y=0$, and ${\cal N}$ is the normalization factor. Assuming $\tilde{\Delta},v_2>0$, we find two normalizable zero-mode solutions, $\psi_1(y) = {\cal N} f(y)[1,i,0,0]^T$ and $\psi_2(y) = {\cal N} f(y)[0,0, 1, -i]^T$, where $f(y)=\text{exp}[-\tilde{\Delta}|y|/v_2]$. Upon a projection onto the zero-mode basis, spanned by the Pauli matrix $\rho_{x,y,z}$, the dispersion of the domain-wall modes is found to be
\begin{equation}
    \tilde{h}_s(k_z,\mu) = v_1 k_z \rho_x + v_3 \mu \rho_z.  
\end{equation}
As a 2D massless Dirac fermion, $\tilde{h}_s$ describes a VLT criticality when $k_z=\mu=0$. For the on-TRIM criticalities, we thus have
\begin{equation}
    \mu_c^{(v)} = \mu_c^{(J)},
    \label{eq:on-TRIM critical}
\end{equation}
up to some higher-order corrections. This quantitative relation between vortex and Josephson topological critical points justifies our initial expectation. 

\subsection{Off-TRIM JT transitions}
\label{sec:off-TRIM JT}

Now let us turn to a pair of off-TRIM JT critical points at $(\pm {\bf k}_c,\mu_c)$. Since $k_z$ remains a good quantum number during dimensional reduction, we can classify the JT critical pair into two distinct cases based on their $k_z$ labels. 

\subsubsection{Opposite $k_z$}

When two JT gap closures happen simultaneously at opposite $k_z\notin\{0,\pi\}$, they do not change $\nu_0$. Instead, we expect them to change the BdG mirror Chern number ${\cal C}_M$ by $\pm 2$, when the system respects $M_x$. Since $k_z$ is conserved during dimensional reduction, each JT critical point will independently contribute to VLT at $k_z=\pm k_0$, respectively. Notably, such finite-$k_z$ vortex gap closures are known to change the chiral winding number ${\cal W_S}$ by $\pm 2$ while leaving the Kitaev $\mathbb{Z}_2$ index invariant~\cite{chen2014magnetic}, which is congruent with the ${\cal W_S}$-${\cal C}_M$ relation in Eq.~\ref{eq:mirror_dimensional relation}. Following the discussion in Sec.~\ref{sec:on-TRIM JT}, we conclude that the mapping relation in Eq.~\ref{eq:on-TRIM critical} should still hold. 

\subsubsection{Same $k_z$}
\label{sec:finite-ky transition}

Meanwhile, the $\pi$-junction can feature two critical points at opposite $k_y$, yet the same $k_z$. For example, ${\bf k}_c=(k_0,0)$. It is convenient to assign a new valley index to the low-energy Dirac fermions associated with the JT gap closures at $k_y=\pm k_0$, respectively. We employ the Pauli matrix $\sigma_i$ to denote the valley degree of freedom. Then an effective Hamiltonian for the critical pair is given by
\begin{equation}
    h_{p,0} ({\bf k}_\parallel, \mu) = \sigma_0 \otimes h_{s,0}({\bf k}_\parallel, \mu),
    \label{hdf2}
\end{equation}
which is compatible with the valley-flipping TRS $\Theta=i\sigma_x \otimes \tau_0 \otimes s_y\mathcal{K}$ and PHS $\Xi=\sigma_x \otimes \tau_x \otimes s_0\mathcal{K}$.  

The next step is to complete the dimensional reduction. Under the low-energy Dirac basis, the $\hat{y}$ pairing domain generally takes the form of $h_{p,1} = \tilde{\Delta}\text{sgn}(y)\Omega$. Notably, the $\hat{y}$ translational symmetry breaking of $h_{p,1}$ arises from the domain-wall structure of the sign function $\text{sgn}(y)$, instead of the projected pairing matrix $\Omega$. Physically, this implies $\Omega$ to be {\it intra-valley} and hence $\Omega\sim \sigma_{0,z}$. Following both symmetries constraints and Clifford algebra, it is quite straightforward to show that the only reasonable choice is
\begin{equation}
    h_{p,1} = \tilde{\Delta} \text{sgn}(y) \sigma_0 \otimes \tau_x \otimes s_y.
\end{equation}
Remarkably, $h_p = h_{p,0} + h_{p,1}$ exactly consists of two identical copies of $h_{s}=h_{s,0}+h_{s,1}$. Hence, the corresponding VLT transition is given by  
\begin{equation}
    \tilde{h}_p(k_z,\mu) = \sigma_0 \otimes \tilde{h}_s(k_z,\mu),
    \label{eq-sec3VLT}
\end{equation}
which describes two independent CdGM gap closings at the same $\mu_c^{(J)}$.   

From a different perspective, $\tilde{h}_p(k_z,\mu)$ is a 4-fold-degenerate 2D Dirac fermion that is generally unstable. Indeed, we can perturb $\tilde{h}_p(k_z,\mu)$ with $\delta \sigma_x \otimes \rho_z$ and split its high degeneracy without (i) breaking any of the key symmetries such as PHS. The CdGM states then disperse following $E(k_z,\mu)=\pm\sqrt{k_z^2+(\mu\pm \delta)^2}$, where we find two VLT transitions at
\begin{eqnarray}
    \mu_{c,+/-}^{(v)} = \mu_c^{(J)} \pm  \delta. 
    \label{eq:off-TRIM critical}
\end{eqnarray}
Here, the splitting $\delta$ is a material-dependent parameter, which should be small compared to other energy scales such as Fermi energy and bandwidth.

\subsection{Effect of rotation symmetries}
\label{sec:critical-rotation}

When the $\pi$-junction and the vortex line both respect a two-fold rotation symmetry $C_{2z}$, we could make a stronger statement regarding the nature of the above $\mu$-splitting VLT transitions from an off-TRIM JT critical pair. To see this, we can simply repeat the dimensional reduction procedure in Sec.~\ref{sec:finite-ky transition} by including an additional $C_{2z}$ operation. 

Starting from the JT effective theory in Eq.~\ref{hdf2}, we note that a proper choice of $C_{2z}$ for the $\pi$-junction is given by $C_{2z}=i\sigma_x\otimes \tau_z\otimes s_y$, which not only exchanges the valley indices but also fulfills $[C_{2z},\Theta]=\{C_{2z},\Xi\}=0$. Crucially, the $\hat{y}$-direction domain-wall term $h_{p,1}$ also commutes with this representation of $C_{2z}$. After the dimensional reduction, we arrive at the VLT transitions described by Eq.~\ref{eq-sec3VLT}. Upon a similar zero-mode projection to that in Sec.~\ref{sec:on-TRIM JT}, we find that the two-fold rotation for the vortex modes is $C_{2z}=\sigma_x \otimes \rho_0$. Since the vortex phase winding introduces a $\pi$-twist in the fermionic boundary condition, we have dropped a factor of $i$ in the representation of $C_{2z}$ to ensure $(C_{2z})^2=1$ and $[C_{2z}, \Xi]=0$. Together with Eq.~\ref{eq:off-TRIM critical}, we find that each of the VLT transitions manifests as the crossing of two CdGM bands with the same $C_{2z}$ index, one with $C_{2z}=1$ and another with $C_{2z}=-1$. 

If the vortex line only respects $C_{2z}$ (but not a higher-fold rotation), a VLT transition with $C_{2z}=+1$ or $-1$ triggers a change of $\mathbb{Z}_2$ VLT topology in the $J_z=0$ or $1$ sector, respectively. In other words, with $C_{2z}$, the off-TRIM JT critical pair will correspond to {\it two subsequent Kitaev vortex transitions}, one in the $J_z=0$ sector and another in $J_z=1$. Meanwhile,  when the vortex line is $C_{4z}$-symmetric, the VLT transition with $C_{2z}=1$ could occur in $J_z=0$ or $2$ sector, both of which describe a Kitaev vortex transition. Meanwhile, the VLT transition with $C_{2z}=-1$ is a nodal vortex transition with $J_z=\pm 1$. Therefore, we arrive at a remarkable conclusion that will play a crucial role in our later discussions:
\begin{itemize}
    \item An off-TRIM JT critical pair necessarily implies a Kitaev VLT transition and a consecutive nodal VLT transition for a $C_{4z}$-symmetric vortex line.
\end{itemize}
Finally, when the vortex line respects $C_{6z}$, the VLT with $C_{2z}=1$ can arise from either $J_z=0$ (a Kitaev transition) or $J_z=\pm 2$ (a nodal transition) sector. Similarly, the VLT with $C_{2z}=-1$ can arise from either $J_z=3$ (a Kitaev transition) or $J_z=\pm 1$ (a nodal transition) sector. In this case, however, we are unable to precisely identify the nature of VLT transitions based on the knowledge of JT critical points.

\subsection{Criticality relations between JT and VLT}
\label{sec:criticality summary}

Let us now summarize our findings. The analytical results in Secs.~\ref{sec:on-TRIM JT} and~\ref{sec:off-TRIM JT} have together implied a {\it criticality conservation} relation that generally holds:
\begin{itemize}
    \item Every transition of the Josephson topology corresponds to a vortex topological transition.   
\end{itemize}
Besides, we can deduce the nature of vortex criticality based on the momentum-space locations of JT phase transitions:    
\begin{enumerate}
    \item[(a)] on-TRIM: indicating one Kitaev or nodal vortex phase transition at around $\mu_c$. 
    \item[(b)] off-TRIM \& opposite $k_z$: indicating two simultaneous vortex gap closures at around $\mu_c$, which together change the chiral winding number by $\pm 2$.
    \item[(c)] off-TRIM \& same $k_z$: indicating two subsequent vortex phase transitions around $\mu_c$, each changing either Kitaev or nodal vortex topology. The nature of vortex transitions can be further predicted when the system features $C_{2z}$ or $C_{4z}$, following Sec.~\ref{sec:critical-rotation}. 
\end{enumerate}
The above {\it criticality correspondence} relations provide concrete guidelines to quantitatively reconstruct the vortex topological phase diagram without performing large-scale vortex simulations.

\section{Josephson-vortex relations in Effective Models} 
\label{sec:effective_models}

In this section, we will numerically investigate a set of effective models for 3D conventional superconductors that exhibit diverse manifestations of JT and VLT physics. For each model, we begin by constructing the topological phase diagram of the $\pi$-junction through comprehensive simulations. Leveraging the JVC relations, we then infer the corresponding vortex-line topological phase diagram. Finally, we directly compute the vortex phase diagram and compare it against the JVC-based prediction, thereby benchmarking the accuracy and robustness of our correspondence framework.

\subsection{Doped topological insulator}
\label{sec:doped TI}

\begin{figure}[tb!]
\includegraphics[width=1.0\columnwidth]{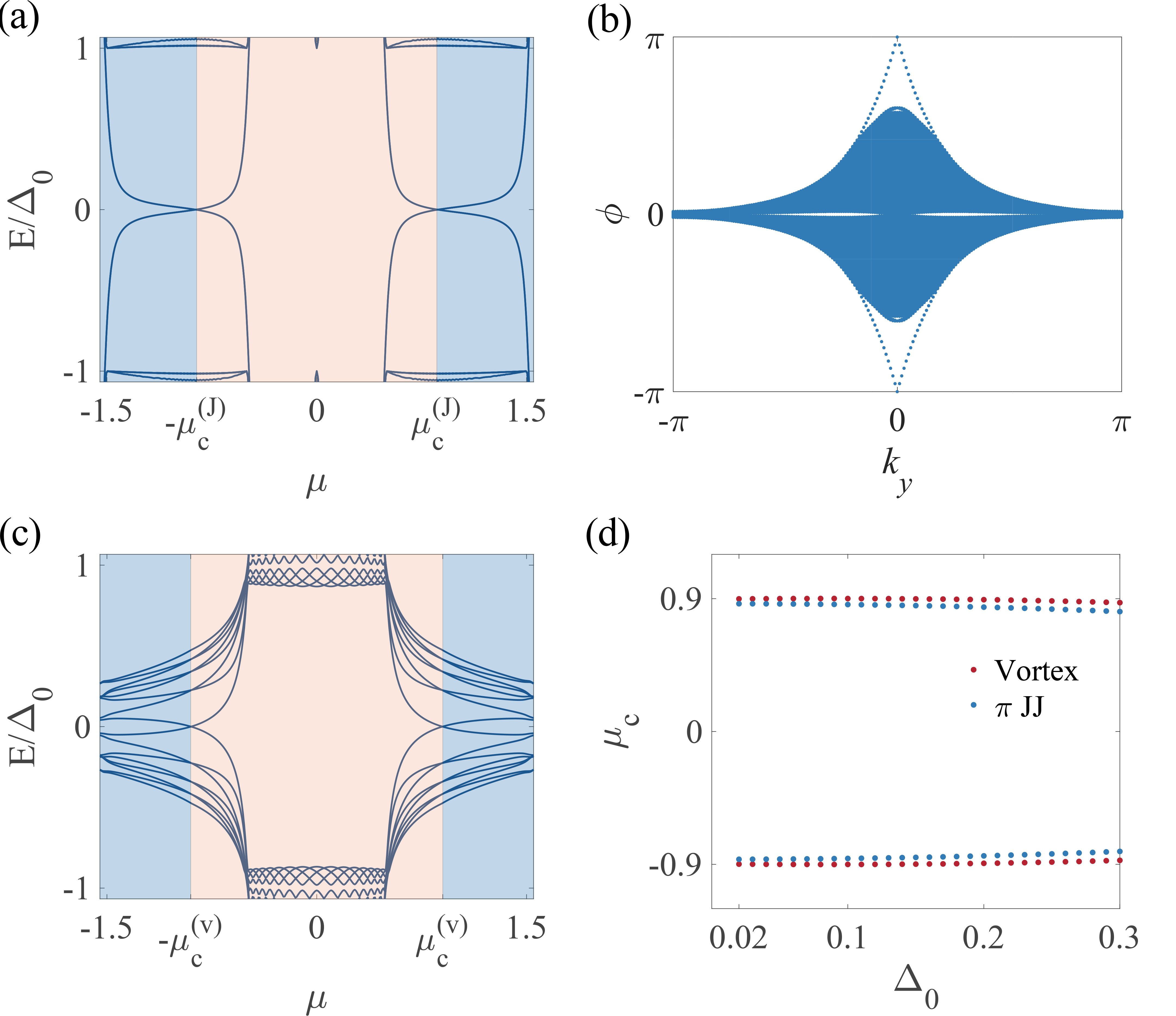}
    \caption{JT and VLT phase diagrams of a superconducting TI. (a) $\mu$-dependent Andreev spectrum in a (100)-directional $\pi$-junction at $k_{y,z} = 0$. Andreev band crossings at $E=0$ manifest as Josephson topological phase transitions. Regions in blue are $\mathbb{Z}_2$ trivial, while the pink region is $\mathbb{Z}_2$ topological. (b) Wilson loop characterization of Andreev bound states at $\mu=0.7$, where the helical winding pattern suggests a nontrivial JT $\mathbb{Z}_2$ index $\nu_0=1$. (c) $\mu$-dependent CdGM spectrum of a (001)-directional vortex line at $k_z=0$. The pink region shows a topological Kitaev vortex phase with end MZMs. (d) Evolution of $\mu_c^{(J)}$ and $\mu_c^{(v)}$ as a function of pairing order $\Delta_0$, which confirms the $\pi$-junction and vortex line are simultaneously topological.} 
    \label{fig:effective_model_TI}
\end{figure}

We start with the standard Fu-Kane system, i.e., a doped 3D class-AII topological insulator (TI) with conventional $s$-wave Cooper pairing~\cite{fu2008vortex}. The normal-state Hamiltonian for such a TI can be minimally described by a massive Dirac model on a cubic lattice with $h_{TI}({\bf k}) = \sum_{i=1}^5 d_i({\bf k}) \gamma_i$. Here, we have defined ${\bf d}=(v\sin k_x, v\sin k_y, v_z\sin k_z, 0, m({\bf k}))$ with the mass term $m({\bf k}) = m_0+m_1(\cos k_x + \cos k_y) + m_2 \cos k_z$. Our choice of the Dirac $\gamma$ matrices are 
\begin{eqnarray}
    && \gamma_1 = \sigma_x \otimes s_x, \gamma_2 = \sigma_x \otimes s_y, \gamma_3 = \sigma_x \otimes s_z, \nonumber \\
    && \gamma_4 = \sigma_y \otimes s_0, \gamma_5 = \sigma_z \otimes s_0.
    \label{eq:gamma matrices}
\end{eqnarray}
We denote $\sigma_\alpha$ and $s_\alpha$ as the Pauli matrices for the orbital and spin degrees of freedom, respectively. Notably, $h_{TI}$ features the spatial inversion symmetry ${\cal P}=\gamma_5$, the TRS $\Theta= -i \gamma_{13} {\cal K}$, and a $4$-fold around-$\hat{z}$ rotation symmetry $C_{4z} = \text{exp}(i J_z \pi/2)$ with the generator $J_z=\gamma_{12}/2$. For our purpose, we choose the model parameter set to be $m_0=2.5$ and $v= v_z=-m_1 = -m_2 = 1$, with which the normal state at $\mu=0$ achieves a strong $\mathbb{Z}_2$ TI phase with a topological band inversion at $\Gamma$ point.  

The BdG Hamiltonian for the doped TI system directly follows the formalism in Eq.~\ref{eq:BdG_form}, where a 
bulk conventional $s$-wave pairing is assumed. We further construct a thick-slab Hamiltonian ${\cal H}_x(k_y, k_z)$ for this BdG system, which consists of 300 unit cells along $\hat{x}$ direction and respects the periodic boundary conditions along other directions. The $\pi$-junction geometry is implemented by taking the left half slab to carry a positive pairing amplitude $\Delta_0 = 0.03$, while the electrons on the other half slab are negatively paired with $-\Delta_0$.

The JT in this setup can be intuitively understood by taking two limits: $\mu=0$ and $\mu=\infty$. With zero doping, the Fermi level crosses only the Dirac surface states, and an effective surface theory will suffice. Specifically, on either $(010)$ or $(001)$ surface, the surface Dirac fermion will develop a pairing-induced mass that flips its sign exactly at the $\pi$-junction, thus leading to a pair of helical Majorana modes bound to the domain~\cite{fu2008vortex,liu2011helical}. To numerically check the JT, we have calculated the inversion eigenvalues of all negative-energy states at $k_{y,z}=0,\pi$ for the $\pi$-junction slab with $\mu=0$. Following Eq.~\ref{eq:inversion_indicator}, we find the BdG inversion-symmetry indicator to be $\kappa=1$. Meanwhile, we also exploit the Wilson-loop technique to track the evolution of 1D Berry phase $\lambda_z(k_y)$ as a function of $k_y$. As shown in Fig.~\ref{fig:effective_model_TI} (b), the Wilson loop spectrum displays a gapless helical winding pattern with robust Kramers degeneracies at $k_y=0$ and $\pi$. Both topological invariant calculations have thus established an emergent 2D $\mathbb{Z}_2$ TSC phase at the pairing domain, agreeing with this intuitive boundary picture. 

On the other hand, the infinite $\mu$ limit should always yield a trivial JT, which we have numerically confirmed for a reasonably large $\mu$ of our setup. Consequently, as we increase $\mu$ from $0$ to $\infty$, there must exist a critical doping $\mu_c^{(J)}$ where JT changes from $\nu_0=1$ to $\nu_0=0$, further manifesting as a zero-energy gap closure at one of the TRIMs. In Fig.~\ref{fig:effective_model_TI} (a), we numerically plot the dispersion relation of Andreev bound states at $\tilde{\Gamma}$, i.e., $k_{y,z}=0$, as a function of $\mu$. Indeed, a pair of JT transitions at $\pm \mu_c^{(J)}$ are found, with
\begin{eqnarray}
    \mu_c^{(J)} \approx 0.86.
\end{eqnarray}
Notably, the Fermi level at the JT transition is outside the normal-state TI gap $[-0.5,0.5]$, where the Dirac surface physics is no longer valid. 

Combining the above $\pi$-junction results and the criticality relation in Sec.~\ref{sec:criticality summary}, we expect the $\mu$-dependent {\it vortex} topological phase diagram for the doped TI to feature two VLT critical points at $\pm \mu_c^{(v)}$, within which there should exist a topological nontrivial phase for the CdGM states. This prediction holds for vortex lines parallel to the $(100)$ plane, e.g., the ones along $\hat{z}$. Since the system respects $C_4$ symmetry, the VLT phase can be either a gapped $\mathbb{Z}_2$ Kitaev vortex phase or a gapless nodal vortex phase, the distinction of which is beyond JVC. When $|\mu|>\mu_c^{(v)}$, we expect a trivial vortex phase.

As a confirmation, we carry out a large-scale vortex simulation for the doped TI by regularizing the BdG Hamiltonian on a 200 by 200 lattice with $\Delta_0 = 0.03$ and including an in-plane phase winding for the pairing order. In Fig.~\ref{fig:effective_model_TI} (c), we calculate the CdGM spectrum at $k_z=0$ as a function of $\mu$.  The vortex line is further found to be a Kitaev vortex phase with one MZM at each end for $|\mu|<\mu_c^{(v)}$ with $\mu_c^{(v)}=0.9$, which agrees well with Ref.~\cite{hosur2011vortex}. Remarkably, we find 
\begin{equation}
    \mu_c^{(v)} = \mu_c^{(J)} + {\cal O}(\Delta_0). 
\end{equation}
directly following the criticality correspondence that we have analytically constructed. In Fig.~\ref{fig:effective_model_TI}(d), we numerically track how $\mu_c^{(v)}$ and $\mu_c^{(J)}$ evolve as a function of $\Delta_0$. We indeed find that both $\mu_c^{(v)}$ and $\mu_c^{(J)}$ will gradually converge as we decrease $\Delta_0$, just as expected.

\subsection{Superconducting Dirac semimetal}
\label{sec:doped DSM}

As a gapless cousin for class-AII TIs, the 3D Dirac semimetals (DSM) is known for hosting symmetry-protected four-fold band crossings~\cite{young2012DSM,wang2012DSM}, whose low-energy kinetics resemble that of a massless Dirac fermion. The Abrikosov vortex line of a DSM, if respecting $C_{4z}$, is known to be a nodal phase~\cite{konig2019nodal,qin2019nodal}. So what is the corresponding JT for DSMs? 

We start by noting that a minimal lattice model for DSM is given by 
\begin{equation}
   h_{DSM}({\bf k}) \equiv h_{TI}({\bf k})|_{v_z=0},
\end{equation}
where the generator of $C_{4z}$ operation is updated to $J_z=\gamma_{34}+\frac{1}{2}\gamma_{12}$. Keeping other parameters unchanged, $h_{DSM}$ features a pair of $C_4$-protected Dirac nodes at $K_\pm = (0,0,\pm \arccos(1/2))$. Coupling the DSM with an $\hat{x}$-directional pairing domain, it is easy to see that the JT phase boundaries, i.e., where the Andreev spectrum closes the gap at $k_z=0$ and $\pi$, are irrelevant to the value of $v_z$. So, the JT phase diagram of $h_{DSM}$ would be quantitatively the same as Fig.~\ref{fig:effective_model_TI} (a), with critical chemical potentials at $\mu \approx \pm 0.86$. 

When $\mu$ is infinite, we always have a trivial JT. When $\mu$ vanishes, a four-fold degenerate BdG Dirac node will appear as the massless domain-wall mode for each normal-state Dirac fermion. Notably, Ref.~\cite{zhang2023gapless} has also reported such gapless Andreev bound states for a DSM-based $\pi$ junction. However, we find that the nodal nature of these Andreev bound states is, in fact, {\it accidental} and lacks symmetry protection. 

To see this, we consider a 2D effective theory to describe the nodal $\pi$-junction as 
\begin{equation}
    h_\text{eff}^{(0)}(k_y, k_z) = v_y k_y \gamma_3 + [M_0 - M_1 (k_y^2 + k_z^2)] \gamma_5,
\end{equation}
where the BdG nodes live at $(0,0,\pm \sqrt{M_0/M_1})$ for $M_0M_1>0$. Notably, we have adopted the definitions of $\gamma$ matrices in Eq.~\ref{eq:gamma matrices} to formulate the matrix algebra of $h_\text{eff}$, while the physical meaning of Pauli matrices here is fundamentally different from that in $H_{TI}$ or $H_{DSM}$. The important symmetries in the low-energy subspace include the particle-hole symmetry $\Xi = \gamma_{45}{\cal K}$, the TRS $\Theta = i\gamma_{13}{\cal K}$, the inversion symmetry ${\cal P} = \gamma_5$. Crucially, the $\pi$-junction geometry breaks the bulk symmetry $C_{4z}$ down to a $C_{2z}=i\gamma_{23}$. The odd-parity nature of the junction is encoded in the anti-commutation relations $\{\Xi, C_{2z}\} = \{\Xi, {\cal P}\} = 0$. We then find that the following perturbation
\begin{eqnarray}
    h_\text{eff}^{(1)} \sim k_z \gamma_4,
\end{eqnarray}
will gap out $h_\text{eff}^{(0)}$ while respecting $\Xi$, $\Theta$, $C_{2z}$, and ${\cal P}$. We highlight that $h_\text{eff} = h_\text{eff}^{(0)} + h_\text{eff}^{(1)}$ is exactly a 2D class DIII TSC in the continuum limit. Since $h_\text{eff}^{(1)}$ is likely small and pertubative in practice, we thus conclude that $\pi$-junction of a weakly-doped DSM will generally host a ``quasi-nodal" $\mathbb{Z}_2$ JT with $\nu_0=1$.

\subsection{Topological iron-based superconductor}
\label{sec:TFeSCs}

\begin{figure}[tb!]
    \includegraphics[width=1.0\columnwidth]{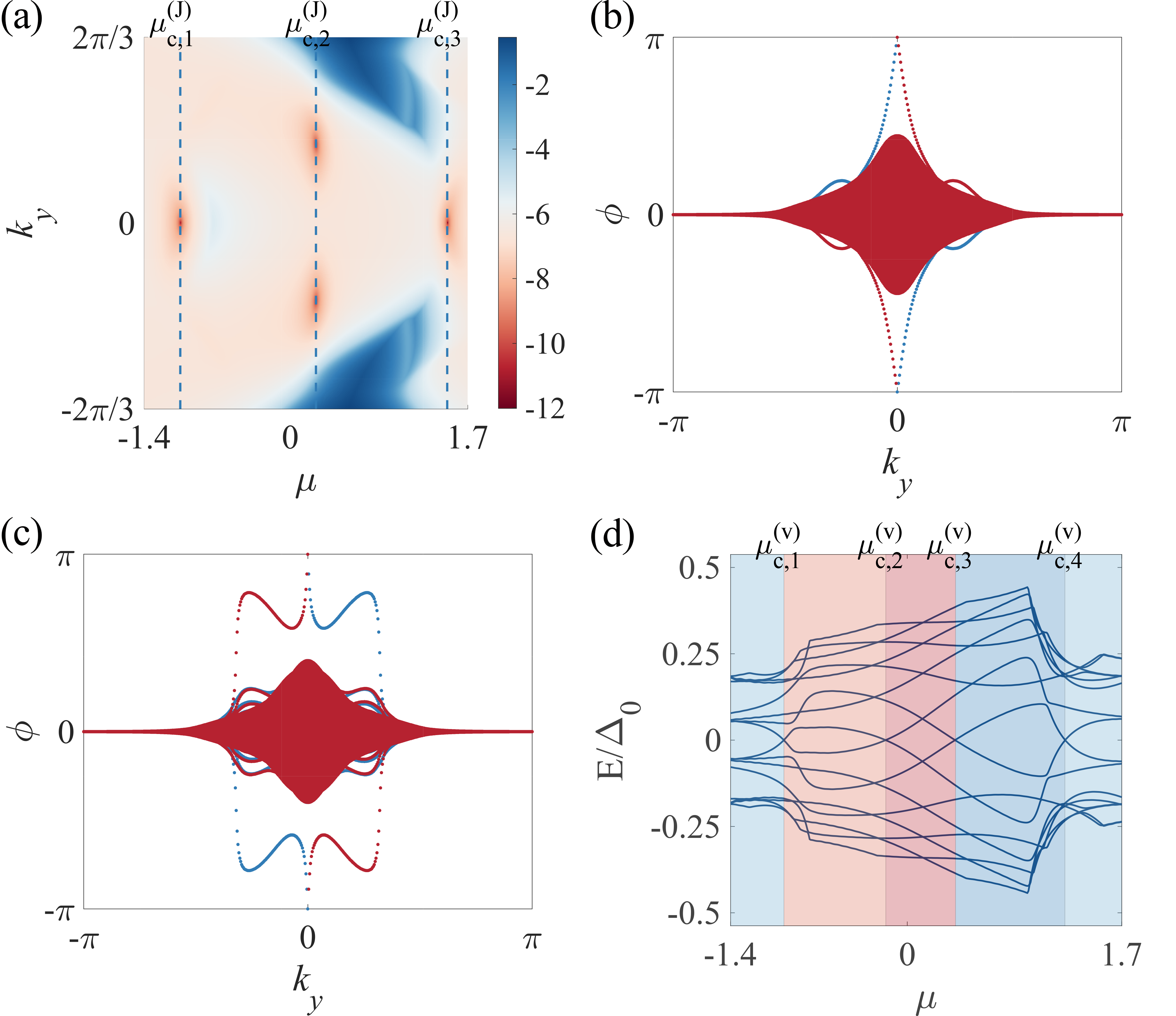}
    \caption{JT and VLT phase diagrams for the tFeSC model. (a) Gap distribution of Andreev bound states in a (100)-directional $\pi$-junction in the $(\mu,k_y)$ parameter space, with $k_z=\pi$. We find two on-TRIM JT critical points and one off-TRIM JT critical pair. (b-c) Mirror-indexed Wilson loop as a function of $k_y$ at $\mu=-0.10$ (b) and $\mu=0.30$ (c). The $M_x=+i$ sector is labeled in red and the negative mirror sector is labeled in blue. (d) Evolution of CdGM states for a (001)-directional vortex-line in the same system. Topologically distinct regions are shown in different colors.} 
    \label{fig:effective_model_FE}
\end{figure}

We proceed to investigate a minimal effective model of the topological iron-based superconductor (tFeSC), which was first proposed in Ref.~\cite{hu2022competing} to describe the intriguing vortex physics in high-$T_c$ iron-based systems such as LiFeAs and Fe(Te,Se). While Ref.~\cite{hu2022competing} has numerically mapped out the VLT phase diagram of this tFeSC model by conducting explicit vortex modeling, we will, however, attempt to establish this phase diagram with a completely different approach, exploiting only (i) $\pi$-junction simulations and (ii) JVC relations.

The normal state of this minimal model consists of six electron bands, as described by the basis function $\Psi_{\mathbf{k}}=$ $(\ket{p_z, \uparrow},$ $\ket{p_z, \downarrow},$ $ \ket{d_+, \downarrow},$ $\ket{d_-, \uparrow},$ $\ket{d_+, \uparrow},$ $\ket{d_-, \downarrow})^T$. Here, $d_\pm$ is short for $d_{xz\pm i yz}$. This effective Hamiltonian is thus given by
\begin{equation}
h_\text{tFeSC}(\mathbf{k}) = 
\begin{bmatrix}
h_0^{(1/2)}(\mathbf{k}) & h_1(\mathbf{k}) \\
h_1^\dagger(\mathbf{k}) & h_0^{(3/2)}(\mathbf{k})
\end{bmatrix},
\label{eq:tFeSCshamiltonian}
\end{equation}
with $h_0^{(1/2)}(\mathbf{k})=h_{\rm TI}(\mathbf{k})$, as defined in Sec.~\ref{sec:doped TI}. $h_0^{(3/2)}(\mathbf{k}) = \left[\delta-m(\mathbf{k}) \right]s_0$, where $\delta$ controls the spin-orbit splitting among the $d$-orbital bands. Besides,
\begin{equation}
h_1^\dagger(\mathbf{k}) = 
\begin{bmatrix}
v k_- & 0 & 0 & b^*(\mathbf{k}) \\
0 & -v k_+ & b(\mathbf{k}) & 0
\end{bmatrix},
\end{equation}
where $k_{\pm} = k_x \pm i k_y$, and $b(\mathbf{k}) = b_1(k_x^2 - k_y^2) - i b_2 k_x k_y$. $h_\text{tFeSC}(\mathbf{k})$ preserves TRS $\Theta = {\rm diag}[-i, i, -i] \otimes s_y \mathcal{K}$, spatial inversion symmetry $\mathcal{P} = {\rm diag}[-1, -1, 1, 1, 1, 1]$, and a fourfold rotation symmetry $C_{4z} = e^{i(\pi/2)J_z}$, with $J_z = {\rm diag}[\frac{1}{2}, -\frac{1}{2}, \frac{1}{2}, -\frac{1}{2}, \frac{3}{2}, -\frac{3}{2}]$, $M_x=iI_3\otimes s_x$. Our choice of parameters are $v=0.5$, $v_z = 0.1$, $b_{1,2} = 0$, $m_0=2m_1=-2m_2=2$, and $\delta=0.5$. This parameter set ensures that the $p_z$ bands cross the bands of $d$-electrons to form two consecutive band inversions along $\Gamma$-$Z$, as the case in both LiFeAs and Fe(Te,Se). One band inversion leads to $\mathbb{Z}_2$ topological bands that mimic the physics of a strong TI, while the other creates a pair of $C_{4z}$-protected bulk Dirac nodes like those that can be found in a DSM. We assume the TI gap to occur at $E=0$, while the Dirac nodes live at a higher energy $E=\delta$. Namely, the value of $\delta$ controls the coupling strength between the TI and DSM bands, with $\delta\rightarrow\infty$ being the fully decoupling limit. 

The $\pi$-junction modeling for the tFeSC system is conducted in the BdG formalism with an $s$-wave spin-singlet pairing potential $\Delta_0=0.08$ and a slab geometry comprising $300$ unit cells along the $\hat{x}$ direction. Periodic boundary conditions are applied for both $\hat{y}$ and $\hat{z}$ directions. In Fig.~\ref{fig:effective_model_FE} (a), we fix $k_z=\pi$ and plot the gap distribution of junction-trapped Andreev bound states as a function of $\mu$ and $k_y$, where the red dots denote where the Andreev gap vanishes and a JT transition takes place. Remarkably, we find two on-TRIM JT transitions at $\mu_{c,1}^{\rm (J)}=-1.05$ and $\mu_{c,3}^{\rm (J)}=1.50$, as well as one off-TRIM, same $k_z$ JT critical pair at $\mu_{c,2}^{\rm (J)}=0.25$. For $k_z=0$, no JT phase transition has been found.

To clarify the nature of each JT transition, we calculate both the $\mathbb{Z}_4$ inversion-symmetry indicator $\kappa$ and the BdG mirror Chern number ${\cal C}_M$ for ${\cal M}_x$ for all occupied states of the $\pi$-junction. In particular, we find that
\begin{equation}
    (\kappa, {\cal C}_M)=\begin{cases} 
      (0, 0) & \mu >1.5,  \\
      (3, 1) & \mu \in  (0.25,1.5),  \\
      (3, -1) & \mu \in   (-1.05,0.25),  \\
      (0, 0) & \mu <-1.05.
   \end{cases}
   \label{eq:tFeSC topo indices}
\end{equation}
Here, we extract the value of ${\cal C}_M$ by plotting the mirror-indexed Wilson loop spectra in Figs.~\ref{fig:effective_model_FE} (b) with $\mu=-0.1$ and (c) with $\mu=0.3$. As expected, the off-TRIM JT critical pair will leave $\kappa$ invariant, but it can change ${\cal C}_M$ by $\pm2$. Meanwhile, the on-TRIM JT transitions always change the $\mathbb{Z}_2$ JT of the $\pi$-junction, as $\kappa=3$ implies $\nu_0=1$. 

We are now ready to sketch the VLT phase diagram based on Fig.~\ref{fig:effective_model_FE} (a) and Sec.~\ref{sec:criticality summary}. This vortex phase diagram should consist of five topologically distinct regions along $\mu$, separated by four critical points of the vortex line at $\mu_{c,i}^{(v)}$ ($i=1,2,3,4$). Specifically, 
\begin{enumerate}
    \item We expect the on-TRIM JT transitions to induce $\mu_{c,1}^{(v)}\approx -1$ and $\mu_{c,4}^{(v)}\approx 1.5$. They each will represent either a Kitaev vortex transition or a nodal vortex transition. 
    \item Both $\mu_{c,2}^{(v)}$ and $\mu_{c,3}^{(v)}$ arise from the JT critical pair and they should obey $\mu_{c,2}^{(v)}<0.25<\mu_{c,3}^{(v)}$. Due to $C_4$, $\mu_{c,2(3)}^{(v)}$ must represent a Kitaev transition or a nodal transition, respectively.   
\end{enumerate}
Together with Eq.~\ref{eq:tFeSC topo indices}, we thus expect the system to undergo the following VLT phase transitions as we increase $\mu$ from $-\infty$:
\begin{eqnarray}
    {\rm trivial} \stackrel{\mu_{c,1}^{(v)}} {\longrightarrow} {\rm Kitaev} \stackrel{\mu_{c,2}^{(v)}}{\longrightarrow} \text{hybrid} \stackrel{\mu_{c,3}^{(v)}}{\longrightarrow} {\rm nodal} \stackrel{\mu_{c,4}^{(v)}}{\longrightarrow} {\rm trivial}, \nonumber \\
    \label{eq:tFeSC-PD}
\end{eqnarray}
where ``trivial", ``Kitaev", ``nodal" are short for trivial, Kitaev, and nodal vortex phases, respectively. The ``hybrid" vortex phase implies the coexistence of both Kitaev and nodal phases, which is protected by $C_4$ symmetry. Remarkably, Eq.~\ref{eq:tFeSC-PD} qualitatively reproduces with the VLT phase diagram found in Ref.~\cite{hu2022competing}. 

Numerically, we further evaluate the VLT phase diagram of the tFeSC by simulating a $\hat{z}$-directional vortex line on a $150 \times 150$ lattice in the $x$-$y$ plane. Periodic boundary condition is imposed for $\hat{z}$, and the pairing potential is $\Delta_0 = 0.08$. As shown in Fig.~\ref{fig:effective_model_FE} (d), the energy spectrum of the vortex is calculated as a function of $\mu$ at $k_z = \pi$, where we find four VLT transitions:
\begin{eqnarray}    
(\mu_{c,1}^{(v)},\mu_{c,2}^{(v)},\mu_{c,3}^{(v)},\mu_{c,4}^{(v)})=(-0.98, -0.17, 0.38, 1.25), \nonumber
\end{eqnarray}
which quantitatively agree with $\mu_c^{(J)}$ for JT, up to ${\cal O}(\Delta_0)$.


\subsection{Dirac octet}
\label{sec:dirac_octet}
\begin{figure*}[ht] 
    \centering
    \includegraphics[width=\textwidth]{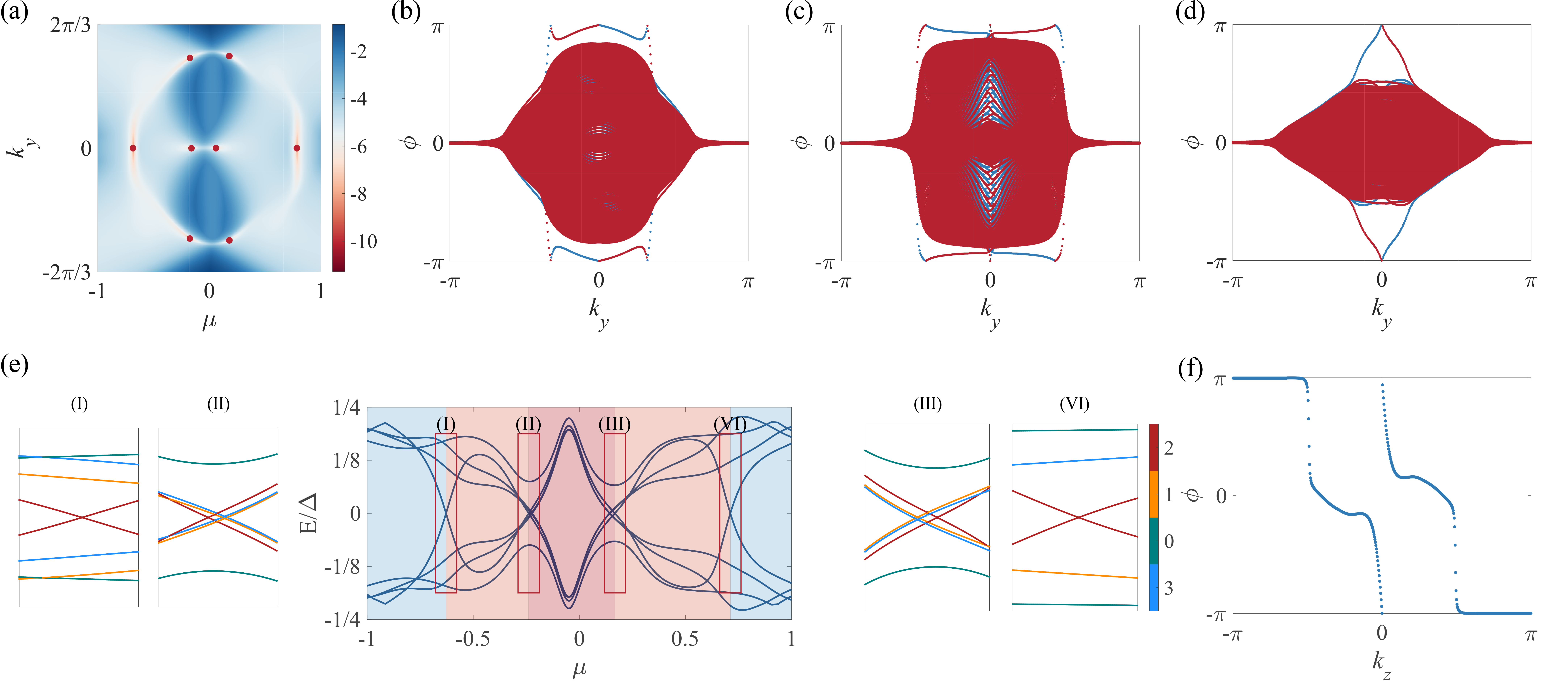}
    \caption{Josephson and vortex-line topological properties of the Dirac octet model. (a) Gap distribution of Andreev bound states for a (100)-directional $\pi$-junction at $k_{z}=0$. Red dots in the figure mark the gap closing points, i.e., JT transitions. (b-d) Spectra of mirror-indexed Wilson loop of Andreev bound states at $\mu=-0.4$ (b), $\mu=0$ (c), and $\mu=0.6$ (d), respectively. The positive (negative) mirror sector is labeled in red (blue). (e) CdGM spectrum for a vortex line along $\hat{z}$ direction at $k_z=0$, where the insets offer a $J_z$-labeled zoom-in dispersions of the CdGM modes at every transition point. Topologically distinct regions are shaded in different colors. (f) Evolution of chiral winding phase along $k_z$ at $\mu=0$, which suggests ${\cal W_S}=-2$.
    }
    \label{fig:SrSnO_model}
\end{figure*}
   
We move on to discuss the eight-band ``Dirac octet" model, which captures the low-energy topological physics of a class of anti-perovskite compounds such as Sr$_3$SnO and Pb$_3$SnO~\cite{kariyado2011three,kariyado2012low,hsieh2014topological,kawakami2018topological,fang2020SSO}. 
This family features two sets of $J=3/2$ bands, one from $p$ orbitals and another from $d$ orbitals, that are inverted near the Fermi level. Each set of $J=3/2$ bands forms a four-fold degenerate quadratic band touching at $\Gamma$, similar to the ones found in HgTe and $\alpha$-Sn~\cite{zhu2024delicate}. Notably, while hope-doped Sr$_3$SnO is reported to be intrinsically superconducting~\cite{oudah2016superconductivity}, there are no theoretical studies on either JT or VLT physics that may emerge from its Dirac octet bands. We aim to fill this theoretical gap in this part.  

We now introduce the Dirac octet model,
\begin{equation}
     h_{\rm DO}(\mathbf{k})=\tau_zm(\mathbf{k}) + \tau_0\alpha(\mathbf{k}) + \tau_x h_1(\mathbf{k}),
\end{equation}
where $\mathbf{J}=\{J_x,J_y,J_z\}$ are the spin-$3/2$ matrices and $\tilde{\mathbf{J}}=\{\tilde{J}_x,\tilde{J}_y,\tilde{J}_z\}$ are defined as $\tilde{J}_i\equiv\frac{5}{3}\sum_{j\neq i} J_jJ_iJ_j-\frac{7}{6}J_i$ $(i,j=x,y,z)$. In particular, $m(\mathbf{k}) = m_0 + \frac{m_1}{2}(3 - \cos k_x - \cos k_y - \cos k_z)$ and $\alpha(\mathbf{k})=\frac{2}{3}\alpha_0 \sum_{i=\{x,y,z\}} {\rm cos}k_iJ_{i}\cdot\tilde{J}_i$. The hybridization term between the $J=3/2$ bands is $h_1(\mathbf{k}) = v_1 \mathbf{J}\cdot{\rm sin}(\mathbf{k}) + v_2 \tilde{\mathbf{J}}\cdot{\rm sin}(\mathbf{k})$, where $v_{1,2}$ control the nature of the band inversion. As a cubic system, the Dirac octet model respects many important symmetries, including TRS $\Theta=-i\tau_0\otimes\gamma_2\mathcal{K}$, inversion $\mathcal{P}=\tau_z$, and four-fold around $\hat{z}$ rotation $C_{4z}={\rm exp}(-i\pi J_z/2)\otimes \tau_0$, where $\gamma$ matrices are defined following Eq.~\ref{eq:gamma matrices}. The system also features two inequivalent mirror symmetries $M_{100}=i\tau_z\otimes \gamma_1$ and $M_{110}=i\tau_z\otimes(\gamma_{15}+\gamma_{25})/\sqrt{2}$, where $M_{100}:(x,y,z)\rightarrow (-x,y,z)$ and $M_{110}:(x,y,z)\rightarrow (y,x,z)$. 

As discussed in Ref.~\cite{fang2020SSO}, the electronic topology of the Dirac octet can be characterized by a 3D class-AII $\mathbb{Z}_4$ indicator $\omega$ for inversion symmetry, as well as two independent mirror Chern numbers of $C_{100}$ and $C_{110}$ defined for $M_{100}$ and $M_{110}$. Please note that $\omega$ is fundamentally distinct from $\kappa$, the 2D class-DIII $\mathbb{Z}_4$ inversion indicator we defined earlier for the Andreev bound states. Our choice of parameter is $(m_0,m_1,\alpha_0)=(-1,2,-0.1,0.12)$, which ensures the $p$-$d$ band inversion around $\Gamma$ and a nontrivial indicator $\omega=2$. 

Meanwhile, both $C_{100}$ and $C_{110}$ are dictated by $v_{1,2}$. For example, we have $(C_{100}, C_{110})=(-2,0)$ when $v_2>3v_1$. At the critical point with $v_2=3v_1$, the band gap closes simultaneously at six different $k$-space locations along the high-symmetry lines, as enforced by the cubic symmetry. This multiple Dirac transition leaves $\omega$ invariant but will lead to a net change of mirror Chern numbers by $\pm 6$. Specifically, we have $(C_{100}, C_{110})=(2,2)$ for $v_2<3v_1$, which exactly matches our ab-initio results for Sr$_3$SnO in Sec.~\ref{sec:Sr3SnO}. Therefore, we will use $v_2=2.7 v_1$ throughout our later discussion. A detailed study of Dirac octet with $v_2>3v_1$ is provided in Appendix {\bf D}.                 

We are now ready to update the Dirac octet to a BdG form with $s$-wave spin-singlet superconductivity and explore its JT and VLT physics. Similar to the construction in Sec.~\ref{sec:TFeSCs}, we consider a $\pi$-junction Hamiltonian with 300 unit cells along $\hat{x}$ with a pairing amplitude $\Delta_0=0.08$. For $k_z=0$, we plot the energy gap of the junction system as a function of $k_y$ and $\mu$, which offers a visualization of the JT phase diagram. As shown in Fig. \ref{fig:SrSnO_model} (a), this phase diagram harbors four on-TRIM JT transitions and two pairs of off-TRIM JT transitions, dividing the phase diagram into seven topologically distinct regions along $\mu$.   

Similar to the tFeSC system, the JT physics here can be captured through the combination of the inversion indicator $\kappa\in\mathbb{Z}_4$ and the BdG mirror Chern number ${\cal C}_M\in\mathbb{Z}$ for $M_x$. Again, ${\cal C}_M$ is calculated with the mirror-indexed Wilson loop technique for all negative-energy states of the $\pi$-junction, where examples with $\mu=0.5,\mu=0$, and $\mu=-0.5$ are shown in Figs.~\ref{fig:SrSnO_model} (b–d). Numerically, we find that 
\begin{equation}
    (\kappa, \mathcal{C}_M)=\begin{cases} 
      (0, 0) & \mu > 0.79,  \\
      (1, -1) & \mu \in  (0.18,0.79),  \\
      (1, 1) & \mu \in   (0.06,0.18),  \\
      (2, 2) & \mu \in   (-0.16,0.06),  \\
      (1, 1) & \mu \in   (-0.18,-0.16),  \\
      (1, -1) & \mu \in  (-0.69,-0.18),  \\
      (0, 0) & \mu < -0.69.
   \end{cases}
   \label{eq.DLSMJTPD}
\end{equation}
As expected, an on-TRIM transition will change both $\kappa$ and ${\cal C}_M$ by $\pm 1$, while an off-TRIM critical pair only changes ${\cal C}_M$ by $\pm 2$.  

Given the information of the JT phase diagrams, we now sketch a ``deduced" vortex phase diagram:
\begin{enumerate}
    \item There will be eight VLT transitions by varying $\mu$.
    \item On-TRIM JT transitions at $\mu=0.79$ and $\mu=-0.69$ will lead to two VLT transitions around the same $\mu$ values.
    \item Two off-TRIM critical pairs at $\mu=\pm 0.18$, along with two on-TRIM transition at $\mu=-0.16$ and $\mu=0.06$, will together lead to six neighboring VLT transitions $\mu=0$.
    \item Since $\kappa=1$ for $\mu\in(-0.69,-0.18)\cup(0.06,0.79)$, the surface vortex core should host a single MZM for a large electron or hole doping regime, following Eq.~\ref{eq:JVC_z2}. 
    \item Since ${\cal C}_M=2$ for $\mu\in(-0.16,0.06)$, we expect the undoped system to feature {\it a pair of decoupled vortex MZMs}, following Eq.~\ref{eq:mirror_dimensional relation}. 
\end{enumerate}
We hope to highlight that the only caveat to the above predictions, in particular \#4 and \#5, is the existence of $C_{4z}$ symmetry. As discussed in Sec.~\ref{sec:rotation sym mismatch}, a nontrivial $\kappa$ or ${\cal C}_M$ may also correspond to nodal vortex phases, rather than the gapped Majorana-carrying Kitaev vortex phases. Slightly breaking $C_{4z}$ in our system will rule out the possibility of nodal vortices, making the above predictions unambiguous.    

With the predictions in mind, we now proceed to conduct vortex modeling of the Dirac octet. We consider an in-plane lattice geometry with $150 \times 150$ unit cells and a $\hat{z}$-directional vortex line penetrating the center of the lattice. By plotting the CdGM spectrum at $k_z=0$, we indeed find eight VLT transitions along $\mu$, as shown in Fig.~\ref{fig:SrSnO_model} (e). The critical chemical potentials of VLT transitions are given by
\begin{eqnarray}
    \mu_{c}^{(v)}\in && \{0.712, 0.168, 0.144, 0.144, \nonumber \\
    &&-0.225, -0.225, -0.239, -0.628\},
\end{eqnarray}
quantitatively agreeing with our predictions \#2 and \#3. 

To understand the nature of the VLT transitions, we employ different colors to label $J_z$ of every CdGM band in Fig.~\ref{fig:SrSnO_model} (e). As discussed in Sec.~\ref{sec:rotation sym mismatch}, $C_{4z}$ enables a $(\mathbb{Z}_2)^2\times\mathbb{Z}$ VLT classification with three topological indices: $\zeta_0\in\mathbb{Z}_2$, $\zeta_2\in\mathbb{Z}_2$, and $\mathcal{Q}_1\in\mathbb{Z}$. Therefore, a $J_z=0(2)$ transition will change $\nu_{0(2)}$ by $1$, while one with $J_z=\pm1$ will change ${\cal Q}_1$ by $\pm 1$. 

In addition to the $C_{4z}$-related indices, the VLT also features a chiral winding number ${\cal W_S}\in\mathbb{Z}$, as discussed in Sec.~\ref{sec:mirror-Chern JT}. The chiral symmetry is ${\cal S}=M_x\Theta\Xi$, with which the Hamiltonian can always be reshaped into an off-block-diagonal form,
\begin{equation}
    \tilde{\mathcal{H}}(k_z) = 
    \begin{pmatrix}
        0 & h(k_z) \\
        h^\dagger(k_z) & 0
    \end{pmatrix}.
\end{equation}
We further define a chiral winding phase $\phi(k_z) ={\rm arg}({\rm det}[ U(k_z) V^\dagger(k_z)])$. Here, the unitary matrices $U$ and $V$ arise from a singular value decomposition with $h(k_z) = U D V^\dagger$. By tracking the evolution of $\phi(k_z)$, we can conclude ${\cal W_S}$ by counting the winding number of $\phi$. For example, Fig. \ref{fig:SrSnO_model} (g) shows the evolution of $\phi(k_z)$ at $\mu=0$, clearly showing ${\cal W_S}=2$. Notably, this double-MZM vortex phase has also been predicted to exist in superconducting SnTe~\cite{chen2014magnetic,liu2024signatures}.

Notably, there exist two pairs of degenerate VLT transitions, one at $\mu=0.144$ and another at $\mu=-0.225$. Interestingly, we find that these degenerate transitions fail to trigger any change in vortex topology. Hence, we will ignore these ``fake transitions" from now on. Below, we summarize the VLT indices for each region,
\begin{equation}
    ({\cal W_S}, \zeta_0, \zeta_2, \mathcal{Q}_1)=\begin{cases} 
      (0, 0, 0, 0) & \mu > 0.712,  \\
      (1, 0, 1, 0) & \mu \in  (0.168,0.712),  \\
      (2, 0, 0, 0) & \mu \in   (-0.239,0.168),  \\
      (1, 0, 1, 0) & \mu \in  (-0.628,-0.239),  \\
      (0, 0, 0, 0) & \mu < -0.628.
   \end{cases}
   \label{eq.DLSMVLTPD}
\end{equation}
This result not only agrees well with our $\pi$-junction-based predictions, but also establish the Dirac octet as a new playground for various Majorana physics.

\section{Application to Real-World Superconductors}
\label{sec:ab-initio}
Thus far, we have theoretically proposed and established a topological correspondence relation between $\pi$-phase junctions and vortex lines, as numerically confirmed by comprehensive model studies. In this section, we will develop a JVC-based {\it ab-initio} workflow to diagnose JT and further extrapolate VLT for real-world candidate superconductors, without performing challenging large-scale vortex simulations. We will first discuss the general construction scheme of $\pi$-junction Hamiltonians at the first-principles level. As concrete examples, we will proceed to apply our theoretical strategy to investigate and clarify the Majorana possibilities of two superconducting materials that have been experimentally synthesized, 2M-WS$_2$ and Sr$_3$SnO. Remarkably, both systems are numerically confirmed to be promising playgrounds for discovering and engineering JT and VLT.    

\subsection{Ab-initio simulations of a $\pi$-junction}

For a candidate superconductor $\chi$, we will first conduct the density-functional theory (DFT)~\cite{hohenberg1964inhomogeneous}  to calculate the full electronic band structure of its normal state within the generalized gradient approximation (GGA)~\cite{dal1996generalized}. This will be followed by a Wannier-function projection~\cite{marzari2012RMP} to obtain a multi-orbital electronic tight-binding Hamiltonian $h_w({\bf k})$ that accurately captures the low-energy dispersions of $\chi$. We will then construct a slab Hamiltonian ${\cal H}_{\alpha}({\bf k}_\parallel)$ to describe an $\hat{\alpha}$-directional $\pi$-junction geometry by updating $h_w({\bf k})$ to a BdG form $H({\bf k})$. Although the above procedure looks straightforward, there are several technical aspects that can be tricky in practice.  

First of all, $h_w({\bf k})$ must respect the full crystalline and time-reversal symmetries of $\chi$, which, however, may not always hold in conventional Wannierization procedures. To understand this stringent symmetry requirement, we note that the pairing matrix for the conventional s-wave channel is exactly the unitary matrix $U_T$ of the TRS operation $\Theta=U_T {\cal K}$, up to a possible minus sign. For example, $U_T = i s_y$ in Eq.~\ref{eq:BdG_form}, and so is the $s$-wave pairing matrix. Consequently, we will be able to write down the correct BdG matrix $H({\bf k})$ {\it if and only if} (i) $h_w({\bf k})$ respects TRS; and (ii) the exact matrix form of $U_T$ is known for $h_w({\bf k})$. Besides, a knowledge of lattice-symmetry representations for $h_w({\bf k})$ will also be essential for evaluating JT symmetry indicators for ${\cal H}_{\alpha}({\bf k}_\parallel)$.     

To cope with this challenge, we employ the full-potential local-orbital ({\small FPLO}) package~\cite{koepernik1999full} to conduct first-principles simulations, which features a built-in module that can wannierize the DFT band structure based on {\it symmetry-conserving} maximally localized Wannier functions (MLWFs)~\cite{koepernik2023symmetry}. Specifically, the real-space hopping matrix element between two Wannier orbitals is given by,  
\begin{equation}
    t_{\alpha\beta}(\mathbf{R-R'}) = \langle {\mathbf{R}'\alpha} | \hat{h}_0 | {\mathbf{R}\beta} \rangle, 
\end{equation}
where $\hat{h}_0$ formally represents the electronic Hamiltonian operator for $\chi$. Here, $| {\mathbf{R}\alpha}\rangle$ represents the symmetry-preserving Wannier function at the real-space lattice vector $\mathbf{R}=m\mathbf{a}+n\mathbf{b}+l\mathbf{c}$, where $\alpha$ denotes the orbital or spin index and $\mathbf{a}$, $\mathbf{b}$ and $\mathbf{c}$ denote the primitive lattice vectors of the crystal. Upon a Fourier transform, we have
\begin{equation}
    [h_0({\bf k})]_{\alpha\beta}  = \sum_{\mathbf{R}} e^{i\mathbf{k \cdot R}} t_{\alpha\beta}(\mathbf{R}).
\end{equation}

Secondly, we are often interested in simulating a $\pi$-junction that does not align with any of the primitive lattice vectors, as will be the case for 2M-WS$_2$. Hence, we need to develop a formalism that allows us to construct a slab Hamiltonian  ${\cal H}_{\alpha}({\bf k}_\parallel)$ along any direction.
This process involves two main steps.
First, for a given Miller-index plane $(hkl)$,
we identify all lattice vectors 
$\mathbf{v} = u\mathbf{a} + v\mathbf{b} + w\mathbf{c}$
 lying within the plane by satisfying the condition
 $\mathbf{v} \cdot [hkl] = 0$.
 Next, a pair of in-plane lattice vectors ($\mathbf{a}^{\prime},\mathbf{b}^{\prime}$) is
 selected to minimize the enclosed area while 
 maximizing their linear independence.
An out-of-plane vector $\mathbf{c}^{\prime}$ is further
 determined by minimizing the slab volume, which ensures our updated choice of lattice parameters is the smallest possible one. $(\mathbf{a}^\prime,\mathbf{b}^\prime,\mathbf{c}^\prime)$ thus defines a new unit cell for the slab-geometry construction. 
Notably, the new cell may have an enlarged volume compared to the original one.

We now construct a rotation matrix
$U=$
$(\mathbf{a}^\prime,\mathbf{b}^\prime,\mathbf{c}^\prime)(\mathbf{a},\mathbf{b},\mathbf{c})^{-1}$
which transforms the hopping parameters as:
\begin{equation}
    t_{\alpha\beta}(\mathbf{R'})=t_{\alpha\beta}(U\mathbf{R}),
\end{equation}
where $\mathbf{R}^{'}$=$m\mathbf{a}^{\prime}+n\mathbf{b}^{\prime}+l\mathbf{c}^{\prime}$ is the new real-space lattice vector in the rotated coordinate system.
With open boundary conditions along ${\bf c}'$, the hopping matrix elements of real-space slab Hamiltonian along the $\mathbf{R}^{'}_3$=$(s-s')\mathbf{c}^\prime$ can be written as:
\begin{equation}
t^{\prime}_{N\cdot s+\alpha,N\cdot s^{\prime}+\beta}(\mathbf{R}'_{\parallel})
=t_{\alpha\beta}(\mathbf{R}'_{\parallel}+(s'-s)\mathbf{c}^{\prime}) 
\end{equation}
where $s$ and $s'$ are the layer indices for the slab model. The reciprocal-space Hamiltonian of the slab model via a Fourier transform,
\begin{equation}\label{eq-hslab}
\hat{H}_{slab}(\mathbf{k_{\parallel}})=
\sum_{\mathbf{R}^\prime_{\parallel}}\sum_{\alpha,\beta}t^{\prime}_{\alpha,\beta}(\mathbf{R}'_{\parallel}) e^{i\mathbf{k_{||}\cdot\mathbf{R^\prime_{1,2}}}}
c^{\dagger}_{\alpha,\mathbf{k}_{\parallel}}c_{\beta,\mathbf{k}_{\parallel}},
\end{equation}
where $\mathbf{k_{||}}$ are the in-plane crystal momenta. Denoting the Hamiltonian matrix of $\hat{H}_{slab}$ as $H_{slab}({\bf k}_{||})$, we construct the $\pi$-junction Hamiltonian ${\cal H}_{{\bf c}'}$ as,
\begin{equation}\label{eq-hbdgjp}
{\cal H}_{{\bf c}'}(\mathbf{k}_{||}) = \begin{pmatrix}
        H_{slab}(\mathbf{k}_{||})-\mu & -i \mathrm{sgn}(s-s_{\pi})\Delta_0\sigma_y \\
        i\mathrm{sgn}(s-s_{\pi})\Delta_0\sigma_y & \mu -H_{slab}^T({\mathbf{-k}}_{||})
    \end{pmatrix},
\end{equation}
where $\mu$ is the chemical potential and 
$\Delta_0$ represents the strength of the $s$-wave superconducting pairing. $s_\pi$ defines the location of the pairing domain. 

\begin{figure*}[htb!]
\includegraphics[width=2.0\columnwidth]{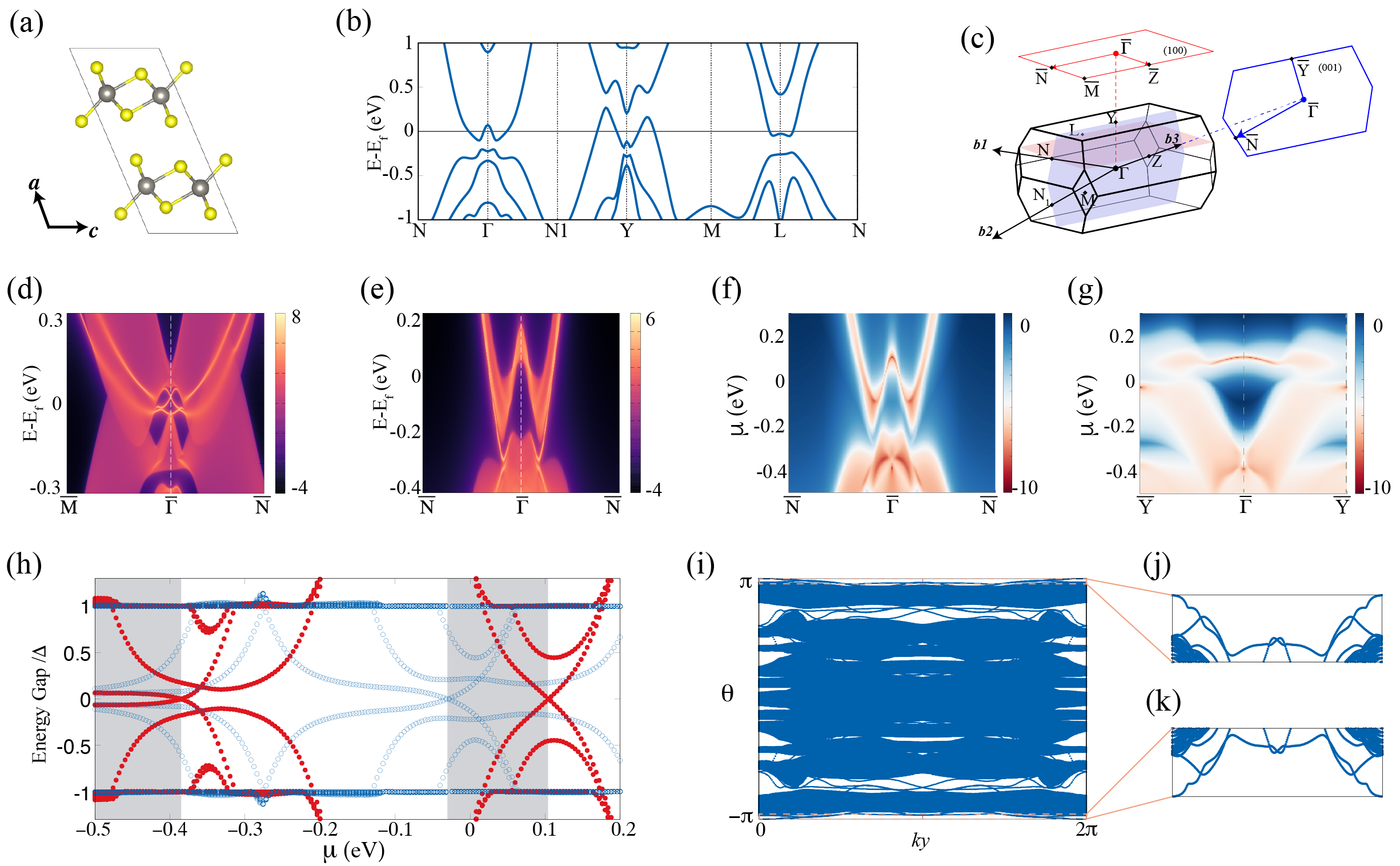}
\caption{Bulk and Josephson topological characterizations for 2M WS$_2$. 
(a) The side view of conventional cell. 
(b) The bulk band structure along a high symmetry path in ${\bf k}$ space. 
(c) The bulk Brillouin zones (BZs) for primitive cell and the projected BZs for (100) and (001) planes. 
The topological surface states for (d) (100) plane and (e) (001) plane. 
$\mu$-dependent gap distributions of Andreev bound states in a (001) Josephson $\pi$ junction along 
(f) $\overline{\mathrm{N}}$-$\overline{\mathrm{\Gamma}}$-$\overline{\mathrm{N}}$ and 
(g) $\overline{\mathrm{Y}}$-$\overline{\mathrm{\Gamma}}$-$\overline{\mathrm{Y}}$ paths.
(h) The E-$\mu$ diagram of (001) $\pi$ junction for
$\overline{\mathrm{\Gamma}}$ point (red circles) and $\overline{\mathrm{M}}$ point (blue circles).
The inversion symmetry indicator $\kappa$ is 1 in the gray region (topologically nontrivial zone) and 0 in the other regions (topologically trivial zone).
(i) Wilson-loop spectrum for the $\pi$ junction at $\mu=0$ along $k_y$, where clearly shows a helical winding pattern that agrees with $\kappa=1$.  Zoom-in plots are shown in (j) for $\theta =\pi$ and in (k) for $\theta =-\pi$, where $\theta$ denotes the Wannier centers.
}\label{fig-ws2} 
\end{figure*}

\subsection{2M-WS$_2$}

Our first system of interest is the 2M phase of WS$_2$. This van der Waals compound has recently attracted significant research attention for intrinsically combining electronic band topology and superconductivity~\cite{fang2019discovery}.
The superconducting transition temperature of 2M-WS$_2$ is reported to be $T_c\approx 8.8$ K at ambient pressure, one of the highest among transition metal dichalcogenides. Above $T_c$, multiple angle-resolved photoemission spectroscopy studies have revealed the existence of a Dirac surface state at the Fermi level~\cite{fang2019discovery,li2021observation,xu2023topological,li2024evidence}, consistent with the first-principles band simulations~\cite{guguchia2019nodeless,lian2020anisotropic,paudyal2022superconducting}. Below $T_c$, non-split zero-bias peaks (ZBPs) in the scanning tunneling spectroscopy (STM) are found to show up around the Abrikosov vortex cores~\cite{yuan2019evidence,fan2024stripe}. While observing ZBPs in 2M-WS$_2$ is consistent with a MZM interpretation, other non-topological origins of ZBPs cannot be easily ruled out. This motivates us to clarify this puzzle for 2M-WS$_2$ from an ab initio computational perspective. 

As shown in Fig.~\ref{fig-ws2} (a), 2M-WS$_2$ crystallizes under the space group $C2/m$ (No. 12). Our choice of optimized lattice constants are $a=12.87$\AA, $b=3.22$\AA, $c=5.27$\AA, with the axial angles $\alpha$=$\gamma$=90$^\circ$, and $\beta$=112.9$^\circ$, which agrees well with experimental results~\cite{fang2019discovery}. There are two monoclinic monolayers in one unit cell, which holds the distorted octahedral building blocks as the monolayer 1T$^\prime$-WTe$_2$~\cite{tang2017quantum} and 1T$^\prime$-WSe$_2$~\cite{chen2018large}.
These two 1T$^\prime$ monolayers stack along the $\mathbf{a}$ direction to form a bulk crystal, thus preserving the inversion symmetry ${\cal P}$. This configuration is different from the well-known T$_d$ phase with Weyl fermions~\cite{soluyanov2015type}.  

Our calculation of the electronic band structure with FPLO is shown in Fig.~\ref{fig-ws2} (b), where we have included the spin-orbital coupling (SOC) effect. We have used Gamma-centered $\mathbf{k}$ meshes of 
$21\times21\times21$ and the convergence criteria for the density and energy are set to 10$^{-6}$ and $10^{-8}$, respectively. Specifically, we find two electron-like pockets at $\Gamma$ and L points and one hole pocket around Y point, where our notation of high symmetry points is given in Fig.~\ref{fig-ws2} (c). We consider a continuous yet energetically curving band gap between conduction and valence bands in the BZ and calculate the occupied-band parity products for all TRIMs. The $\mathbb{Z}_2$ topological indices are found to be $(1;000)$, consistent with calculations in the literature~\cite{fang2019discovery}. 

We now construct a set of symmetry-conserving MLWF from the GGA Bloch wave functions by FPLO. The projected basis of the MLWF involves 20 $d$-orbitals of two W atoms and 24 $p$-orbitals of four S atoms. Therefore, the Wannier model $h_0({\bf k})$ for 2M-WS$_2$ consists of 44 bands, whose symmetry properties have been comprehensively confirmed. With the iterative Green's function method~\cite{Sancho1985green}, the (100) and (001) surface dispersions are shown in Fig.~\ref{fig-ws2} (d) and (e), respectively. 
In particular, the (100) surface shows a single Dirac cone at
$\overline{\mathrm{\Gamma}}$ point around the Fermi level, which is in excellent agreement with prior ARPES results~\cite{li2021observation,li2024evidence}. 
Interestingly, we also find a second Dirac surface state at $\overline{\mathrm{\Gamma}}$,
slightly below the Fermi level, which we attribute to the band inversion at the Y point between the two topmost valance bands. Due to the energetic proximity,
the two Dirac cones anticross when dispersing away from $\overline{\mathrm{\Gamma}}$. As shown in Fig~\ref{fig-ws2} (e), the (001) surface also features a gapless surface state, as expected from the nontrivial $\mathbb{Z}_2$ topology of the system.

In experiments, STM studies of vortices were performed on the cleaved (100) surface. Therefore, we need to study the CdGM physics of a [100]-directional vortex tube, or equivalently a $\pi$-junction along either [001] or [010] direction. We have constructed a BdG model, dubbed ${\cal H}_{[001]}({\bf k}_\parallel)$, which describes a $\pi$-junction geometry with $N_a$ unit cells along [001]. We find that the simulation results have converged when the pairing amplitude is 30 meV and below. Thus, we use $\Delta_0=10$ meV as the default pairing amplitude in all our simulations of 2M-WS$_2$, with which the localization length of the Andreev bound states is around 3 nm. As a result, we set the junction size to be $N_a = 500$, or equivalently 263 nm, to avoid any finite-size effect.     

Upon diagonalizing ${\cal H}_{[001]}$, we map out $\mu$-dependent energy gap distributions for 2M-WS$_2$ along $\overline{\mathrm{N}}$-$\overline{\mathrm{\Gamma}}$-$\overline{\mathrm{N}}$ and $\overline{\mathrm{Y}}$-$\overline{\mathrm{\Gamma}}$-$\overline{\mathrm{Y}}$, as plotted in Figs.~\ref{fig-ws2} (f) and (g), respectively. In particular, we are looking for gap-closure points in the $(k,\mu)$-space that suggest quantum critical points of JT. For example, there exist two gapless nodes in Fig.~\ref{fig-ws2} (f), both at $\overline{\Gamma}$, where the corresponding critical chemical potentials are $\mu_{\Gamma}^{(1)} = 0.104$ eV and $\mu_{\Gamma}^{(2)} = -0.388$ eV. A few other local gap minima are found to exist away from $\overline{\Gamma}$, while none of them represent a true gap closing. Physically, the bulk-state origin of JT transitions at $\overline{\Gamma}$ is exactly the TI-like band inversion at $\Gamma$, which is reminiscent of the Fu-Kane paradigm. 

Surprisingly, Fig.~\ref{fig-ws2} (g) also reveals an unexpected JT transition at $\overline{Y}$ with a $\mu_Y = -0.025$ eV. This JT transition at $\overline{Y}$ is thus {\it not} due to the $\mathbb{Z}_2$ topology of electrons around $\Gamma$. Geometrically, we find that this $\overline{Y}$-transition can be traced back to the bulk electrons at $L$. Remarkably, this JT criticality arises exactly when the Fermi level crosses the band bottom of $L$ pocket. While the microscopic nature of this $\overline{Y}$-transition is an intriguing open question, it is beyond the scope of this work and will be left for future discussions. 

To better visualize the JT phase diagram, we merge the $\mu$-varying energy dispersions for $\overline{\Gamma}$ (red dots) and $\overline{Y}$ (blue dots) into Fig.~\ref{fig-ws2} (h), which displays four regions separated by the JT critical points. To characterize the JT for each region, we compute the $\mathbb{Z}_4$ indicator $\kappa$ by calculating the inversion eigenvalues for all 22,000 states below zero energy for all TRIMs in the (001) projected BZ. Remarkably, we find that  
\begin{equation}
\kappa = \begin{cases} 
      0 & \mu > 0.104,  \\
      1 & \mu \in  (-0.025,0.104),  \\
      0 & \mu \in  (-0.388,-0.025),  \\
      1 & \mu < -0.388,
   \end{cases}
   \label{eq:kappa_WS2}
\end{equation}
where $\mu$ is in the unit of eV. As a complementary check, we further carry out a large-scale Wilson-loop calculation for ${\cal H}_{[001]}$, again taking into account all occupied states of the junction Hamiltonian. The Wilson loop spectrum at $\mu=0$ is shown in Figs.~\ref{fig-ws2} (i) - (k), whose gapless helical winding pattern unambiguously informs $\nu_0=1$. We thus numerically conclude that:
\begin{enumerate}
    \item[(i)] $\pi$-junction of an undoped 2M-WS$_2$ with $\mu=0$ is $\mathbb{Z}_2$ topological.
    \item[(ii)] Undoped 2M-WS$_2$ carries vortex MZMs.
    \item[(iii)] Slight hole doping will destroy the vortex MZMs. 
\end{enumerate}
The second point above is an immediate deduction from the JVC, by noting that the space group of 2M-WS$_2$ does not possess a high-fold rotation symmetry that would support nodal vortices. As a result, a $\mathbb{Z}_2$ topological JT in 2M-WS$_2$ will always imply the existence of a $\mathbb{Z}_2$ Kitaev vortex with end-localized MZMs.      

\begin{figure*}[htb!]
\includegraphics[width=2.0\columnwidth]{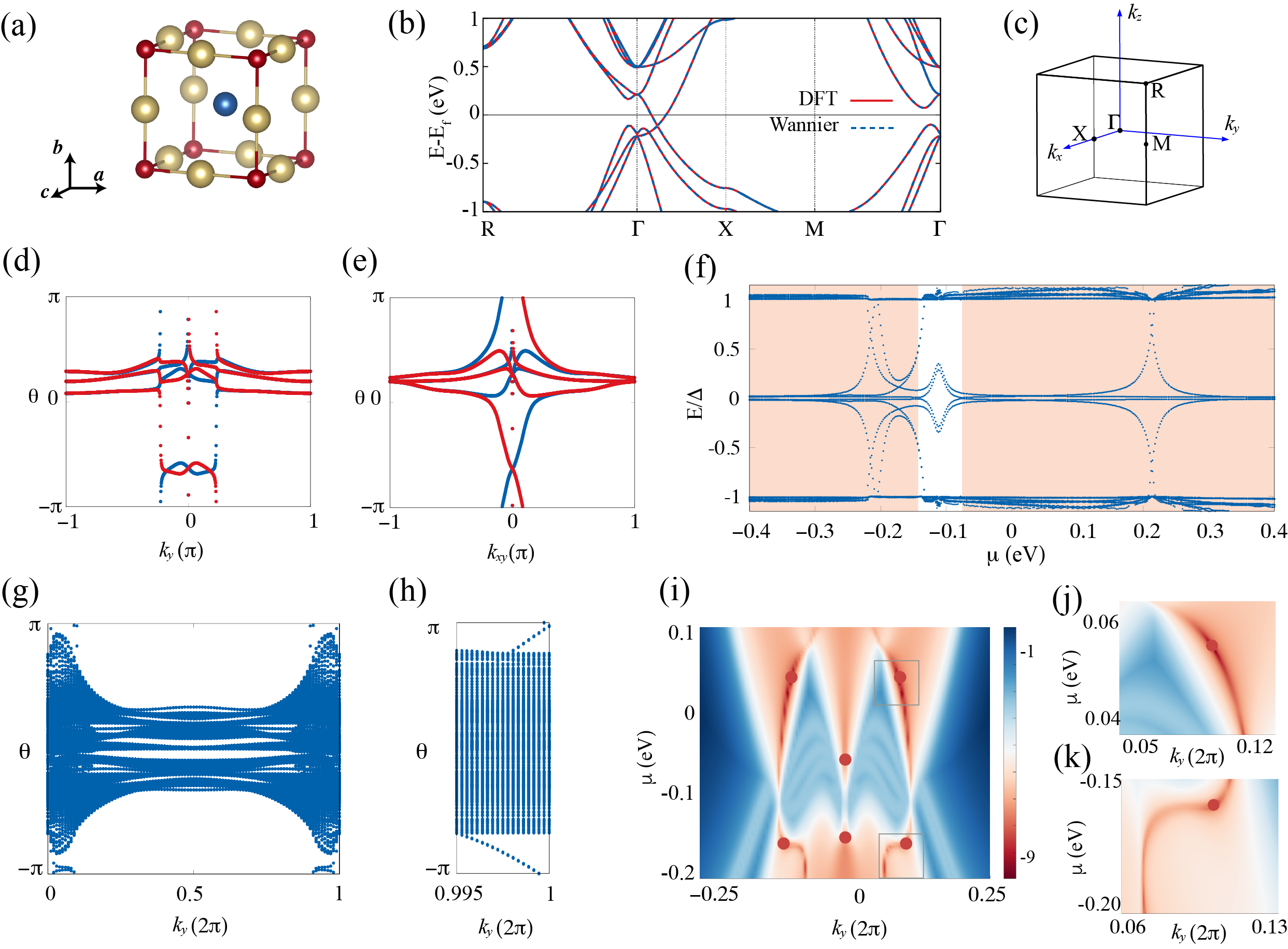}
\caption{Bulk and Josephson topological characterizations for Sr$_3$SnO. 
(a) Crystal structure of Sr$_3$SnO. 
(b) Bulk electronic band structure computed using FPLO (red lines) and Wannier interpolations (blue dots) along high-symmetry lines. 
(c) BZ and high symmetry points.
(d) $M_x$-indexed Wilson loop spectrum for occupied bulk electrons, showing a nontrivial mirror Chern number $C_x=2$. (e) $M_{xy}$-indexed Wilson loop spectrum for occupied bulk electrons, showing a nontrivial mirror Chern number $C_{xy}=2$. The red and blue points denote $\pm i$ mirror sectors, respectively.
(f) The Andreev spectrum of an $\hat{x}$-directional $\pi$ junction at
$\overline{\mathrm{\Gamma}}$ point.
(g) At $\mu=0$, we calculate the Wilson-loop spectrum for Andreev bound states of (001) $\pi$ junction. (h) A zoom-in view around $k_z$ =2$\pi$ shows a helical Wilson loop winding, agreeing with a $\mathbb{Z}_2$ nontrivial JT with $\nu_0$=1.
(i) At $k_z=0$, we plot the gap distribution of the (001) $\pi$ junction as a function of $\mu$ and $k_y$, with the red solid circles marking the JT transitions. We find two on-TRIM JT critical points and two off-TRIM JT critical pairs, which quantitatively agree with Fig.~\ref{fig:SrSnO_model} (a).
(j) and (k) show the zoom-in plots for both off-TRIM critical points.}\label{fig-srsno} 
\end{figure*}

\subsection{Sr$_3$SnO}
\label{sec:Sr3SnO}

Our first-principles diagnosis also enables us to predict new candidate material systems with Majorana physics. Here, we take Sr$_3$SnO as an example, which is an anti-perovskite that has been experimentally synthesized and measured. Intrinsic superconductivity with $T_c\sim 5$K has been reported for the hole-doped compound~\cite{oudah2016superconductivity}. Meanwhile, DFT and effective model analysis have revealed rich topological band features for the low-energy electrons~\cite{kariyado2011three,kariyado2012low,hsieh2014topological,kawakami2018topological,fang2020SSO}. However, it has remained unclear, both theoretically and experimentally, whether MZMs would necessarily arise in Sr$_3$SnO from the coexistence of band topology and superconductivity. 

The cubic antiperovskite Sr$_3$SnO crystallizes in the 
space group of $Pm\hat{3}m$ ($\mathrm{No.} 221$). As shown in Fig.~\ref{fig-srsno}(a), oxygen atoms and Sn atoms are located at the center and corners of the cube, respectively, which are surrounded by six Sr atoms in an octahedral arrangement. The Sr-based $d$ electrons
of Sr atoms and Sn-based $p$ electrons contribute to two sets of $J$=3/2 quartets near the Fermi level, which have been coined as the ``Dirac octet"~\cite{hsieh2014topological}. It is the inter-quartet band inversion that leads to various topological crystalline physics, which is a generic feature for the antiperovskite $A_3BX$ family 
with $A=$(Ca, Sr, La), $B=$(Pb, Sn) and $X=$(C, N, O).

We apply FPLO to calculate the DFT band structure of Sr$_3$SnO, as shown in Fig.~\ref{fig-srsno} (b). Due to the crystal symmetry, there exist six massive Dirac points with a tiny gap of 19 meV along the $\mathrm{\Gamma}$-X line, which is in excellent agreement with previous literature. We further construct a set of symmetry-conserving MLWFs based on
30 $d$-orbitals of Sr atom and 6 $p$-orbitals of Sn atoms. The Wannier model $h_0({\bf k})$ quantitatively reproduces the DFT band structure, as shown by the dotted 
blue line in Fig~\ref{fig-srsno} (b). As discussed in Refs.~\cite{hsieh2014topological,kawakami2018topological,fang2020SSO}, Sr$_3$SnO features two inequivalent mirror symmetries, $M_x(x,y,z)\rightarrow (-x,y,z)$ and $M_{xy}(x,y,z)\rightarrow (y,x,z)$, based on which we can define two electronic mirror Chern numbers~\cite{hsieh2012topological}, $C_x\in\mathbb{Z}$ and $C_{xy}\in\mathbb{Z}$. Besides, the centrosymmetric nature of the lattice allows us to define a $\mathbb{Z}_4$ inversion symmetry indicator $\kappa_0$ for the occupied electrons, with $\kappa_0=2$. 

Interestingly, ab initio values of $C_x$ and $C_{xy}$ are missing in the literature. Thanks to the symmetry-preserving MLWFs, we can feasibly decompose $h_0({\bf k})$ into different mirror sectors for a target mirror symmetry, and numerically evaluate the mirror-indexed Wilson loop spectra on the mirror-invariant high-symmetry plane. Our Wilson-loop results for $M_x$ and $M_{xy}$ are shown in Figs.~\ref{fig-srsno} (d) and (e),  where red and blue dots denote the positive and mirror parities, respectively. In particular, we find that,
\begin{equation}
    C_x = C_{xy} = 2.
    \label{eq:MCN_SSO}
\end{equation}
In Sec.~\ref{sec:dirac_octet}, we have discussed an 8-band effective model $h_{DO}({\bf k})$ for Sr$_3$SnO, whose topological physics is controlled by two parameters $v_{1,2}$. Eq.~\ref{eq:MCN_SSO} directly suggests that $v_2/v_1$$\in(0.5,3)$, which is also our choice of parameters in Sec.~\ref{sec:dirac_octet}. As will be shown below, this effective model $h_{DO}({\bf k})$ has indeed captured the key Josephson topological features for Sr$_3$SnO.  

We proceed to construct the $\pi$-junction Hamiltonian ${\cal H}_{[001]}(k_x,k_y)$ with a 500-layer slab along [001] and $\Delta_0 = 15$ meV. We further diagonalize ${\cal H}_{[001]}$ at $\overline{\mathrm{\Gamma}}$ and track the evolution of zero-momentum Andreev bound states as a function of $\mu$, as shown in
Fig.~\ref{fig-srsno} (d). At $\overline{\Gamma}$, we find two level crossings at  $\mu_\Gamma^{(1)} = -0.075$ eV and $\mu_\Gamma^{(2)} = -0.143$ eV, while other TRIMs do not display any gap closure. We further calculate the $\mathbb{Z}_4$ indicator and find that
\begin{equation}
\kappa = \begin{cases} 
      1 & \mu > -0.075,  \\
      2 & \mu \in  (-0.143,-0.075),  \\
      1 & \mu < -0.143,
   \end{cases}
   \label{eq:kappa_SSO}
\end{equation}
where $\mu$ is in units of eV. In Figs.~\ref{fig-srsno} (g) and (h), we plot the Wilson loop flow for all 18,000 occupied states in the 500-layer $\pi$-junction of Sr$_3$SnO. A gapless helical winding pattern is found at $\mu=0$, consistent with $\kappa=1$.   

For completeness, in Figs.~\ref{fig-srsno} (i) - (k), we proceed to calculate a 2D color map of gap distribution in the $(k_y, \mu)$ space. Notably, we find two pairs of {\it off-TRIM} critical points that are missing in Fig.~\ref{fig-srsno} (f), with one pair at $\mu_k^{(+)} = 0.032$ eV and the other at $\mu_k^{(-)} = -0.160$ eV. As discussed in Sec.~\ref{sec:Criti Corresp}, a pair of off-TRIM criticalities can change the value of certain BdG mirror Chern number by $\pm 2$, but they are invisible to either $\mathbb{Z}_2$ index $\nu_0$ or $\mathbb{Z}_4$ indicator $\kappa$. This is why Eq.~\ref{eq:kappa_SSO} fails to capture them. Thus far, our ab initio junction simulations have revealed two JT criticalities at TRIM (i.e., $\overline{\Gamma}$) and two pairs of off-TRIM gap closures. Quite remarkably, all of the above JT features, including JT critical points and the values of of $\kappa$, have been quantatitvely reproduced by the low-energy sector of Dirac-octet model $h_{DO}({\bf k})$ in Fig.~\ref{fig:SrSnO_model}. 

Combining effective theory (Fig.~\ref{fig:SrSnO_model}) and ab-initio results (Fig.~\ref{fig-srsno}), we expect that the on-TRIM transition at $\mu_\Gamma^{(2)}$ and off-TRIM transition at $\mu_k^{(-)}$ will merge and lead to three VLT transitions very close to $\mu_{c,1}\sim-0.2$ eV. Similarly, the on-TRIM and off-TRIM transitions at $\mu_\Gamma^{(1)}$ and $\mu_k^{(+)}$ will be mapped to three neighboring VLT transitions at $\mu_{c,2}\sim 0$ eV. Now, we can predict and sketch the VLT phase diagram for Sr$_3$SnO as follows: 
\begin{enumerate}
    \item The vortex is $\mathbb{Z}_2$ topological and features a single MZM for both $\mu<-0.2$ (heavily hole doped) and $\mu>0$ (electron doped).
    \item The vortex features two $\Theta_M$-protected MZMs when $\mu\sim[-0.2, 0]$.
\end{enumerate}
In experiments, Sr$_3$SnO becomes intrinsically superconducting when it is hole-doped. This is exactly when we expect the vortices to feature a pair of overlapping yet decoupled Majorana modes.  

\section{Conclusion and Discussions}
\label{sec:conclusion}

To summarize, we have introduced the Josephson-vortex correspondence (JVC) as a novel framework for understanding and diagnosing vortex topological physics in $s$-wave superconductors. This correspondence establishes a direct mapping between the emergent topological properties of 2D $\pi$-junctions and 1D vortex lines in the same bulk superconductor. Using a dimensional reduction approach, we analytically derive JVC relations across multiple symmetry classes and uncover a striking quantitative match between the topological critical points of $\pi$-junctions and vortices. This insight enables the efficient construction of vortex topological phase diagrams directly from their Josephson counterparts, bypassing the need for computationally intensive vortex simulations. Validating the JVC across a range of effective models, we demonstrate consistent agreement between deduced and explicitly computed vortex phase diagrams. By designing an ab initio computational workflow, we identify 2M-WS$_2$ and Sr$_3$SnO as experimentally viable platforms for realizing and probing vortex and junction Majorana physics. 

Our analysis of 2M-WS$_2$ highlights the importance, and in some cases the necessity, of exploring VLT transitions {\it at the ab initio level}. While effective theories such as the Fu-Kane model provide valuable conceptual guidance, they can overlook essential physics, particularly in multi-band superconductors like 2M-WS$_2$. In this system, the Fu-Kane paradigm successfully captures VLT behavior arising from the $\mathbb{Z}_2$ topological bands near the $\Gamma$ point. However, relying solely on this picture to construct a $\Gamma$-centered effective model will lead to a qualitatively incorrect VLT phase diagram. In particular, such an approach would entirely miss the additional VLT transition near $\mu \sim -25$ meV, which originates from the Lifshitz transition at $L$. While first-principles vortex simulations remain computationally demanding, our JVC framework offers a powerful and efficient alternative to fill in this crucial gap.  

Building on these findings, a promising future direction is to apply our ab initio $\pi$-junction framework to other superconductors where vortex zero-bias peaks have been experimentally reported, such as Fe(Te,Se)~\cite{wang2018evidence}, LiFeAs~\cite{kong2021majorana}, and PbTaSe$_2$~\cite{guan2016superconducting}. This approach could play a pivotal role in addressing ongoing controversies surrounding the Majorana nature of these systems. For materials found to lack topological vortex states at their native chemical potential, our phase diagram construction provides a clear roadmap for Fermi-level engineering to induce vortex MZMs, offering critical guidance for future experimental efforts. Although not the primary focus of this work, our framework for Josephson topology, especially its ab initio implementation, also serves as a powerful screening tool for identifying superconductors suitable for Josephson-junction-based devices with nontrivial topological functionalities. For example, a superconductor exhibiting second-order Josephson topology is guaranteed to host Majorana corner modes at its $\pi$-junction~\cite{xun2025}, providing a natural platform for topological qubits. Moreover, the concept of Josephson topology can be readily extended to superconductors of different dimensionalities or symmetry classes. We leave these intriguing directions for future work.

\section{Acknowledgment}

We thank J. Yu,  P. Zhu, L. Kong, Y. Wang, F. Zhang, E. Rossi, J.D. Sau, C.-X. Liu, and S. Das Sarma for their helpful discussions. Z.C. and J.L. are primarily supported by the National Science Foundation Materials Research Science and Engineering Center program through the UT Knoxville Center for Advanced Materials and Manufacturing (DMR-2309083). R.-X.Z is supported by a start-up fund of the University of Tennessee. This research was supported in part by grant NSF PHY-2309135 to the Kavli Institute for Theoretical Physics (KITP).

\bibliography{ref}

\begin{thebibliography}{82}%
\makeatletter
\providecommand \@ifxundefined [1]{%
 \@ifx{#1\undefined}
}%
\providecommand \@ifnum [1]{%
 \ifnum #1\expandafter \@firstoftwo
 \else \expandafter \@secondoftwo
 \fi
}%
\providecommand \@ifx [1]{%
 \ifx #1\expandafter \@firstoftwo
 \else \expandafter \@secondoftwo
 \fi
}%
\providecommand \natexlab [1]{#1}%
\providecommand \enquote  [1]{``#1''}%
\providecommand \bibnamefont  [1]{#1}%
\providecommand \bibfnamefont [1]{#1}%
\providecommand \citenamefont [1]{#1}%
\providecommand \href@noop [0]{\@secondoftwo}%
\providecommand \href [0]{\begingroup \@sanitize@url \@href}%
\providecommand \@href[1]{\@@startlink{#1}\@@href}%
\providecommand \@@href[1]{\endgroup#1\@@endlink}%
\providecommand \@sanitize@url [0]{\catcode `\\12\catcode `\$12\catcode
  `\&12\catcode `\#12\catcode `\^12\catcode `\_12\catcode `\%12\relax}%
\providecommand \@@startlink[1]{}%
\providecommand \@@endlink[0]{}%
\providecommand \url  [0]{\begingroup\@sanitize@url \@url }%
\providecommand \@url [1]{\endgroup\@href {#1}{\urlprefix }}%
\providecommand \urlprefix  [0]{URL }%
\providecommand \Eprint [0]{\href }%
\providecommand \doibase [0]{https://doi.org/}%
\providecommand \selectlanguage [0]{\@gobble}%
\providecommand \bibinfo  [0]{\@secondoftwo}%
\providecommand \bibfield  [0]{\@secondoftwo}%
\providecommand \translation [1]{[#1]}%
\providecommand \BibitemOpen [0]{}%
\providecommand \bibitemStop [0]{}%
\providecommand \bibitemNoStop [0]{.\EOS\space}%
\providecommand \EOS [0]{\spacefactor3000\relax}%
\providecommand \BibitemShut  [1]{\csname bibitem#1\endcsname}%
\let\auto@bib@innerbib\@empty
\bibitem [{\citenamefont {Wilczek}(1982)}]{wilczek1982anyon}%
  \BibitemOpen
  \bibfield  {author} {\bibinfo {author} {\bibfnamefont {F.}~\bibnamefont
  {Wilczek}},\ }\bibfield  {title} {\bibinfo {title} {Quantum mechanics of
  fractional-spin particles},\ }\href
  {https://doi.org/10.1103/PhysRevLett.49.957} {\bibfield  {journal} {\bibinfo
  {journal} {Phys. Rev. Lett.}\ }\textbf {\bibinfo {volume} {49}},\ \bibinfo
  {pages} {957} (\bibinfo {year} {1982})}\BibitemShut {NoStop}%
\bibitem [{\citenamefont {Kitaev}(2006)}]{kitaev2006anyons}%
  \BibitemOpen
  \bibfield  {author} {\bibinfo {author} {\bibfnamefont {A.}~\bibnamefont
  {Kitaev}},\ }\bibfield  {title} {\bibinfo {title} {Anyons in an exactly
  solved model and beyond},\ }\href {https://doi.org/10.1016/j.aop.2005.10.005}
  {\bibfield  {journal} {\bibinfo  {journal} {Annals of Physics}\ }\textbf
  {\bibinfo {volume} {321}},\ \bibinfo {pages} {2} (\bibinfo {year}
  {2006})}\BibitemShut {NoStop}%
\bibitem [{\citenamefont {Hasan}\ and\ \citenamefont
  {Kane}(2010)}]{hasan2010RMP}%
  \BibitemOpen
  \bibfield  {author} {\bibinfo {author} {\bibfnamefont {M.~Z.}\ \bibnamefont
  {Hasan}}\ and\ \bibinfo {author} {\bibfnamefont {C.~L.}\ \bibnamefont
  {Kane}},\ }\bibfield  {title} {\bibinfo {title} {Colloquium: Topological
  insulators},\ }\href {https://doi.org/10.1103/RevModPhys.82.3045} {\bibfield
  {journal} {\bibinfo  {journal} {Rev. Mod. Phys.}\ }\textbf {\bibinfo {volume}
  {82}},\ \bibinfo {pages} {3045} (\bibinfo {year} {2010})}\BibitemShut
  {NoStop}%
\bibitem [{\citenamefont {Qi}\ and\ \citenamefont {Zhang}(2011)}]{qi2011RMP}%
  \BibitemOpen
  \bibfield  {author} {\bibinfo {author} {\bibfnamefont {X.-L.}\ \bibnamefont
  {Qi}}\ and\ \bibinfo {author} {\bibfnamefont {S.-C.}\ \bibnamefont {Zhang}},\
  }\bibfield  {title} {\bibinfo {title} {Topological insulators and
  superconductors},\ }\href {https://doi.org/10.1103/RevModPhys.83.1057}
  {\bibfield  {journal} {\bibinfo  {journal} {Rev. Mod. Phys.}\ }\textbf
  {\bibinfo {volume} {83}},\ \bibinfo {pages} {1057} (\bibinfo {year}
  {2011})}\BibitemShut {NoStop}%
\bibitem [{\citenamefont {Bradlyn}\ \emph {et~al.}(2016)\citenamefont
  {Bradlyn}, \citenamefont {Cano}, \citenamefont {Wang}, \citenamefont
  {Vergniory}, \citenamefont {Felser}, \citenamefont {Cava},\ and\
  \citenamefont {Bernevig}}]{barry2016science}%
  \BibitemOpen
  \bibfield  {author} {\bibinfo {author} {\bibfnamefont {B.}~\bibnamefont
  {Bradlyn}}, \bibinfo {author} {\bibfnamefont {J.}~\bibnamefont {Cano}},
  \bibinfo {author} {\bibfnamefont {Z.}~\bibnamefont {Wang}}, \bibinfo {author}
  {\bibfnamefont {M.~G.}\ \bibnamefont {Vergniory}}, \bibinfo {author}
  {\bibfnamefont {C.}~\bibnamefont {Felser}}, \bibinfo {author} {\bibfnamefont
  {R.~J.}\ \bibnamefont {Cava}},\ and\ \bibinfo {author} {\bibfnamefont
  {B.~A.}\ \bibnamefont {Bernevig}},\ }\bibfield  {title} {\bibinfo {title}
  {Beyond dirac and weyl fermions: Unconventional quasiparticles in
  conventional crystals},\ }\href {https://doi.org/10.1126/science.aaf5037}
  {\bibfield  {journal} {\bibinfo  {journal} {Science}\ }\textbf {\bibinfo
  {volume} {353}},\ \bibinfo {pages} {aaf5037} (\bibinfo {year}
  {2016})}\BibitemShut {NoStop}%
\bibitem [{\citenamefont {Read}\ and\ \citenamefont
  {Green}(2000)}]{read2000paired}%
  \BibitemOpen
  \bibfield  {author} {\bibinfo {author} {\bibfnamefont {N.}~\bibnamefont
  {Read}}\ and\ \bibinfo {author} {\bibfnamefont {D.}~\bibnamefont {Green}},\
  }\bibfield  {title} {\bibinfo {title} {Paired states of fermions in two
  dimensions with breaking of parity and time-reversal symmetries and the
  fractional quantum hall effect},\ }\href
  {https://doi.org/10.1103/PhysRevB.61.10267} {\bibfield  {journal} {\bibinfo
  {journal} {Phys. Rev. B}\ }\textbf {\bibinfo {volume} {61}},\ \bibinfo
  {pages} {10267} (\bibinfo {year} {2000})}\BibitemShut {NoStop}%
\bibitem [{\citenamefont {Kitaev}(2001)}]{kitaev2001unpaired}%
  \BibitemOpen
  \bibfield  {author} {\bibinfo {author} {\bibfnamefont {A.~Y.}\ \bibnamefont
  {Kitaev}},\ }\bibfield  {title} {\bibinfo {title} {Unpaired majorana fermions
  in quantum wires},\ }\href {https://doi.org/10.1070/1063-7869/44/10S/S29}
  {\bibfield  {journal} {\bibinfo  {journal} {Physics-uspekhi}\ }\textbf
  {\bibinfo {volume} {44}},\ \bibinfo {pages} {131} (\bibinfo {year}
  {2001})}\BibitemShut {NoStop}%
\bibitem [{\citenamefont {Sau}\ \emph {et~al.}(2010)\citenamefont {Sau},
  \citenamefont {Lutchyn}, \citenamefont {Tewari},\ and\ \citenamefont
  {Das~Sarma}}]{sau2010generic}%
  \BibitemOpen
  \bibfield  {author} {\bibinfo {author} {\bibfnamefont {J.~D.}\ \bibnamefont
  {Sau}}, \bibinfo {author} {\bibfnamefont {R.~M.}\ \bibnamefont {Lutchyn}},
  \bibinfo {author} {\bibfnamefont {S.}~\bibnamefont {Tewari}},\ and\ \bibinfo
  {author} {\bibfnamefont {S.}~\bibnamefont {Das~Sarma}},\ }\bibfield  {title}
  {\bibinfo {title} {Generic new platform for topological quantum computation
  using semiconductor heterostructures},\ }\href
  {https://doi.org/10.1103/PhysRevLett.104.040502} {\bibfield  {journal}
  {\bibinfo  {journal} {Phys. Rev. Lett.}\ }\textbf {\bibinfo {volume} {104}},\
  \bibinfo {pages} {040502} (\bibinfo {year} {2010})}\BibitemShut {NoStop}%
\bibitem [{\citenamefont {Lutchyn}\ \emph {et~al.}(2010)\citenamefont
  {Lutchyn}, \citenamefont {Sau},\ and\ \citenamefont
  {Das~Sarma}}]{lutchyn2010majorana}%
  \BibitemOpen
  \bibfield  {author} {\bibinfo {author} {\bibfnamefont {R.~M.}\ \bibnamefont
  {Lutchyn}}, \bibinfo {author} {\bibfnamefont {J.~D.}\ \bibnamefont {Sau}},\
  and\ \bibinfo {author} {\bibfnamefont {S.}~\bibnamefont {Das~Sarma}},\
  }\bibfield  {title} {\bibinfo {title} {Majorana fermions and a topological
  phase transition in semiconductor-superconductor heterostructures},\ }\href
  {https://doi.org/10.1103/PhysRevLett.105.077001} {\bibfield  {journal}
  {\bibinfo  {journal} {Phys. Rev. Lett.}\ }\textbf {\bibinfo {volume} {105}},\
  \bibinfo {pages} {077001} (\bibinfo {year} {2010})}\BibitemShut {NoStop}%
\bibitem [{\citenamefont {Oreg}\ \emph {et~al.}(2010)\citenamefont {Oreg},
  \citenamefont {Refael},\ and\ \citenamefont {von Oppen}}]{oreg2010helical}%
  \BibitemOpen
  \bibfield  {author} {\bibinfo {author} {\bibfnamefont {Y.}~\bibnamefont
  {Oreg}}, \bibinfo {author} {\bibfnamefont {G.}~\bibnamefont {Refael}},\ and\
  \bibinfo {author} {\bibfnamefont {F.}~\bibnamefont {von Oppen}},\ }\bibfield
  {title} {\bibinfo {title} {Helical liquids and majorana bound states in
  quantum wires},\ }\href {https://doi.org/10.1103/PhysRevLett.105.177002}
  {\bibfield  {journal} {\bibinfo  {journal} {Phys. Rev. Lett.}\ }\textbf
  {\bibinfo {volume} {105}},\ \bibinfo {pages} {177002} (\bibinfo {year}
  {2010})}\BibitemShut {NoStop}%
\bibitem [{\citenamefont {Alicea}(2012)}]{alicea2012new}%
  \BibitemOpen
  \bibfield  {author} {\bibinfo {author} {\bibfnamefont {J.}~\bibnamefont
  {Alicea}},\ }\bibfield  {title} {\bibinfo {title} {New directions in the
  pursuit of majorana fermions in solid state systems},\ }\href
  {https://iopscience.iop.org/article/10.1088/0034-4885/75/7/076501} {\bibfield
   {journal} {\bibinfo  {journal} {Reports on progress in physics}\ }\textbf
  {\bibinfo {volume} {75}},\ \bibinfo {pages} {076501} (\bibinfo {year}
  {2012})}\BibitemShut {NoStop}%
\bibitem [{\citenamefont {Nayak}\ \emph {et~al.}(2008)\citenamefont {Nayak},
  \citenamefont {Simon}, \citenamefont {Stern}, \citenamefont {Freedman},\ and\
  \citenamefont {Das~Sarma}}]{nayak2008RMP}%
  \BibitemOpen
  \bibfield  {author} {\bibinfo {author} {\bibfnamefont {C.}~\bibnamefont
  {Nayak}}, \bibinfo {author} {\bibfnamefont {S.~H.}\ \bibnamefont {Simon}},
  \bibinfo {author} {\bibfnamefont {A.}~\bibnamefont {Stern}}, \bibinfo
  {author} {\bibfnamefont {M.}~\bibnamefont {Freedman}},\ and\ \bibinfo
  {author} {\bibfnamefont {S.}~\bibnamefont {Das~Sarma}},\ }\bibfield  {title}
  {\bibinfo {title} {Non-abelian anyons and topological quantum computation},\
  }\href {https://doi.org/10.1103/RevModPhys.80.1083} {\bibfield  {journal}
  {\bibinfo  {journal} {Rev. Mod. Phys.}\ }\textbf {\bibinfo {volume} {80}},\
  \bibinfo {pages} {1083} (\bibinfo {year} {2008})}\BibitemShut {NoStop}%
\bibitem [{\citenamefont {Das~Sarma}(2023)}]{dassarma2023search}%
  \BibitemOpen
  \bibfield  {author} {\bibinfo {author} {\bibfnamefont {S.}~\bibnamefont
  {Das~Sarma}},\ }\bibfield  {title} {\bibinfo {title} {In search of
  majorana},\ }\href {https://doi.org/10.1038/s41567-022-01900-9} {\bibfield
  {journal} {\bibinfo  {journal} {Nature Physics}\ }\textbf {\bibinfo {volume}
  {19}},\ \bibinfo {pages} {165} (\bibinfo {year} {2023})}\BibitemShut
  {NoStop}%
\bibitem [{\citenamefont {Wang}\ \emph {et~al.}(2018)\citenamefont {Wang},
  \citenamefont {Kong}, \citenamefont {Fan}, \citenamefont {Chen},
  \citenamefont {Zhu}, \citenamefont {Liu}, \citenamefont {Cao}, \citenamefont
  {Sun}, \citenamefont {Du}, \citenamefont {Schneeloch} \emph
  {et~al.}}]{wang2018evidence}%
  \BibitemOpen
  \bibfield  {author} {\bibinfo {author} {\bibfnamefont {D.}~\bibnamefont
  {Wang}}, \bibinfo {author} {\bibfnamefont {L.}~\bibnamefont {Kong}}, \bibinfo
  {author} {\bibfnamefont {P.}~\bibnamefont {Fan}}, \bibinfo {author}
  {\bibfnamefont {H.}~\bibnamefont {Chen}}, \bibinfo {author} {\bibfnamefont
  {S.}~\bibnamefont {Zhu}}, \bibinfo {author} {\bibfnamefont {W.}~\bibnamefont
  {Liu}}, \bibinfo {author} {\bibfnamefont {L.}~\bibnamefont {Cao}}, \bibinfo
  {author} {\bibfnamefont {Y.}~\bibnamefont {Sun}}, \bibinfo {author}
  {\bibfnamefont {S.}~\bibnamefont {Du}}, \bibinfo {author} {\bibfnamefont
  {J.}~\bibnamefont {Schneeloch}}, \emph {et~al.},\ }\bibfield  {title}
  {\bibinfo {title} {Evidence for majorana bound states in an iron-based
  superconductor},\ }\href
  {https://www.science.org/doi/10.1126/science.aao1797} {\bibfield  {journal}
  {\bibinfo  {journal} {Science}\ }\textbf {\bibinfo {volume} {362}},\ \bibinfo
  {pages} {333} (\bibinfo {year} {2018})}\BibitemShut {NoStop}%
\bibitem [{\citenamefont {Machida}\ \emph {et~al.}(2019)\citenamefont
  {Machida}, \citenamefont {Sun}, \citenamefont {Pyon}, \citenamefont {Takeda},
  \citenamefont {Kohsaka}, \citenamefont {Hanaguri}, \citenamefont {Sasagawa},\
  and\ \citenamefont {Tamegai}}]{machida2019zero}%
  \BibitemOpen
  \bibfield  {author} {\bibinfo {author} {\bibfnamefont {T.}~\bibnamefont
  {Machida}}, \bibinfo {author} {\bibfnamefont {Y.}~\bibnamefont {Sun}},
  \bibinfo {author} {\bibfnamefont {S.}~\bibnamefont {Pyon}}, \bibinfo {author}
  {\bibfnamefont {S.}~\bibnamefont {Takeda}}, \bibinfo {author} {\bibfnamefont
  {Y.}~\bibnamefont {Kohsaka}}, \bibinfo {author} {\bibfnamefont
  {T.}~\bibnamefont {Hanaguri}}, \bibinfo {author} {\bibfnamefont
  {T.}~\bibnamefont {Sasagawa}},\ and\ \bibinfo {author} {\bibfnamefont
  {T.}~\bibnamefont {Tamegai}},\ }\bibfield  {title} {\bibinfo {title}
  {Zero-energy vortex bound state in the superconducting topological surface
  state of fe (se, te)},\ }\href {https://doi.org/10.1038/s41563-019-0397-1}
  {\bibfield  {journal} {\bibinfo  {journal} {Nature materials}\ }\textbf
  {\bibinfo {volume} {18}},\ \bibinfo {pages} {811} (\bibinfo {year}
  {2019})}\BibitemShut {NoStop}%
\bibitem [{\citenamefont {Kong}\ \emph {et~al.}(2019)\citenamefont {Kong},
  \citenamefont {Zhu}, \citenamefont {Papaj}, \citenamefont {Chen},
  \citenamefont {Cao}, \citenamefont {Isobe}, \citenamefont {Xing},
  \citenamefont {Liu}, \citenamefont {Wang}, \citenamefont {Fan} \emph
  {et~al.}}]{kong2019half}%
  \BibitemOpen
  \bibfield  {author} {\bibinfo {author} {\bibfnamefont {L.}~\bibnamefont
  {Kong}}, \bibinfo {author} {\bibfnamefont {S.}~\bibnamefont {Zhu}}, \bibinfo
  {author} {\bibfnamefont {M.}~\bibnamefont {Papaj}}, \bibinfo {author}
  {\bibfnamefont {H.}~\bibnamefont {Chen}}, \bibinfo {author} {\bibfnamefont
  {L.}~\bibnamefont {Cao}}, \bibinfo {author} {\bibfnamefont {H.}~\bibnamefont
  {Isobe}}, \bibinfo {author} {\bibfnamefont {Y.}~\bibnamefont {Xing}},
  \bibinfo {author} {\bibfnamefont {W.}~\bibnamefont {Liu}}, \bibinfo {author}
  {\bibfnamefont {D.}~\bibnamefont {Wang}}, \bibinfo {author} {\bibfnamefont
  {P.}~\bibnamefont {Fan}}, \emph {et~al.},\ }\bibfield  {title} {\bibinfo
  {title} {Half-integer level shift of vortex bound states in an iron-based
  superconductor},\ }\href {https://doi.org/10.1038/s41567-019-0630-5}
  {\bibfield  {journal} {\bibinfo  {journal} {Nature Physics}\ }\textbf
  {\bibinfo {volume} {15}},\ \bibinfo {pages} {1181} (\bibinfo {year}
  {2019})}\BibitemShut {NoStop}%
\bibitem [{\citenamefont {Zhu}\ \emph {et~al.}(2020)\citenamefont {Zhu},
  \citenamefont {Kong}, \citenamefont {Cao}, \citenamefont {Chen},
  \citenamefont {Papaj}, \citenamefont {Du}, \citenamefont {Xing},
  \citenamefont {Liu}, \citenamefont {Wang}, \citenamefont {Shen} \emph
  {et~al.}}]{zhu2020nearly}%
  \BibitemOpen
  \bibfield  {author} {\bibinfo {author} {\bibfnamefont {S.}~\bibnamefont
  {Zhu}}, \bibinfo {author} {\bibfnamefont {L.}~\bibnamefont {Kong}}, \bibinfo
  {author} {\bibfnamefont {L.}~\bibnamefont {Cao}}, \bibinfo {author}
  {\bibfnamefont {H.}~\bibnamefont {Chen}}, \bibinfo {author} {\bibfnamefont
  {M.}~\bibnamefont {Papaj}}, \bibinfo {author} {\bibfnamefont
  {S.}~\bibnamefont {Du}}, \bibinfo {author} {\bibfnamefont {Y.}~\bibnamefont
  {Xing}}, \bibinfo {author} {\bibfnamefont {W.}~\bibnamefont {Liu}}, \bibinfo
  {author} {\bibfnamefont {D.}~\bibnamefont {Wang}}, \bibinfo {author}
  {\bibfnamefont {C.}~\bibnamefont {Shen}}, \emph {et~al.},\ }\bibfield
  {title} {\bibinfo {title} {Nearly quantized conductance plateau of vortex
  zero mode in an iron-based superconductor},\ }\href
  {https://www.science.org/doi/10.1126/science.aax0274} {\bibfield  {journal}
  {\bibinfo  {journal} {Science}\ }\textbf {\bibinfo {volume} {367}},\ \bibinfo
  {pages} {189} (\bibinfo {year} {2020})}\BibitemShut {NoStop}%
\bibitem [{\citenamefont {Liu}\ \emph {et~al.}(2018)\citenamefont {Liu},
  \citenamefont {Chen}, \citenamefont {Zhang}, \citenamefont {Peng},
  \citenamefont {Yan}, \citenamefont {Wen}, \citenamefont {Lou}, \citenamefont
  {Huang}, \citenamefont {Tian}, \citenamefont {Dong}, \citenamefont {Wang},
  \citenamefont {Bao}, \citenamefont {Wang}, \citenamefont {Yin}, \citenamefont
  {Zhao},\ and\ \citenamefont {Feng}}]{liu2018LiFeOH}%
  \BibitemOpen
  \bibfield  {author} {\bibinfo {author} {\bibfnamefont {Q.}~\bibnamefont
  {Liu}}, \bibinfo {author} {\bibfnamefont {C.}~\bibnamefont {Chen}}, \bibinfo
  {author} {\bibfnamefont {T.}~\bibnamefont {Zhang}}, \bibinfo {author}
  {\bibfnamefont {R.}~\bibnamefont {Peng}}, \bibinfo {author} {\bibfnamefont
  {Y.-J.}\ \bibnamefont {Yan}}, \bibinfo {author} {\bibfnamefont {C.-H.-P.}\
  \bibnamefont {Wen}}, \bibinfo {author} {\bibfnamefont {X.}~\bibnamefont
  {Lou}}, \bibinfo {author} {\bibfnamefont {Y.-L.}\ \bibnamefont {Huang}},
  \bibinfo {author} {\bibfnamefont {J.-P.}\ \bibnamefont {Tian}}, \bibinfo
  {author} {\bibfnamefont {X.-L.}\ \bibnamefont {Dong}}, \bibinfo {author}
  {\bibfnamefont {G.-W.}\ \bibnamefont {Wang}}, \bibinfo {author}
  {\bibfnamefont {W.-C.}\ \bibnamefont {Bao}}, \bibinfo {author} {\bibfnamefont
  {Q.-H.}\ \bibnamefont {Wang}}, \bibinfo {author} {\bibfnamefont {Z.-P.}\
  \bibnamefont {Yin}}, \bibinfo {author} {\bibfnamefont {Z.-X.}\ \bibnamefont
  {Zhao}},\ and\ \bibinfo {author} {\bibfnamefont {D.-L.}\ \bibnamefont
  {Feng}},\ }\bibfield  {title} {\bibinfo {title} {Robust and clean majorana
  zero mode in the vortex core of high-temperature superconductor
  $\mathbf{(}{\mathrm{li}}_{0.84}{\mathrm{fe}}_{0.16}\mathbf{)}\mathrm{OHFeSe}$},\
  }\href {https://doi.org/10.1103/PhysRevX.8.041056} {\bibfield  {journal}
  {\bibinfo  {journal} {Phys. Rev. X}\ }\textbf {\bibinfo {volume} {8}},\
  \bibinfo {pages} {041056} (\bibinfo {year} {2018})}\BibitemShut {NoStop}%
\bibitem [{\citenamefont {Yuan}\ \emph {et~al.}(2019)\citenamefont {Yuan},
  \citenamefont {Pan}, \citenamefont {Wang}, \citenamefont {Fang},
  \citenamefont {Song}, \citenamefont {Wang}, \citenamefont {He}, \citenamefont
  {Ma}, \citenamefont {Zhang}, \citenamefont {Huang} \emph
  {et~al.}}]{yuan2019evidence}%
  \BibitemOpen
  \bibfield  {author} {\bibinfo {author} {\bibfnamefont {Y.}~\bibnamefont
  {Yuan}}, \bibinfo {author} {\bibfnamefont {J.}~\bibnamefont {Pan}}, \bibinfo
  {author} {\bibfnamefont {X.}~\bibnamefont {Wang}}, \bibinfo {author}
  {\bibfnamefont {Y.}~\bibnamefont {Fang}}, \bibinfo {author} {\bibfnamefont
  {C.}~\bibnamefont {Song}}, \bibinfo {author} {\bibfnamefont {L.}~\bibnamefont
  {Wang}}, \bibinfo {author} {\bibfnamefont {K.}~\bibnamefont {He}}, \bibinfo
  {author} {\bibfnamefont {X.}~\bibnamefont {Ma}}, \bibinfo {author}
  {\bibfnamefont {H.}~\bibnamefont {Zhang}}, \bibinfo {author} {\bibfnamefont
  {F.}~\bibnamefont {Huang}}, \emph {et~al.},\ }\bibfield  {title} {\bibinfo
  {title} {Evidence of anisotropic majorana bound states in 2m-ws2},\ }\href
  {https://doi.org/10.1038/s41567-019-0576-7} {\bibfield  {journal} {\bibinfo
  {journal} {Nature Physics}\ }\textbf {\bibinfo {volume} {15}},\ \bibinfo
  {pages} {1046} (\bibinfo {year} {2019})}\BibitemShut {NoStop}%
\bibitem [{\citenamefont {Fan}\ \emph {et~al.}(2024)\citenamefont {Fan},
  \citenamefont {Sun}, \citenamefont {Zhu}, \citenamefont {Fang}, \citenamefont
  {Ju}, \citenamefont {Yuan}, \citenamefont {Yan}, \citenamefont {Huang},
  \citenamefont {Hughes}, \citenamefont {Tang}, \citenamefont {Xue},\ and\
  \citenamefont {Li}}]{fan2024stripe}%
  \BibitemOpen
  \bibfield  {author} {\bibinfo {author} {\bibfnamefont {X.}~\bibnamefont
  {Fan}}, \bibinfo {author} {\bibfnamefont {X.-Q.}\ \bibnamefont {Sun}},
  \bibinfo {author} {\bibfnamefont {P.}~\bibnamefont {Zhu}}, \bibinfo {author}
  {\bibfnamefont {Y.}~\bibnamefont {Fang}}, \bibinfo {author} {\bibfnamefont
  {Y.}~\bibnamefont {Ju}}, \bibinfo {author} {\bibfnamefont {Y.}~\bibnamefont
  {Yuan}}, \bibinfo {author} {\bibfnamefont {J.}~\bibnamefont {Yan}}, \bibinfo
  {author} {\bibfnamefont {F.}~\bibnamefont {Huang}}, \bibinfo {author}
  {\bibfnamefont {T.~L.}\ \bibnamefont {Hughes}}, \bibinfo {author}
  {\bibfnamefont {P.}~\bibnamefont {Tang}}, \bibinfo {author} {\bibfnamefont
  {Q.-K.}\ \bibnamefont {Xue}},\ and\ \bibinfo {author} {\bibfnamefont
  {W.}~\bibnamefont {Li}},\ }\bibfield  {title} {\bibinfo {title} {{Stripe
  charge order and its interaction with Majorana bound states in 2M-WS2
  topological superconductor}},\ }\href {https://doi.org/10.1093/nsr/nwae312}
  {\bibfield  {journal} {\bibinfo  {journal} {National Science Review}\ ,\
  \bibinfo {pages} {nwae312}} (\bibinfo {year} {2024})}\BibitemShut {NoStop}%
\bibitem [{\citenamefont {Kong}\ \emph {et~al.}(2021)\citenamefont {Kong},
  \citenamefont {Cao}, \citenamefont {Zhu}, \citenamefont {Papaj},
  \citenamefont {Dai}, \citenamefont {Li}, \citenamefont {Fan}, \citenamefont
  {Liu}, \citenamefont {Yang}, \citenamefont {Wang} \emph
  {et~al.}}]{kong2021majorana}%
  \BibitemOpen
  \bibfield  {author} {\bibinfo {author} {\bibfnamefont {L.}~\bibnamefont
  {Kong}}, \bibinfo {author} {\bibfnamefont {L.}~\bibnamefont {Cao}}, \bibinfo
  {author} {\bibfnamefont {S.}~\bibnamefont {Zhu}}, \bibinfo {author}
  {\bibfnamefont {M.}~\bibnamefont {Papaj}}, \bibinfo {author} {\bibfnamefont
  {G.}~\bibnamefont {Dai}}, \bibinfo {author} {\bibfnamefont {G.}~\bibnamefont
  {Li}}, \bibinfo {author} {\bibfnamefont {P.}~\bibnamefont {Fan}}, \bibinfo
  {author} {\bibfnamefont {W.}~\bibnamefont {Liu}}, \bibinfo {author}
  {\bibfnamefont {F.}~\bibnamefont {Yang}}, \bibinfo {author} {\bibfnamefont
  {X.}~\bibnamefont {Wang}}, \emph {et~al.},\ }\bibfield  {title} {\bibinfo
  {title} {Majorana zero modes in impurity-assisted vortex of lifeas
  superconductor},\ }\href {https://doi.org/10.1038/s41467-021-24372-6}
  {\bibfield  {journal} {\bibinfo  {journal} {Nature Communications}\ }\textbf
  {\bibinfo {volume} {12}},\ \bibinfo {pages} {4146} (\bibinfo {year}
  {2021})}\BibitemShut {NoStop}%
\bibitem [{\citenamefont {Liu}\ \emph {et~al.}(2022)\citenamefont {Liu},
  \citenamefont {Hu}, \citenamefont {Wang}, \citenamefont {Zhong},
  \citenamefont {Yang}, \citenamefont {Kong}, \citenamefont {Cao},
  \citenamefont {Li}, \citenamefont {Peng}, \citenamefont {Okazaki} \emph
  {et~al.}}]{liu2022tunable}%
  \BibitemOpen
  \bibfield  {author} {\bibinfo {author} {\bibfnamefont {W.}~\bibnamefont
  {Liu}}, \bibinfo {author} {\bibfnamefont {Q.}~\bibnamefont {Hu}}, \bibinfo
  {author} {\bibfnamefont {X.}~\bibnamefont {Wang}}, \bibinfo {author}
  {\bibfnamefont {Y.}~\bibnamefont {Zhong}}, \bibinfo {author} {\bibfnamefont
  {F.}~\bibnamefont {Yang}}, \bibinfo {author} {\bibfnamefont {L.}~\bibnamefont
  {Kong}}, \bibinfo {author} {\bibfnamefont {L.}~\bibnamefont {Cao}}, \bibinfo
  {author} {\bibfnamefont {G.}~\bibnamefont {Li}}, \bibinfo {author}
  {\bibfnamefont {Y.}~\bibnamefont {Peng}}, \bibinfo {author} {\bibfnamefont
  {K.}~\bibnamefont {Okazaki}}, \emph {et~al.},\ }\bibfield  {title} {\bibinfo
  {title} {Tunable vortex majorana modes controlled by strain in homogeneous
  lifeas},\ }\href {https://doi.org/10.1007/s44214-022-00022-w} {\bibfield
  {journal} {\bibinfo  {journal} {Quantum Frontiers}\ }\textbf {\bibinfo
  {volume} {1}},\ \bibinfo {pages} {20} (\bibinfo {year} {2022})}\BibitemShut
  {NoStop}%
\bibitem [{\citenamefont {Li}\ \emph {et~al.}(2022)\citenamefont {Li},
  \citenamefont {Li}, \citenamefont {Cao}, \citenamefont {Zhou}, \citenamefont
  {Wang}, \citenamefont {Jin}, \citenamefont {Chiu}, \citenamefont {Pennycook},
  \citenamefont {Wang},\ and\ \citenamefont {Gao}}]{li2022ordered}%
  \BibitemOpen
  \bibfield  {author} {\bibinfo {author} {\bibfnamefont {M.}~\bibnamefont
  {Li}}, \bibinfo {author} {\bibfnamefont {G.}~\bibnamefont {Li}}, \bibinfo
  {author} {\bibfnamefont {L.}~\bibnamefont {Cao}}, \bibinfo {author}
  {\bibfnamefont {X.}~\bibnamefont {Zhou}}, \bibinfo {author} {\bibfnamefont
  {X.}~\bibnamefont {Wang}}, \bibinfo {author} {\bibfnamefont {C.}~\bibnamefont
  {Jin}}, \bibinfo {author} {\bibfnamefont {C.-K.}\ \bibnamefont {Chiu}},
  \bibinfo {author} {\bibfnamefont {S.~J.}\ \bibnamefont {Pennycook}}, \bibinfo
  {author} {\bibfnamefont {Z.}~\bibnamefont {Wang}},\ and\ \bibinfo {author}
  {\bibfnamefont {H.-J.}\ \bibnamefont {Gao}},\ }\bibfield  {title} {\bibinfo
  {title} {Ordered and tunable majorana-zero-mode lattice in naturally strained
  lifeas},\ }\href {https://doi.org/10.1038/s41586-022-04744-8} {\bibfield
  {journal} {\bibinfo  {journal} {Nature}\ }\textbf {\bibinfo {volume} {606}},\
  \bibinfo {pages} {890} (\bibinfo {year} {2022})}\BibitemShut {NoStop}%
\bibitem [{\citenamefont {Fu}\ and\ \citenamefont {Kane}(2008)}]{fu2008vortex}%
  \BibitemOpen
  \bibfield  {author} {\bibinfo {author} {\bibfnamefont {L.}~\bibnamefont
  {Fu}}\ and\ \bibinfo {author} {\bibfnamefont {C.~L.}\ \bibnamefont {Kane}},\
  }\bibfield  {title} {\bibinfo {title} {Superconducting proximity effect and
  majorana fermions at the surface of a topological insulator},\ }\href
  {https://doi.org/10.1103/PhysRevLett.100.096407} {\bibfield  {journal}
  {\bibinfo  {journal} {Phys. Rev. Lett.}\ }\textbf {\bibinfo {volume} {100}},\
  \bibinfo {pages} {096407} (\bibinfo {year} {2008})}\BibitemShut {NoStop}%
\bibitem [{\citenamefont {Zhang}\ \emph {et~al.}(2019)\citenamefont {Zhang},
  \citenamefont {Wang}, \citenamefont {Wu}, \citenamefont {Yaji}, \citenamefont
  {Ishida}, \citenamefont {Kohama}, \citenamefont {Dai}, \citenamefont {Sun},
  \citenamefont {Bareille}, \citenamefont {Kuroda} \emph
  {et~al.}}]{zhang2019multiple}%
  \BibitemOpen
  \bibfield  {author} {\bibinfo {author} {\bibfnamefont {P.}~\bibnamefont
  {Zhang}}, \bibinfo {author} {\bibfnamefont {Z.}~\bibnamefont {Wang}},
  \bibinfo {author} {\bibfnamefont {X.}~\bibnamefont {Wu}}, \bibinfo {author}
  {\bibfnamefont {K.}~\bibnamefont {Yaji}}, \bibinfo {author} {\bibfnamefont
  {Y.}~\bibnamefont {Ishida}}, \bibinfo {author} {\bibfnamefont
  {Y.}~\bibnamefont {Kohama}}, \bibinfo {author} {\bibfnamefont
  {G.}~\bibnamefont {Dai}}, \bibinfo {author} {\bibfnamefont {Y.}~\bibnamefont
  {Sun}}, \bibinfo {author} {\bibfnamefont {C.}~\bibnamefont {Bareille}},
  \bibinfo {author} {\bibfnamefont {K.}~\bibnamefont {Kuroda}}, \emph
  {et~al.},\ }\bibfield  {title} {\bibinfo {title} {Multiple topological states
  in iron-based superconductors},\ }\href
  {https://doi.org/10.1038/s41567-018-0280-z} {\bibfield  {journal} {\bibinfo
  {journal} {Nature Physics}\ }\textbf {\bibinfo {volume} {15}},\ \bibinfo
  {pages} {41} (\bibinfo {year} {2019})}\BibitemShut {NoStop}%
\bibitem [{\citenamefont {Chiu}\ \emph {et~al.}(2012)\citenamefont {Chiu},
  \citenamefont {Ghaemi},\ and\ \citenamefont
  {Hughes}}]{chiu2012stabilization}%
  \BibitemOpen
  \bibfield  {author} {\bibinfo {author} {\bibfnamefont {C.-K.}\ \bibnamefont
  {Chiu}}, \bibinfo {author} {\bibfnamefont {P.}~\bibnamefont {Ghaemi}},\ and\
  \bibinfo {author} {\bibfnamefont {T.~L.}\ \bibnamefont {Hughes}},\ }\bibfield
   {title} {\bibinfo {title} {Stabilization of majorana modes in magnetic
  vortices in the superconducting phase of topological insulators using
  topologically trivial bands},\ }\href
  {https://doi.org/10.1103/PhysRevLett.109.237009} {\bibfield  {journal}
  {\bibinfo  {journal} {Phys. Rev. Lett.}\ }\textbf {\bibinfo {volume} {109}},\
  \bibinfo {pages} {237009} (\bibinfo {year} {2012})}\BibitemShut {NoStop}%
\bibitem [{\citenamefont {Xu}\ \emph {et~al.}(2016)\citenamefont {Xu},
  \citenamefont {Lian}, \citenamefont {Tang}, \citenamefont {Qi},\ and\
  \citenamefont {Zhang}}]{xu2016topological}%
  \BibitemOpen
  \bibfield  {author} {\bibinfo {author} {\bibfnamefont {G.}~\bibnamefont
  {Xu}}, \bibinfo {author} {\bibfnamefont {B.}~\bibnamefont {Lian}}, \bibinfo
  {author} {\bibfnamefont {P.}~\bibnamefont {Tang}}, \bibinfo {author}
  {\bibfnamefont {X.-L.}\ \bibnamefont {Qi}},\ and\ \bibinfo {author}
  {\bibfnamefont {S.-C.}\ \bibnamefont {Zhang}},\ }\bibfield  {title} {\bibinfo
  {title} {Topological superconductivity on the surface of fe-based
  superconductors},\ }\href {https://doi.org/10.1103/PhysRevLett.117.047001}
  {\bibfield  {journal} {\bibinfo  {journal} {Phys. Rev. Lett.}\ }\textbf
  {\bibinfo {volume} {117}},\ \bibinfo {pages} {047001} (\bibinfo {year}
  {2016})}\BibitemShut {NoStop}%
\bibitem [{\citenamefont {Hu}\ \emph {et~al.}(2022)\citenamefont {Hu},
  \citenamefont {Wu}, \citenamefont {Liu},\ and\ \citenamefont
  {Zhang}}]{hu2022competing}%
  \BibitemOpen
  \bibfield  {author} {\bibinfo {author} {\bibfnamefont {L.-H.}\ \bibnamefont
  {Hu}}, \bibinfo {author} {\bibfnamefont {X.}~\bibnamefont {Wu}}, \bibinfo
  {author} {\bibfnamefont {C.-X.}\ \bibnamefont {Liu}},\ and\ \bibinfo {author}
  {\bibfnamefont {R.-X.}\ \bibnamefont {Zhang}},\ }\bibfield  {title} {\bibinfo
  {title} {Competing vortex topologies in iron-based superconductors},\ }\href
  {https://doi.org/10.1103/PhysRevLett.129.277001} {\bibfield  {journal}
  {\bibinfo  {journal} {Phys. Rev. Lett.}\ }\textbf {\bibinfo {volume} {129}},\
  \bibinfo {pages} {277001} (\bibinfo {year} {2022})}\BibitemShut {NoStop}%
\bibitem [{\citenamefont {Hu}\ and\ \citenamefont
  {Zhang}(2023)}]{hu2023topological}%
  \BibitemOpen
  \bibfield  {author} {\bibinfo {author} {\bibfnamefont {L.-H.}\ \bibnamefont
  {Hu}}\ and\ \bibinfo {author} {\bibfnamefont {R.-X.}\ \bibnamefont {Zhang}},\
  }\bibfield  {title} {\bibinfo {title} {Topological superconducting vortex
  from trivial electronic bands},\ }\href
  {https://doi.org/10.1038/s41467-023-36347-w} {\bibfield  {journal} {\bibinfo
  {journal} {Nature Communications}\ }\textbf {\bibinfo {volume} {14}},\
  \bibinfo {pages} {640} (\bibinfo {year} {2023})}\BibitemShut {NoStop}%
\bibitem [{\citenamefont {Pan}\ and\ \citenamefont
  {Das~Sarma}(2020)}]{pan2020physical}%
  \BibitemOpen
  \bibfield  {author} {\bibinfo {author} {\bibfnamefont {H.}~\bibnamefont
  {Pan}}\ and\ \bibinfo {author} {\bibfnamefont {S.}~\bibnamefont
  {Das~Sarma}},\ }\bibfield  {title} {\bibinfo {title} {Physical mechanisms for
  zero-bias conductance peaks in majorana nanowires},\ }\href
  {https://doi.org/10.1103/PhysRevResearch.2.013377} {\bibfield  {journal}
  {\bibinfo  {journal} {Phys. Rev. Res.}\ }\textbf {\bibinfo {volume} {2}},\
  \bibinfo {pages} {013377} (\bibinfo {year} {2020})}\BibitemShut {NoStop}%
\bibitem [{\citenamefont {Caroli}\ \emph {et~al.}(1964)\citenamefont {Caroli},
  \citenamefont {De~Gennes},\ and\ \citenamefont {Matricon}}]{caroli1964bound}%
  \BibitemOpen
  \bibfield  {author} {\bibinfo {author} {\bibfnamefont {C.}~\bibnamefont
  {Caroli}}, \bibinfo {author} {\bibfnamefont {P.}~\bibnamefont {De~Gennes}},\
  and\ \bibinfo {author} {\bibfnamefont {J.}~\bibnamefont {Matricon}},\
  }\bibfield  {title} {\bibinfo {title} {Bound fermion states on a vortex line
  in a type ii superconductor},\ }\href
  {https://doi.org/10.1016/0031-9163(64)90375-0} {\bibfield  {journal}
  {\bibinfo  {journal} {Physics Letters}\ }\textbf {\bibinfo {volume} {9}},\
  \bibinfo {pages} {307} (\bibinfo {year} {1964})}\BibitemShut {NoStop}%
\bibitem [{\citenamefont {Hosur}\ \emph {et~al.}(2011)\citenamefont {Hosur},
  \citenamefont {Ghaemi}, \citenamefont {Mong},\ and\ \citenamefont
  {Vishwanath}}]{hosur2011vortex}%
  \BibitemOpen
  \bibfield  {author} {\bibinfo {author} {\bibfnamefont {P.}~\bibnamefont
  {Hosur}}, \bibinfo {author} {\bibfnamefont {P.}~\bibnamefont {Ghaemi}},
  \bibinfo {author} {\bibfnamefont {R.~S.~K.}\ \bibnamefont {Mong}},\ and\
  \bibinfo {author} {\bibfnamefont {A.}~\bibnamefont {Vishwanath}},\ }\bibfield
   {title} {\bibinfo {title} {Majorana modes at the ends of superconductor
  vortices in doped topological insulators},\ }\href
  {https://doi.org/10.1103/PhysRevLett.107.097001} {\bibfield  {journal}
  {\bibinfo  {journal} {Phys. Rev. Lett.}\ }\textbf {\bibinfo {volume} {107}},\
  \bibinfo {pages} {097001} (\bibinfo {year} {2011})}\BibitemShut {NoStop}%
\bibitem [{\citenamefont {Qi}\ \emph {et~al.}(2008)\citenamefont {Qi},
  \citenamefont {Hughes},\ and\ \citenamefont {Zhang}}]{qi2008TQFT}%
  \BibitemOpen
  \bibfield  {author} {\bibinfo {author} {\bibfnamefont {X.-L.}\ \bibnamefont
  {Qi}}, \bibinfo {author} {\bibfnamefont {T.~L.}\ \bibnamefont {Hughes}},\
  and\ \bibinfo {author} {\bibfnamefont {S.-C.}\ \bibnamefont {Zhang}},\
  }\bibfield  {title} {\bibinfo {title} {Topological field theory of
  time-reversal invariant insulators},\ }\href
  {https://doi.org/10.1103/PhysRevB.78.195424} {\bibfield  {journal} {\bibinfo
  {journal} {Phys. Rev. B}\ }\textbf {\bibinfo {volume} {78}},\ \bibinfo
  {pages} {195424} (\bibinfo {year} {2008})}\BibitemShut {NoStop}%
\bibitem [{\citenamefont {Huang}\ \emph {et~al.}(2024)\citenamefont {Huang},
  \citenamefont {Yu},\ and\ \citenamefont {Zhang}}]{huang2024class}%
  \BibitemOpen
  \bibfield  {author} {\bibinfo {author} {\bibfnamefont {S.-J.}\ \bibnamefont
  {Huang}}, \bibinfo {author} {\bibfnamefont {J.}~\bibnamefont {Yu}},\ and\
  \bibinfo {author} {\bibfnamefont {R.-X.}\ \bibnamefont {Zhang}},\ }\bibfield
  {title} {\bibinfo {title} {Classification of interacting dirac semimetals},\
  }\href {https://doi.org/10.1103/PhysRevB.110.035134} {\bibfield  {journal}
  {\bibinfo  {journal} {Phys. Rev. B}\ }\textbf {\bibinfo {volume} {110}},\
  \bibinfo {pages} {035134} (\bibinfo {year} {2024})}\BibitemShut {NoStop}%
\bibitem [{\citenamefont {Zhang}(2022)}]{zhang2022bulk}%
  \BibitemOpen
  \bibfield  {author} {\bibinfo {author} {\bibfnamefont {R.-X.}\ \bibnamefont
  {Zhang}},\ }\bibfield  {title} {\bibinfo {title} {Bulk-vortex correspondence
  of higher-order topological superconductors},\ }\href
  {https://doi.org/10.48550/arXiv.2208.01652} {\bibfield  {journal} {\bibinfo
  {journal} {arXiv preprint arXiv:2208.01652}\ } (\bibinfo {year}
  {2022})}\BibitemShut {NoStop}%
\bibitem [{\citenamefont {Fu}\ \emph {et~al.}(2007)\citenamefont {Fu},
  \citenamefont {Kane},\ and\ \citenamefont {Mele}}]{fu20073dTI}%
  \BibitemOpen
  \bibfield  {author} {\bibinfo {author} {\bibfnamefont {L.}~\bibnamefont
  {Fu}}, \bibinfo {author} {\bibfnamefont {C.~L.}\ \bibnamefont {Kane}},\ and\
  \bibinfo {author} {\bibfnamefont {E.~J.}\ \bibnamefont {Mele}},\ }\bibfield
  {title} {\bibinfo {title} {Topological insulators in three dimensions},\
  }\href {https://doi.org/10.1103/PhysRevLett.98.106803} {\bibfield  {journal}
  {\bibinfo  {journal} {Phys. Rev. Lett.}\ }\textbf {\bibinfo {volume} {98}},\
  \bibinfo {pages} {106803} (\bibinfo {year} {2007})}\BibitemShut {NoStop}%
\bibitem [{\citenamefont {Moore}\ and\ \citenamefont
  {Balents}(2007)}]{moore20073dTI}%
  \BibitemOpen
  \bibfield  {author} {\bibinfo {author} {\bibfnamefont {J.~E.}\ \bibnamefont
  {Moore}}\ and\ \bibinfo {author} {\bibfnamefont {L.}~\bibnamefont
  {Balents}},\ }\bibfield  {title} {\bibinfo {title} {Topological invariants of
  time-reversal-invariant band structures},\ }\href
  {https://doi.org/10.1103/PhysRevB.75.121306} {\bibfield  {journal} {\bibinfo
  {journal} {Phys. Rev. B}\ }\textbf {\bibinfo {volume} {75}},\ \bibinfo
  {pages} {121306} (\bibinfo {year} {2007})}\BibitemShut {NoStop}%
\bibitem [{\citenamefont {Roy}(2009)}]{roy20093dTI}%
  \BibitemOpen
  \bibfield  {author} {\bibinfo {author} {\bibfnamefont {R.}~\bibnamefont
  {Roy}},\ }\bibfield  {title} {\bibinfo {title} {Topological phases and the
  quantum spin hall effect in three dimensions},\ }\href
  {https://doi.org/10.1103/PhysRevB.79.195322} {\bibfield  {journal} {\bibinfo
  {journal} {Phys. Rev. B}\ }\textbf {\bibinfo {volume} {79}},\ \bibinfo
  {pages} {195322} (\bibinfo {year} {2009})}\BibitemShut {NoStop}%
\bibitem [{\citenamefont {Armitage}\ \emph {et~al.}(2018)\citenamefont
  {Armitage}, \citenamefont {Mele},\ and\ \citenamefont
  {Vishwanath}}]{armitage2018RMP}%
  \BibitemOpen
  \bibfield  {author} {\bibinfo {author} {\bibfnamefont {N.~P.}\ \bibnamefont
  {Armitage}}, \bibinfo {author} {\bibfnamefont {E.~J.}\ \bibnamefont {Mele}},\
  and\ \bibinfo {author} {\bibfnamefont {A.}~\bibnamefont {Vishwanath}},\
  }\bibfield  {title} {\bibinfo {title} {Weyl and dirac semimetals in
  three-dimensional solids},\ }\href
  {https://doi.org/10.1103/RevModPhys.90.015001} {\bibfield  {journal}
  {\bibinfo  {journal} {Rev. Mod. Phys.}\ }\textbf {\bibinfo {volume} {90}},\
  \bibinfo {pages} {015001} (\bibinfo {year} {2018})}\BibitemShut {NoStop}%
\bibitem [{\citenamefont {K\"onig}\ and\ \citenamefont
  {Coleman}(2019)}]{konig2019nodal}%
  \BibitemOpen
  \bibfield  {author} {\bibinfo {author} {\bibfnamefont {E.~J.}\ \bibnamefont
  {K\"onig}}\ and\ \bibinfo {author} {\bibfnamefont {P.}~\bibnamefont
  {Coleman}},\ }\bibfield  {title} {\bibinfo {title}
  {Crystalline-symmetry-protected helical majorana modes in the iron
  pnictides},\ }\href {https://doi.org/10.1103/PhysRevLett.122.207001}
  {\bibfield  {journal} {\bibinfo  {journal} {Phys. Rev. Lett.}\ }\textbf
  {\bibinfo {volume} {122}},\ \bibinfo {pages} {207001} (\bibinfo {year}
  {2019})}\BibitemShut {NoStop}%
\bibitem [{\citenamefont {Qin}\ \emph {et~al.}(2019)\citenamefont {Qin},
  \citenamefont {Hu}, \citenamefont {Le}, \citenamefont {Zeng}, \citenamefont
  {Zhang}, \citenamefont {Fang},\ and\ \citenamefont {Hu}}]{qin2019nodal}%
  \BibitemOpen
  \bibfield  {author} {\bibinfo {author} {\bibfnamefont {S.}~\bibnamefont
  {Qin}}, \bibinfo {author} {\bibfnamefont {L.}~\bibnamefont {Hu}}, \bibinfo
  {author} {\bibfnamefont {C.}~\bibnamefont {Le}}, \bibinfo {author}
  {\bibfnamefont {J.}~\bibnamefont {Zeng}}, \bibinfo {author} {\bibfnamefont
  {F.-c.}\ \bibnamefont {Zhang}}, \bibinfo {author} {\bibfnamefont
  {C.}~\bibnamefont {Fang}},\ and\ \bibinfo {author} {\bibfnamefont
  {J.}~\bibnamefont {Hu}},\ }\bibfield  {title} {\bibinfo {title} {Quasi-1d
  topological nodal vortex line phase in doped superconducting 3d dirac
  semimetals},\ }\href {https://doi.org/10.1103/PhysRevLett.123.027003}
  {\bibfield  {journal} {\bibinfo  {journal} {Phys. Rev. Lett.}\ }\textbf
  {\bibinfo {volume} {123}},\ \bibinfo {pages} {027003} (\bibinfo {year}
  {2019})}\BibitemShut {NoStop}%
\bibitem [{\citenamefont {Hsieh}\ \emph {et~al.}(2014)\citenamefont {Hsieh},
  \citenamefont {Liu},\ and\ \citenamefont {Fu}}]{hsieh2014topological}%
  \BibitemOpen
  \bibfield  {author} {\bibinfo {author} {\bibfnamefont {T.~H.}\ \bibnamefont
  {Hsieh}}, \bibinfo {author} {\bibfnamefont {J.}~\bibnamefont {Liu}},\ and\
  \bibinfo {author} {\bibfnamefont {L.}~\bibnamefont {Fu}},\ }\bibfield
  {title} {\bibinfo {title} {Topological crystalline insulators and dirac
  octets in antiperovskites},\ }\href
  {https://doi.org/10.1103/PhysRevB.90.081112} {\bibfield  {journal} {\bibinfo
  {journal} {Phys. Rev. B}\ }\textbf {\bibinfo {volume} {90}},\ \bibinfo
  {pages} {081112} (\bibinfo {year} {2014})}\BibitemShut {NoStop}%
\bibitem [{\citenamefont {Kariyado}\ and\ \citenamefont
  {Ogata}(2011)}]{kariyado2011three}%
  \BibitemOpen
  \bibfield  {author} {\bibinfo {author} {\bibfnamefont {T.}~\bibnamefont
  {Kariyado}}\ and\ \bibinfo {author} {\bibfnamefont {M.}~\bibnamefont
  {Ogata}},\ }\bibfield  {title} {\bibinfo {title} {Three-dimensional dirac
  electrons at the fermi energy in cubic inverse perovskites: Ca3pbo and its
  family},\ }\href {https://doi.org/10.1143/JPSJ.80.083704} {\bibfield
  {journal} {\bibinfo  {journal} {Journal of the Physical Society of Japan}\
  }\textbf {\bibinfo {volume} {80}},\ \bibinfo {pages} {083704} (\bibinfo
  {year} {2011})}\BibitemShut {NoStop}%
\bibitem [{\citenamefont {Kariyado}\ and\ \citenamefont
  {Ogata}(2012)}]{kariyado2012low}%
  \BibitemOpen
  \bibfield  {author} {\bibinfo {author} {\bibfnamefont {T.}~\bibnamefont
  {Kariyado}}\ and\ \bibinfo {author} {\bibfnamefont {M.}~\bibnamefont
  {Ogata}},\ }\bibfield  {title} {\bibinfo {title} {Low-energy effective
  hamiltonian and the surface states of ca3pbo},\ }\href
  {https://doi.org/10.1143/JPSJ.81.064701} {\bibfield  {journal} {\bibinfo
  {journal} {Journal of the Physical Society of Japan}\ }\textbf {\bibinfo
  {volume} {81}},\ \bibinfo {pages} {064701} (\bibinfo {year}
  {2012})}\BibitemShut {NoStop}%
\bibitem [{\citenamefont {Kawakami}\ \emph {et~al.}(2018)\citenamefont
  {Kawakami}, \citenamefont {Okamura}, \citenamefont {Kobayashi},\ and\
  \citenamefont {Sato}}]{kawakami2018topological}%
  \BibitemOpen
  \bibfield  {author} {\bibinfo {author} {\bibfnamefont {T.}~\bibnamefont
  {Kawakami}}, \bibinfo {author} {\bibfnamefont {T.}~\bibnamefont {Okamura}},
  \bibinfo {author} {\bibfnamefont {S.}~\bibnamefont {Kobayashi}},\ and\
  \bibinfo {author} {\bibfnamefont {M.}~\bibnamefont {Sato}},\ }\bibfield
  {title} {\bibinfo {title} {Topological crystalline materials of $j=3/2$
  electrons: Antiperovskites, dirac points, and high winding topological
  superconductivity},\ }\href {https://doi.org/10.1103/PhysRevX.8.041026}
  {\bibfield  {journal} {\bibinfo  {journal} {Phys. Rev. X}\ }\textbf {\bibinfo
  {volume} {8}},\ \bibinfo {pages} {041026} (\bibinfo {year}
  {2018})}\BibitemShut {NoStop}%
\bibitem [{\citenamefont {Fang}\ and\ \citenamefont
  {Cano}(2020)}]{fang2020SSO}%
  \BibitemOpen
  \bibfield  {author} {\bibinfo {author} {\bibfnamefont {Y.}~\bibnamefont
  {Fang}}\ and\ \bibinfo {author} {\bibfnamefont {J.}~\bibnamefont {Cano}},\
  }\bibfield  {title} {\bibinfo {title} {Higher-order topological insulators in
  antiperovskites},\ }\href {https://doi.org/10.1103/PhysRevB.101.245110}
  {\bibfield  {journal} {\bibinfo  {journal} {Phys. Rev. B}\ }\textbf {\bibinfo
  {volume} {101}},\ \bibinfo {pages} {245110} (\bibinfo {year}
  {2020})}\BibitemShut {NoStop}%
\bibitem [{\citenamefont {Oudah}\ \emph {et~al.}(2016)\citenamefont {Oudah},
  \citenamefont {Ikeda}, \citenamefont {Hausmann}, \citenamefont {Yonezawa},
  \citenamefont {Fukumoto}, \citenamefont {Kobayashi}, \citenamefont {Sato},\
  and\ \citenamefont {Maeno}}]{oudah2016superconductivity}%
  \BibitemOpen
  \bibfield  {author} {\bibinfo {author} {\bibfnamefont {M.}~\bibnamefont
  {Oudah}}, \bibinfo {author} {\bibfnamefont {A.}~\bibnamefont {Ikeda}},
  \bibinfo {author} {\bibfnamefont {J.~N.}\ \bibnamefont {Hausmann}}, \bibinfo
  {author} {\bibfnamefont {S.}~\bibnamefont {Yonezawa}}, \bibinfo {author}
  {\bibfnamefont {T.}~\bibnamefont {Fukumoto}}, \bibinfo {author}
  {\bibfnamefont {S.}~\bibnamefont {Kobayashi}}, \bibinfo {author}
  {\bibfnamefont {M.}~\bibnamefont {Sato}},\ and\ \bibinfo {author}
  {\bibfnamefont {Y.}~\bibnamefont {Maeno}},\ }\bibfield  {title} {\bibinfo
  {title} {Superconductivity in the antiperovskite dirac-metal oxide
  sr3-xsno},\ }\href {https://doi.org/10.1038/ncomms13617} {\bibfield
  {journal} {\bibinfo  {journal} {Nature Communications}\ }\textbf {\bibinfo
  {volume} {7}},\ \bibinfo {pages} {13617} (\bibinfo {year}
  {2016})}\BibitemShut {NoStop}%
\bibitem [{\citenamefont {Haim}\ \emph {et~al.}(2016)\citenamefont {Haim},
  \citenamefont {Berg}, \citenamefont {Flensberg},\ and\ \citenamefont
  {Oreg}}]{haim2016no-go}%
  \BibitemOpen
  \bibfield  {author} {\bibinfo {author} {\bibfnamefont {A.}~\bibnamefont
  {Haim}}, \bibinfo {author} {\bibfnamefont {E.}~\bibnamefont {Berg}}, \bibinfo
  {author} {\bibfnamefont {K.}~\bibnamefont {Flensberg}},\ and\ \bibinfo
  {author} {\bibfnamefont {Y.}~\bibnamefont {Oreg}},\ }\bibfield  {title}
  {\bibinfo {title} {No-go theorem for a time-reversal invariant topological
  phase in noninteracting systems coupled to conventional superconductors},\
  }\href {https://doi.org/10.1103/PhysRevB.94.161110} {\bibfield  {journal}
  {\bibinfo  {journal} {Phys. Rev. B}\ }\textbf {\bibinfo {volume} {94}},\
  \bibinfo {pages} {161110} (\bibinfo {year} {2016})}\BibitemShut {NoStop}%
\bibitem [{\citenamefont {Qi}\ \emph {et~al.}(2009)\citenamefont {Qi},
  \citenamefont {Hughes}, \citenamefont {Raghu},\ and\ \citenamefont
  {Zhang}}]{qi2009TSC}%
  \BibitemOpen
  \bibfield  {author} {\bibinfo {author} {\bibfnamefont {X.-L.}\ \bibnamefont
  {Qi}}, \bibinfo {author} {\bibfnamefont {T.~L.}\ \bibnamefont {Hughes}},
  \bibinfo {author} {\bibfnamefont {S.}~\bibnamefont {Raghu}},\ and\ \bibinfo
  {author} {\bibfnamefont {S.-C.}\ \bibnamefont {Zhang}},\ }\bibfield  {title}
  {\bibinfo {title} {Time-reversal-invariant topological superconductors and
  superfluids in two and three dimensions},\ }\href
  {https://doi.org/10.1103/PhysRevLett.102.187001} {\bibfield  {journal}
  {\bibinfo  {journal} {Phys. Rev. Lett.}\ }\textbf {\bibinfo {volume} {102}},\
  \bibinfo {pages} {187001} (\bibinfo {year} {2009})}\BibitemShut {NoStop}%
\bibitem [{\citenamefont {Zhang}\ \emph {et~al.}(2013)\citenamefont {Zhang},
  \citenamefont {Kane},\ and\ \citenamefont {Mele}}]{zhang2013TSC}%
  \BibitemOpen
  \bibfield  {author} {\bibinfo {author} {\bibfnamefont {F.}~\bibnamefont
  {Zhang}}, \bibinfo {author} {\bibfnamefont {C.~L.}\ \bibnamefont {Kane}},\
  and\ \bibinfo {author} {\bibfnamefont {E.~J.}\ \bibnamefont {Mele}},\
  }\bibfield  {title} {\bibinfo {title} {Time-reversal-invariant topological
  superconductivity and majorana kramers pairs},\ }\href
  {https://doi.org/10.1103/PhysRevLett.111.056402} {\bibfield  {journal}
  {\bibinfo  {journal} {Phys. Rev. Lett.}\ }\textbf {\bibinfo {volume} {111}},\
  \bibinfo {pages} {056402} (\bibinfo {year} {2013})}\BibitemShut {NoStop}%
\bibitem [{\citenamefont {Skurativska}\ \emph {et~al.}(2020)\citenamefont
  {Skurativska}, \citenamefont {Neupert},\ and\ \citenamefont
  {Fischer}}]{skurativska2020atomic}%
  \BibitemOpen
  \bibfield  {author} {\bibinfo {author} {\bibfnamefont {A.}~\bibnamefont
  {Skurativska}}, \bibinfo {author} {\bibfnamefont {T.}~\bibnamefont
  {Neupert}},\ and\ \bibinfo {author} {\bibfnamefont {M.~H.}\ \bibnamefont
  {Fischer}},\ }\bibfield  {title} {\bibinfo {title} {Atomic limit and
  inversion-symmetry indicators for topological superconductors},\ }\href
  {https://doi.org/10.1103/PhysRevResearch.2.013064} {\bibfield  {journal}
  {\bibinfo  {journal} {Phys. Rev. Res.}\ }\textbf {\bibinfo {volume} {2}},\
  \bibinfo {pages} {013064} (\bibinfo {year} {2020})}\BibitemShut {NoStop}%
\bibitem [{\citenamefont {Huang}\ and\ \citenamefont
  {Hsu}(2021)}]{huang2021faithful}%
  \BibitemOpen
  \bibfield  {author} {\bibinfo {author} {\bibfnamefont {S.-J.}\ \bibnamefont
  {Huang}}\ and\ \bibinfo {author} {\bibfnamefont {Y.-T.}\ \bibnamefont
  {Hsu}},\ }\bibfield  {title} {\bibinfo {title} {Faithful derivation of
  symmetry indicators: A case study for topological superconductors with
  time-reversal and inversion symmetries},\ }\href
  {https://doi.org/10.1103/PhysRevResearch.3.013243} {\bibfield  {journal}
  {\bibinfo  {journal} {Phys. Rev. Res.}\ }\textbf {\bibinfo {volume} {3}},\
  \bibinfo {pages} {013243} (\bibinfo {year} {2021})}\BibitemShut {NoStop}%
\bibitem [{\citenamefont {Khalaf}(2018)}]{khalaf2018higher}%
  \BibitemOpen
  \bibfield  {author} {\bibinfo {author} {\bibfnamefont {E.}~\bibnamefont
  {Khalaf}},\ }\bibfield  {title} {\bibinfo {title} {Higher-order topological
  insulators and superconductors protected by inversion symmetry},\ }\href
  {https://doi.org/10.1103/PhysRevB.97.205136} {\bibfield  {journal} {\bibinfo
  {journal} {Phys. Rev. B}\ }\textbf {\bibinfo {volume} {97}},\ \bibinfo
  {pages} {205136} (\bibinfo {year} {2018})}\BibitemShut {NoStop}%
\bibitem [{\citenamefont {Hsu}\ \emph {et~al.}(2020)\citenamefont {Hsu},
  \citenamefont {Cole}, \citenamefont {Zhang},\ and\ \citenamefont
  {Sau}}]{hsu2020inversion}%
  \BibitemOpen
  \bibfield  {author} {\bibinfo {author} {\bibfnamefont {Y.-T.}\ \bibnamefont
  {Hsu}}, \bibinfo {author} {\bibfnamefont {W.~S.}\ \bibnamefont {Cole}},
  \bibinfo {author} {\bibfnamefont {R.-X.}\ \bibnamefont {Zhang}},\ and\
  \bibinfo {author} {\bibfnamefont {J.~D.}\ \bibnamefont {Sau}},\ }\bibfield
  {title} {\bibinfo {title} {Inversion-protected higher-order topological
  superconductivity in monolayer ${\mathrm{wte}}_{2}$},\ }\href
  {https://doi.org/10.1103/PhysRevLett.125.097001} {\bibfield  {journal}
  {\bibinfo  {journal} {Phys. Rev. Lett.}\ }\textbf {\bibinfo {volume} {125}},\
  \bibinfo {pages} {097001} (\bibinfo {year} {2020})}\BibitemShut {NoStop}%
\bibitem [{\citenamefont {Teo}\ \emph {et~al.}(2008)\citenamefont {Teo},
  \citenamefont {Fu},\ and\ \citenamefont {Kane}}]{teo2008surface}%
  \BibitemOpen
  \bibfield  {author} {\bibinfo {author} {\bibfnamefont {J.~C.~Y.}\
  \bibnamefont {Teo}}, \bibinfo {author} {\bibfnamefont {L.}~\bibnamefont
  {Fu}},\ and\ \bibinfo {author} {\bibfnamefont {C.~L.}\ \bibnamefont {Kane}},\
  }\bibfield  {title} {\bibinfo {title} {Surface states and topological
  invariants in three-dimensional topological insulators: Application to
  ${\text{bi}}_{1\ensuremath{-}x}{\text{sb}}_{x}$},\ }\href
  {https://doi.org/10.1103/PhysRevB.78.045426} {\bibfield  {journal} {\bibinfo
  {journal} {Phys. Rev. B}\ }\textbf {\bibinfo {volume} {78}},\ \bibinfo
  {pages} {045426} (\bibinfo {year} {2008})}\BibitemShut {NoStop}%
\bibitem [{\citenamefont {Hsieh}\ \emph {et~al.}(2012)\citenamefont {Hsieh},
  \citenamefont {Lin}, \citenamefont {Liu}, \citenamefont {Duan}, \citenamefont
  {Bansil},\ and\ \citenamefont {Fu}}]{hsieh2012topological}%
  \BibitemOpen
  \bibfield  {author} {\bibinfo {author} {\bibfnamefont {T.~H.}\ \bibnamefont
  {Hsieh}}, \bibinfo {author} {\bibfnamefont {H.}~\bibnamefont {Lin}}, \bibinfo
  {author} {\bibfnamefont {J.}~\bibnamefont {Liu}}, \bibinfo {author}
  {\bibfnamefont {W.}~\bibnamefont {Duan}}, \bibinfo {author} {\bibfnamefont
  {A.}~\bibnamefont {Bansil}},\ and\ \bibinfo {author} {\bibfnamefont
  {L.}~\bibnamefont {Fu}},\ }\bibfield  {title} {\bibinfo {title} {Topological
  crystalline insulators in the snte material class},\ }\href
  {https://doi.org/10.1038/ncomms1969} {\bibfield  {journal} {\bibinfo
  {journal} {Nature communications}\ }\textbf {\bibinfo {volume} {3}},\
  \bibinfo {pages} {982} (\bibinfo {year} {2012})}\BibitemShut {NoStop}%
\bibitem [{\citenamefont {Fang}\ \emph {et~al.}(2014)\citenamefont {Fang},
  \citenamefont {Gilbert},\ and\ \citenamefont {Bernevig}}]{chen2014magnetic}%
  \BibitemOpen
  \bibfield  {author} {\bibinfo {author} {\bibfnamefont {C.}~\bibnamefont
  {Fang}}, \bibinfo {author} {\bibfnamefont {M.~J.}\ \bibnamefont {Gilbert}},\
  and\ \bibinfo {author} {\bibfnamefont {B.~A.}\ \bibnamefont {Bernevig}},\
  }\bibfield  {title} {\bibinfo {title} {New class of topological
  superconductors protected by magnetic group symmetries},\ }\href
  {https://doi.org/10.1103/PhysRevLett.112.106401} {\bibfield  {journal}
  {\bibinfo  {journal} {Phys. Rev. Lett.}\ }\textbf {\bibinfo {volume} {112}},\
  \bibinfo {pages} {106401} (\bibinfo {year} {2014})}\BibitemShut {NoStop}%
\bibitem [{\citenamefont {Kitaev}(2009)}]{kitaev2009periodic}%
  \BibitemOpen
  \bibfield  {author} {\bibinfo {author} {\bibfnamefont {A.}~\bibnamefont
  {Kitaev}},\ }\bibfield  {title} {\bibinfo {title} {Periodic table for
  topological insulators and superconductors},\ }in\ \href
  {https://doi.org/10.1063/1.3149495} {\emph {\bibinfo {booktitle} {AIP
  conference proceedings}}},\ Vol.\ \bibinfo {volume} {1134}\ (\bibinfo
  {organization} {American Institute of Physics},\ \bibinfo {year} {2009})\
  pp.\ \bibinfo {pages} {22--30}\BibitemShut {NoStop}%
\bibitem [{\citenamefont {Liu}\ and\ \citenamefont
  {Trauzettel}(2011)}]{liu2011helical}%
  \BibitemOpen
  \bibfield  {author} {\bibinfo {author} {\bibfnamefont {C.-X.}\ \bibnamefont
  {Liu}}\ and\ \bibinfo {author} {\bibfnamefont {B.}~\bibnamefont
  {Trauzettel}},\ }\bibfield  {title} {\bibinfo {title} {Helical dirac-majorana
  interferometer in a superconductor/topological insulator sandwich
  structure},\ }\href {https://doi.org/10.1103/PhysRevB.83.220510} {\bibfield
  {journal} {\bibinfo  {journal} {Phys. Rev. B}\ }\textbf {\bibinfo {volume}
  {83}},\ \bibinfo {pages} {220510} (\bibinfo {year} {2011})}\BibitemShut
  {NoStop}%
\bibitem [{\citenamefont {Young}\ \emph {et~al.}(2012)\citenamefont {Young},
  \citenamefont {Zaheer}, \citenamefont {Teo}, \citenamefont {Kane},
  \citenamefont {Mele},\ and\ \citenamefont {Rappe}}]{young2012DSM}%
  \BibitemOpen
  \bibfield  {author} {\bibinfo {author} {\bibfnamefont {S.~M.}\ \bibnamefont
  {Young}}, \bibinfo {author} {\bibfnamefont {S.}~\bibnamefont {Zaheer}},
  \bibinfo {author} {\bibfnamefont {J.~C.~Y.}\ \bibnamefont {Teo}}, \bibinfo
  {author} {\bibfnamefont {C.~L.}\ \bibnamefont {Kane}}, \bibinfo {author}
  {\bibfnamefont {E.~J.}\ \bibnamefont {Mele}},\ and\ \bibinfo {author}
  {\bibfnamefont {A.~M.}\ \bibnamefont {Rappe}},\ }\bibfield  {title} {\bibinfo
  {title} {Dirac semimetal in three dimensions},\ }\href
  {https://doi.org/10.1103/PhysRevLett.108.140405} {\bibfield  {journal}
  {\bibinfo  {journal} {Phys. Rev. Lett.}\ }\textbf {\bibinfo {volume} {108}},\
  \bibinfo {pages} {140405} (\bibinfo {year} {2012})}\BibitemShut {NoStop}%
\bibitem [{\citenamefont {Wang}\ \emph {et~al.}(2012)\citenamefont {Wang},
  \citenamefont {Sun}, \citenamefont {Chen}, \citenamefont {Franchini},
  \citenamefont {Xu}, \citenamefont {Weng}, \citenamefont {Dai},\ and\
  \citenamefont {Fang}}]{wang2012DSM}%
  \BibitemOpen
  \bibfield  {author} {\bibinfo {author} {\bibfnamefont {Z.}~\bibnamefont
  {Wang}}, \bibinfo {author} {\bibfnamefont {Y.}~\bibnamefont {Sun}}, \bibinfo
  {author} {\bibfnamefont {X.-Q.}\ \bibnamefont {Chen}}, \bibinfo {author}
  {\bibfnamefont {C.}~\bibnamefont {Franchini}}, \bibinfo {author}
  {\bibfnamefont {G.}~\bibnamefont {Xu}}, \bibinfo {author} {\bibfnamefont
  {H.}~\bibnamefont {Weng}}, \bibinfo {author} {\bibfnamefont {X.}~\bibnamefont
  {Dai}},\ and\ \bibinfo {author} {\bibfnamefont {Z.}~\bibnamefont {Fang}},\
  }\bibfield  {title} {\bibinfo {title} {Dirac semimetal and topological phase
  transitions in ${A}_{3}$bi ($a=\text{Na}$, k, rb)},\ }\href
  {https://doi.org/10.1103/PhysRevB.85.195320} {\bibfield  {journal} {\bibinfo
  {journal} {Phys. Rev. B}\ }\textbf {\bibinfo {volume} {85}},\ \bibinfo
  {pages} {195320} (\bibinfo {year} {2012})}\BibitemShut {NoStop}%
\bibitem [{\citenamefont {Zhang}\ \emph {et~al.}(2023)\citenamefont {Zhang},
  \citenamefont {Qin}, \citenamefont {Jiang},\ and\ \citenamefont
  {Hu}}]{zhang2023gapless}%
  \BibitemOpen
  \bibfield  {author} {\bibinfo {author} {\bibfnamefont {Y.}~\bibnamefont
  {Zhang}}, \bibinfo {author} {\bibfnamefont {S.}~\bibnamefont {Qin}}, \bibinfo
  {author} {\bibfnamefont {K.}~\bibnamefont {Jiang}},\ and\ \bibinfo {author}
  {\bibfnamefont {J.}~\bibnamefont {Hu}},\ }\bibfield  {title} {\bibinfo
  {title} {Gapless vortex bound states in superconducting topological
  semimetals},\ }\href {https://doi.org/10.1093/nsr/nwac121} {\bibfield
  {journal} {\bibinfo  {journal} {National Science Review}\ }\textbf {\bibinfo
  {volume} {10}},\ \bibinfo {pages} {nwac121} (\bibinfo {year}
  {2023})}\BibitemShut {NoStop}%
\bibitem [{\citenamefont {Zhu}\ and\ \citenamefont
  {Zhang}(2024)}]{zhu2024delicate}%
  \BibitemOpen
  \bibfield  {author} {\bibinfo {author} {\bibfnamefont {P.}~\bibnamefont
  {Zhu}}\ and\ \bibinfo {author} {\bibfnamefont {R.-X.}\ \bibnamefont
  {Zhang}},\ }\bibfield  {title} {\bibinfo {title} {Delicate topology of
  luttinger semimetal},\ }\href {https://doi.org/10.1103/PhysRevB.110.165120}
  {\bibfield  {journal} {\bibinfo  {journal} {Phys. Rev. B}\ }\textbf {\bibinfo
  {volume} {110}},\ \bibinfo {pages} {165120} (\bibinfo {year}
  {2024})}\BibitemShut {NoStop}%
\bibitem [{\citenamefont {Liu}\ \emph {et~al.}(2024)\citenamefont {Liu},
  \citenamefont {Wan}, \citenamefont {Yang}, \citenamefont {Zhao},
  \citenamefont {Xie}, \citenamefont {Zheng}, \citenamefont {Yi}, \citenamefont
  {Guan}, \citenamefont {Wang}, \citenamefont {Zheng} \emph
  {et~al.}}]{liu2024signatures}%
  \BibitemOpen
  \bibfield  {author} {\bibinfo {author} {\bibfnamefont {T.}~\bibnamefont
  {Liu}}, \bibinfo {author} {\bibfnamefont {C.~Y.}\ \bibnamefont {Wan}},
  \bibinfo {author} {\bibfnamefont {H.}~\bibnamefont {Yang}}, \bibinfo {author}
  {\bibfnamefont {Y.}~\bibnamefont {Zhao}}, \bibinfo {author} {\bibfnamefont
  {B.}~\bibnamefont {Xie}}, \bibinfo {author} {\bibfnamefont {W.}~\bibnamefont
  {Zheng}}, \bibinfo {author} {\bibfnamefont {Z.}~\bibnamefont {Yi}}, \bibinfo
  {author} {\bibfnamefont {D.}~\bibnamefont {Guan}}, \bibinfo {author}
  {\bibfnamefont {S.}~\bibnamefont {Wang}}, \bibinfo {author} {\bibfnamefont
  {H.}~\bibnamefont {Zheng}}, \emph {et~al.},\ }\bibfield  {title} {\bibinfo
  {title} {Signatures of hybridization of multiple majorana zero modes in a
  vortex},\ }\href {https://doi.org/10.1038/s41586-024-07857-4} {\bibfield
  {journal} {\bibinfo  {journal} {Nature}\ }\textbf {\bibinfo {volume} {633}},\
  \bibinfo {pages} {71} (\bibinfo {year} {2024})}\BibitemShut {NoStop}%
\bibitem [{\citenamefont {Hohenberg}\ and\ \citenamefont
  {Kohn}(1964)}]{hohenberg1964inhomogeneous}%
  \BibitemOpen
  \bibfield  {author} {\bibinfo {author} {\bibfnamefont {P.}~\bibnamefont
  {Hohenberg}}\ and\ \bibinfo {author} {\bibfnamefont {W.}~\bibnamefont
  {Kohn}},\ }\bibfield  {title} {\bibinfo {title} {Inhomogeneous electron
  gas},\ }\href {https://doi.org/10.1103/PhysRev.136.B864} {\bibfield
  {journal} {\bibinfo  {journal} {Phys. Rev.}\ }\textbf {\bibinfo {volume}
  {136}},\ \bibinfo {pages} {B864} (\bibinfo {year} {1964})}\BibitemShut
  {NoStop}%
\bibitem [{\citenamefont {Dal~Corso}\ \emph {et~al.}(1996)\citenamefont
  {Dal~Corso}, \citenamefont {Pasquarello}, \citenamefont {Baldereschi},\ and\
  \citenamefont {Car}}]{dal1996generalized}%
  \BibitemOpen
  \bibfield  {author} {\bibinfo {author} {\bibfnamefont {A.}~\bibnamefont
  {Dal~Corso}}, \bibinfo {author} {\bibfnamefont {A.}~\bibnamefont
  {Pasquarello}}, \bibinfo {author} {\bibfnamefont {A.}~\bibnamefont
  {Baldereschi}},\ and\ \bibinfo {author} {\bibfnamefont {R.}~\bibnamefont
  {Car}},\ }\bibfield  {title} {\bibinfo {title} {Generalized-gradient
  approximations to density-functional theory: A comparative study for atoms
  and solids},\ }\href {https://doi.org/10.1103/PhysRevB.53.1180} {\bibfield
  {journal} {\bibinfo  {journal} {Phys. Rev. B}\ }\textbf {\bibinfo {volume}
  {53}},\ \bibinfo {pages} {1180} (\bibinfo {year} {1996})}\BibitemShut
  {NoStop}%
\bibitem [{\citenamefont {Marzari}\ \emph {et~al.}(2012)\citenamefont
  {Marzari}, \citenamefont {Mostofi}, \citenamefont {Yates}, \citenamefont
  {Souza},\ and\ \citenamefont {Vanderbilt}}]{marzari2012RMP}%
  \BibitemOpen
  \bibfield  {author} {\bibinfo {author} {\bibfnamefont {N.}~\bibnamefont
  {Marzari}}, \bibinfo {author} {\bibfnamefont {A.~A.}\ \bibnamefont
  {Mostofi}}, \bibinfo {author} {\bibfnamefont {J.~R.}\ \bibnamefont {Yates}},
  \bibinfo {author} {\bibfnamefont {I.}~\bibnamefont {Souza}},\ and\ \bibinfo
  {author} {\bibfnamefont {D.}~\bibnamefont {Vanderbilt}},\ }\bibfield  {title}
  {\bibinfo {title} {Maximally localized wannier functions: Theory and
  applications},\ }\href {https://doi.org/10.1103/RevModPhys.84.1419}
  {\bibfield  {journal} {\bibinfo  {journal} {Rev. Mod. Phys.}\ }\textbf
  {\bibinfo {volume} {84}},\ \bibinfo {pages} {1419} (\bibinfo {year}
  {2012})}\BibitemShut {NoStop}%
\bibitem [{\citenamefont {Koepernik}\ and\ \citenamefont
  {Eschrig}(1999)}]{koepernik1999full}%
  \BibitemOpen
  \bibfield  {author} {\bibinfo {author} {\bibfnamefont {K.}~\bibnamefont
  {Koepernik}}\ and\ \bibinfo {author} {\bibfnamefont {H.}~\bibnamefont
  {Eschrig}},\ }\bibfield  {title} {\bibinfo {title} {Full-potential
  nonorthogonal local-orbital minimum-basis band-structure scheme},\ }\href
  {https://doi.org/10.1103/PhysRevB.59.1743} {\bibfield  {journal} {\bibinfo
  {journal} {Phys. Rev. B}\ }\textbf {\bibinfo {volume} {59}},\ \bibinfo
  {pages} {1743} (\bibinfo {year} {1999})}\BibitemShut {NoStop}%
\bibitem [{\citenamefont {Koepernik}\ \emph {et~al.}(2023)\citenamefont
  {Koepernik}, \citenamefont {Janson}, \citenamefont {Sun},\ and\ \citenamefont
  {van~den Brink}}]{koepernik2023symmetry}%
  \BibitemOpen
  \bibfield  {author} {\bibinfo {author} {\bibfnamefont {K.}~\bibnamefont
  {Koepernik}}, \bibinfo {author} {\bibfnamefont {O.}~\bibnamefont {Janson}},
  \bibinfo {author} {\bibfnamefont {Y.}~\bibnamefont {Sun}},\ and\ \bibinfo
  {author} {\bibfnamefont {J.}~\bibnamefont {van~den Brink}},\ }\bibfield
  {title} {\bibinfo {title} {Symmetry-conserving maximally projected wannier
  functions},\ }\href {https://doi.org/10.1103/PhysRevB.107.235135} {\bibfield
  {journal} {\bibinfo  {journal} {Phys. Rev. B}\ }\textbf {\bibinfo {volume}
  {107}},\ \bibinfo {pages} {235135} (\bibinfo {year} {2023})}\BibitemShut
  {NoStop}%
\bibitem [{\citenamefont {Fang}\ \emph {et~al.}(2019)\citenamefont {Fang},
  \citenamefont {Pan}, \citenamefont {Zhang}, \citenamefont {Wang},
  \citenamefont {Hirose}, \citenamefont {Terashima}, \citenamefont {Uji},
  \citenamefont {Yuan}, \citenamefont {Li}, \citenamefont {Tian}, \citenamefont
  {Xue}, \citenamefont {Ma}, \citenamefont {Zhao}, \citenamefont {Xue},
  \citenamefont {Mu}, \citenamefont {Zhang},\ and\ \citenamefont
  {Huang}}]{fang2019discovery}%
  \BibitemOpen
  \bibfield  {author} {\bibinfo {author} {\bibfnamefont {Y.}~\bibnamefont
  {Fang}}, \bibinfo {author} {\bibfnamefont {J.}~\bibnamefont {Pan}}, \bibinfo
  {author} {\bibfnamefont {D.}~\bibnamefont {Zhang}}, \bibinfo {author}
  {\bibfnamefont {D.}~\bibnamefont {Wang}}, \bibinfo {author} {\bibfnamefont
  {H.~T.}\ \bibnamefont {Hirose}}, \bibinfo {author} {\bibfnamefont
  {T.}~\bibnamefont {Terashima}}, \bibinfo {author} {\bibfnamefont
  {S.}~\bibnamefont {Uji}}, \bibinfo {author} {\bibfnamefont {Y.}~\bibnamefont
  {Yuan}}, \bibinfo {author} {\bibfnamefont {W.}~\bibnamefont {Li}}, \bibinfo
  {author} {\bibfnamefont {Z.}~\bibnamefont {Tian}}, \bibinfo {author}
  {\bibfnamefont {J.}~\bibnamefont {Xue}}, \bibinfo {author} {\bibfnamefont
  {Y.}~\bibnamefont {Ma}}, \bibinfo {author} {\bibfnamefont {W.}~\bibnamefont
  {Zhao}}, \bibinfo {author} {\bibfnamefont {Q.}~\bibnamefont {Xue}}, \bibinfo
  {author} {\bibfnamefont {G.}~\bibnamefont {Mu}}, \bibinfo {author}
  {\bibfnamefont {H.}~\bibnamefont {Zhang}},\ and\ \bibinfo {author}
  {\bibfnamefont {F.}~\bibnamefont {Huang}},\ }\bibfield  {title} {\bibinfo
  {title} {Discovery of superconductivity in 2m ws2 with possible topological
  surface states},\ }\href
  {https://doi.org/https://doi.org/10.1002/adma.201901942} {\bibfield
  {journal} {\bibinfo  {journal} {Advanced Materials}\ }\textbf {\bibinfo
  {volume} {31}},\ \bibinfo {pages} {1901942} (\bibinfo {year}
  {2019})}\BibitemShut {NoStop}%
\bibitem [{\citenamefont {Li}\ \emph {et~al.}(2021)\citenamefont {Li},
  \citenamefont {Zheng}, \citenamefont {Fang}, \citenamefont {Zhang},
  \citenamefont {Chen}, \citenamefont {Chen}, \citenamefont {Liang},
  \citenamefont {Shi}, \citenamefont {Pei}, \citenamefont {Xu} \emph
  {et~al.}}]{li2021observation}%
  \BibitemOpen
  \bibfield  {author} {\bibinfo {author} {\bibfnamefont {Y.}~\bibnamefont
  {Li}}, \bibinfo {author} {\bibfnamefont {H.}~\bibnamefont {Zheng}}, \bibinfo
  {author} {\bibfnamefont {Y.}~\bibnamefont {Fang}}, \bibinfo {author}
  {\bibfnamefont {D.}~\bibnamefont {Zhang}}, \bibinfo {author} {\bibfnamefont
  {Y.}~\bibnamefont {Chen}}, \bibinfo {author} {\bibfnamefont {C.}~\bibnamefont
  {Chen}}, \bibinfo {author} {\bibfnamefont {A.}~\bibnamefont {Liang}},
  \bibinfo {author} {\bibfnamefont {W.}~\bibnamefont {Shi}}, \bibinfo {author}
  {\bibfnamefont {D.}~\bibnamefont {Pei}}, \bibinfo {author} {\bibfnamefont
  {L.}~\bibnamefont {Xu}}, \emph {et~al.},\ }\bibfield  {title} {\bibinfo
  {title} {Observation of topological superconductivity in a stoichiometric
  transition metal dichalcogenide 2m-ws2},\ }\href
  {https://doi.org/10.1038/s41467-021-23076-1} {\bibfield  {journal} {\bibinfo
  {journal} {Nature communications}\ }\textbf {\bibinfo {volume} {12}},\
  \bibinfo {pages} {2874} (\bibinfo {year} {2021})}\BibitemShut {NoStop}%
\bibitem [{\citenamefont {Xu}\ \emph {et~al.}(2023)\citenamefont {Xu},
  \citenamefont {Li}, \citenamefont {Fang}, \citenamefont {Zheng},
  \citenamefont {Shi}, \citenamefont {Chen}, \citenamefont {Pei}, \citenamefont
  {Lu}, \citenamefont {Hashimoto}, \citenamefont {Wang}, \citenamefont {Yang},
  \citenamefont {Feng}, \citenamefont {Zhang}, \citenamefont {Huang},
  \citenamefont {Xue}, \citenamefont {He}, \citenamefont {Liu},\ and\
  \citenamefont {Chen}}]{xu2023topological}%
  \BibitemOpen
  \bibfield  {author} {\bibinfo {author} {\bibfnamefont {L.}~\bibnamefont
  {Xu}}, \bibinfo {author} {\bibfnamefont {Y.}~\bibnamefont {Li}}, \bibinfo
  {author} {\bibfnamefont {Y.}~\bibnamefont {Fang}}, \bibinfo {author}
  {\bibfnamefont {H.}~\bibnamefont {Zheng}}, \bibinfo {author} {\bibfnamefont
  {W.}~\bibnamefont {Shi}}, \bibinfo {author} {\bibfnamefont {C.}~\bibnamefont
  {Chen}}, \bibinfo {author} {\bibfnamefont {D.}~\bibnamefont {Pei}}, \bibinfo
  {author} {\bibfnamefont {D.}~\bibnamefont {Lu}}, \bibinfo {author}
  {\bibfnamefont {M.}~\bibnamefont {Hashimoto}}, \bibinfo {author}
  {\bibfnamefont {M.}~\bibnamefont {Wang}}, \bibinfo {author} {\bibfnamefont
  {L.}~\bibnamefont {Yang}}, \bibinfo {author} {\bibfnamefont {X.}~\bibnamefont
  {Feng}}, \bibinfo {author} {\bibfnamefont {H.}~\bibnamefont {Zhang}},
  \bibinfo {author} {\bibfnamefont {F.}~\bibnamefont {Huang}}, \bibinfo
  {author} {\bibfnamefont {Q.}~\bibnamefont {Xue}}, \bibinfo {author}
  {\bibfnamefont {K.}~\bibnamefont {He}}, \bibinfo {author} {\bibfnamefont
  {Z.}~\bibnamefont {Liu}},\ and\ \bibinfo {author} {\bibfnamefont
  {Y.}~\bibnamefont {Chen}},\ }\bibfield  {title} {\bibinfo {title} {Topology
  hierarchy of transition metal dichalcogenides built from quantum spin hall
  layers},\ }\href {https://doi.org/https://doi.org/10.1002/adma.202300227}
  {\bibfield  {journal} {\bibinfo  {journal} {Advanced Materials}\ }\textbf
  {\bibinfo {volume} {35}},\ \bibinfo {pages} {2300227} (\bibinfo {year}
  {2023})}\BibitemShut {NoStop}%
\bibitem [{\citenamefont {Li}\ \emph {et~al.}(2024)\citenamefont {Li},
  \citenamefont {Xu}, \citenamefont {Liu}, \citenamefont {Fang}, \citenamefont
  {Zheng}, \citenamefont {Dai}, \citenamefont {Li}, \citenamefont {Zhu},
  \citenamefont {Zhang}, \citenamefont {Liang} \emph
  {et~al.}}]{li2024evidence}%
  \BibitemOpen
  \bibfield  {author} {\bibinfo {author} {\bibfnamefont {Y.}~\bibnamefont
  {Li}}, \bibinfo {author} {\bibfnamefont {L.}~\bibnamefont {Xu}}, \bibinfo
  {author} {\bibfnamefont {G.}~\bibnamefont {Liu}}, \bibinfo {author}
  {\bibfnamefont {Y.}~\bibnamefont {Fang}}, \bibinfo {author} {\bibfnamefont
  {H.}~\bibnamefont {Zheng}}, \bibinfo {author} {\bibfnamefont
  {S.}~\bibnamefont {Dai}}, \bibinfo {author} {\bibfnamefont {E.}~\bibnamefont
  {Li}}, \bibinfo {author} {\bibfnamefont {G.}~\bibnamefont {Zhu}}, \bibinfo
  {author} {\bibfnamefont {S.}~\bibnamefont {Zhang}}, \bibinfo {author}
  {\bibfnamefont {S.}~\bibnamefont {Liang}}, \emph {et~al.},\ }\bibfield
  {title} {\bibinfo {title} {Evidence of strong and mode-selective
  electron--phonon coupling in the topological superconductor candidate
  2m-ws2},\ }\href {https://doi.org/10.1038/s41467-024-50590-9} {\bibfield
  {journal} {\bibinfo  {journal} {Nature Communications}\ }\textbf {\bibinfo
  {volume} {15}},\ \bibinfo {pages} {6235} (\bibinfo {year}
  {2024})}\BibitemShut {NoStop}%
\bibitem [{\citenamefont {Guguchia}\ \emph {et~al.}(2019)\citenamefont
  {Guguchia}, \citenamefont {Gawryluk}, \citenamefont {Brzezinska},
  \citenamefont {Tsirkin}, \citenamefont {Khasanov}, \citenamefont
  {Pomjakushina}, \citenamefont {von Rohr}, \citenamefont {Verezhak},
  \citenamefont {Hasan}, \citenamefont {Neupert} \emph
  {et~al.}}]{guguchia2019nodeless}%
  \BibitemOpen
  \bibfield  {author} {\bibinfo {author} {\bibfnamefont {Z.}~\bibnamefont
  {Guguchia}}, \bibinfo {author} {\bibfnamefont {D.~J.}\ \bibnamefont
  {Gawryluk}}, \bibinfo {author} {\bibfnamefont {M.}~\bibnamefont
  {Brzezinska}}, \bibinfo {author} {\bibfnamefont {S.~S.}\ \bibnamefont
  {Tsirkin}}, \bibinfo {author} {\bibfnamefont {R.}~\bibnamefont {Khasanov}},
  \bibinfo {author} {\bibfnamefont {E.}~\bibnamefont {Pomjakushina}}, \bibinfo
  {author} {\bibfnamefont {F.~O.}\ \bibnamefont {von Rohr}}, \bibinfo {author}
  {\bibfnamefont {J.~A.}\ \bibnamefont {Verezhak}}, \bibinfo {author}
  {\bibfnamefont {M.~Z.}\ \bibnamefont {Hasan}}, \bibinfo {author}
  {\bibfnamefont {T.}~\bibnamefont {Neupert}}, \emph {et~al.},\ }\bibfield
  {title} {\bibinfo {title} {Nodeless superconductivity and its evolution with
  pressure in the layered dirac semimetal 2m-ws2},\ }\href
  {https://doi.org/10.1038/s41535-019-0189-5} {\bibfield  {journal} {\bibinfo
  {journal} {npj Quantum Materials}\ }\textbf {\bibinfo {volume} {4}},\
  \bibinfo {pages} {50} (\bibinfo {year} {2019})}\BibitemShut {NoStop}%
\bibitem [{\citenamefont {Lian}\ \emph {et~al.}(2020)\citenamefont {Lian},
  \citenamefont {Si},\ and\ \citenamefont {Duan}}]{lian2020anisotropic}%
  \BibitemOpen
  \bibfield  {author} {\bibinfo {author} {\bibfnamefont {C.-S.}\ \bibnamefont
  {Lian}}, \bibinfo {author} {\bibfnamefont {C.}~\bibnamefont {Si}},\ and\
  \bibinfo {author} {\bibfnamefont {W.}~\bibnamefont {Duan}},\ }\bibfield
  {title} {\bibinfo {title} {Anisotropic full-gap superconductivity in 2m-ws2
  topological metal with intrinsic proximity effect},\ }\href
  {https://doi.org/10.1021/acs.nanolett.0c04357} {\bibfield  {journal}
  {\bibinfo  {journal} {Nano Letters}\ }\textbf {\bibinfo {volume} {21}},\
  \bibinfo {pages} {709} (\bibinfo {year} {2020})}\BibitemShut {NoStop}%
\bibitem [{\citenamefont {Paudyal}\ and\ \citenamefont
  {Margine}(2022)}]{paudyal2022superconducting}%
  \BibitemOpen
  \bibfield  {author} {\bibinfo {author} {\bibfnamefont {H.}~\bibnamefont
  {Paudyal}}\ and\ \bibinfo {author} {\bibfnamefont {E.~R.}\ \bibnamefont
  {Margine}},\ }\bibfield  {title} {\bibinfo {title} {Superconducting
  properties in doped 2m-ws 2 from first principles},\ }\href
  {https://doi.org/10.1039/D2TC01173E} {\bibfield  {journal} {\bibinfo
  {journal} {Journal of Materials Chemistry C}\ }\textbf {\bibinfo {volume}
  {10}},\ \bibinfo {pages} {7917} (\bibinfo {year} {2022})}\BibitemShut
  {NoStop}%
\bibitem [{\citenamefont {Tang}\ \emph {et~al.}(2017)\citenamefont {Tang},
  \citenamefont {Zhang}, \citenamefont {Wong}, \citenamefont {Pedramrazi},
  \citenamefont {Tsai}, \citenamefont {Jia}, \citenamefont {Moritz},
  \citenamefont {Claassen}, \citenamefont {Ryu}, \citenamefont {Kahn} \emph
  {et~al.}}]{tang2017quantum}%
  \BibitemOpen
  \bibfield  {author} {\bibinfo {author} {\bibfnamefont {S.}~\bibnamefont
  {Tang}}, \bibinfo {author} {\bibfnamefont {C.}~\bibnamefont {Zhang}},
  \bibinfo {author} {\bibfnamefont {D.}~\bibnamefont {Wong}}, \bibinfo {author}
  {\bibfnamefont {Z.}~\bibnamefont {Pedramrazi}}, \bibinfo {author}
  {\bibfnamefont {H.-Z.}\ \bibnamefont {Tsai}}, \bibinfo {author}
  {\bibfnamefont {C.}~\bibnamefont {Jia}}, \bibinfo {author} {\bibfnamefont
  {B.}~\bibnamefont {Moritz}}, \bibinfo {author} {\bibfnamefont
  {M.}~\bibnamefont {Claassen}}, \bibinfo {author} {\bibfnamefont
  {H.}~\bibnamefont {Ryu}}, \bibinfo {author} {\bibfnamefont {S.}~\bibnamefont
  {Kahn}}, \emph {et~al.},\ }\bibfield  {title} {\bibinfo {title} {Quantum spin
  hall state in monolayer 1t-wte2},\ }\href {https://doi.org/10.1038/nphys4174}
  {\bibfield  {journal} {\bibinfo  {journal} {Nature Physics}\ }\textbf
  {\bibinfo {volume} {13}},\ \bibinfo {pages} {683} (\bibinfo {year}
  {2017})}\BibitemShut {NoStop}%
\bibitem [{\citenamefont {Chen}\ \emph {et~al.}(2018)\citenamefont {Chen},
  \citenamefont {Pai}, \citenamefont {Chan}, \citenamefont {Sun}, \citenamefont
  {Xu}, \citenamefont {Lin}, \citenamefont {Chou}, \citenamefont {Fedorov},\
  and\ \citenamefont {Chiang}}]{chen2018large}%
  \BibitemOpen
  \bibfield  {author} {\bibinfo {author} {\bibfnamefont {P.}~\bibnamefont
  {Chen}}, \bibinfo {author} {\bibfnamefont {W.~W.}\ \bibnamefont {Pai}},
  \bibinfo {author} {\bibfnamefont {Y.-H.}\ \bibnamefont {Chan}}, \bibinfo
  {author} {\bibfnamefont {W.-L.}\ \bibnamefont {Sun}}, \bibinfo {author}
  {\bibfnamefont {C.-Z.}\ \bibnamefont {Xu}}, \bibinfo {author} {\bibfnamefont
  {D.-S.}\ \bibnamefont {Lin}}, \bibinfo {author} {\bibfnamefont
  {M.}~\bibnamefont {Chou}}, \bibinfo {author} {\bibfnamefont {A.-V.}\
  \bibnamefont {Fedorov}},\ and\ \bibinfo {author} {\bibfnamefont {T.-C.}\
  \bibnamefont {Chiang}},\ }\bibfield  {title} {\bibinfo {title} {Large
  quantum-spin-hall gap in single-layer 1t wse2},\ }\href
  {https://doi.org/10.1038/s41467-018-04395-2} {\bibfield  {journal} {\bibinfo
  {journal} {Nature communications}\ }\textbf {\bibinfo {volume} {9}},\
  \bibinfo {pages} {1} (\bibinfo {year} {2018})}\BibitemShut {NoStop}%
\bibitem [{\citenamefont {Soluyanov}\ \emph {et~al.}(2015)\citenamefont
  {Soluyanov}, \citenamefont {Gresch}, \citenamefont {Wang}, \citenamefont
  {Wu}, \citenamefont {Troyer}, \citenamefont {Dai},\ and\ \citenamefont
  {Bernevig}}]{soluyanov2015type}%
  \BibitemOpen
  \bibfield  {author} {\bibinfo {author} {\bibfnamefont {A.~A.}\ \bibnamefont
  {Soluyanov}}, \bibinfo {author} {\bibfnamefont {D.}~\bibnamefont {Gresch}},
  \bibinfo {author} {\bibfnamefont {Z.}~\bibnamefont {Wang}}, \bibinfo {author}
  {\bibfnamefont {Q.}~\bibnamefont {Wu}}, \bibinfo {author} {\bibfnamefont
  {M.}~\bibnamefont {Troyer}}, \bibinfo {author} {\bibfnamefont
  {X.}~\bibnamefont {Dai}},\ and\ \bibinfo {author} {\bibfnamefont {B.~A.}\
  \bibnamefont {Bernevig}},\ }\bibfield  {title} {\bibinfo {title} {Type-ii
  weyl semimetals},\ }\href {https://doi.org/10.1038/nature15768} {\bibfield
  {journal} {\bibinfo  {journal} {Nature}\ }\textbf {\bibinfo {volume} {527}},\
  \bibinfo {pages} {495} (\bibinfo {year} {2015})}\BibitemShut {NoStop}%
\bibitem [{\citenamefont {Sancho}\ \emph {et~al.}(1985)\citenamefont {Sancho},
  \citenamefont {Sancho}, \citenamefont {Sancho},\ and\ \citenamefont
  {Rubio}}]{Sancho1985green}%
  \BibitemOpen
  \bibfield  {author} {\bibinfo {author} {\bibfnamefont {M.~P.~L.}\
  \bibnamefont {Sancho}}, \bibinfo {author} {\bibfnamefont {J.~M.~L.}\
  \bibnamefont {Sancho}}, \bibinfo {author} {\bibfnamefont {J.~M.~L.}\
  \bibnamefont {Sancho}},\ and\ \bibinfo {author} {\bibfnamefont
  {J.}~\bibnamefont {Rubio}},\ }\bibfield  {title} {\bibinfo {title} {Highly
  convergent schemes for the calculation of bulk and surface green functions},\
  }\href {https://doi.org/10.1088/0305-4608/15/4/009} {\bibfield  {journal}
  {\bibinfo  {journal} {Journal of Physics F: Metal Physics}\ }\textbf
  {\bibinfo {volume} {15}},\ \bibinfo {pages} {851} (\bibinfo {year}
  {1985})}\BibitemShut {NoStop}%
\bibitem [{\citenamefont {Guan}\ \emph {et~al.}(2016)\citenamefont {Guan},
  \citenamefont {Chen}, \citenamefont {Chu}, \citenamefont {Sankar},
  \citenamefont {Chou}, \citenamefont {Jeng}, \citenamefont {Chang},\ and\
  \citenamefont {Chuang}}]{guan2016superconducting}%
  \BibitemOpen
  \bibfield  {author} {\bibinfo {author} {\bibfnamefont {S.-Y.}\ \bibnamefont
  {Guan}}, \bibinfo {author} {\bibfnamefont {P.-J.}\ \bibnamefont {Chen}},
  \bibinfo {author} {\bibfnamefont {M.-W.}\ \bibnamefont {Chu}}, \bibinfo
  {author} {\bibfnamefont {R.}~\bibnamefont {Sankar}}, \bibinfo {author}
  {\bibfnamefont {F.}~\bibnamefont {Chou}}, \bibinfo {author} {\bibfnamefont
  {H.-T.}\ \bibnamefont {Jeng}}, \bibinfo {author} {\bibfnamefont {C.-S.}\
  \bibnamefont {Chang}},\ and\ \bibinfo {author} {\bibfnamefont {T.-M.}\
  \bibnamefont {Chuang}},\ }\bibfield  {title} {\bibinfo {title}
  {Superconducting topological surface states in the noncentrosymmetric bulk
  superconductor pbtase2},\ }\href
  {https://www.science.org/doi/10.1126/sciadv.1600894} {\bibfield  {journal}
  {\bibinfo  {journal} {Science advances}\ }\textbf {\bibinfo {volume} {2}},\
  \bibinfo {pages} {e1600894} (\bibinfo {year} {2016})}\BibitemShut {NoStop}%
\bibitem [{\citenamefont {Xun}\ and\ \citenamefont {Zhang}()}]{xun2025}%
  \BibitemOpen
  \bibfield  {author} {\bibinfo {author} {\bibfnamefont {Y.}~\bibnamefont
  {Xun}}\ and\ \bibinfo {author} {\bibfnamefont {R.-X.}\ \bibnamefont
  {Zhang}},\ }\href@noop {} {\bibinfo  {journal} {in preparation}\
  }\BibitemShut {NoStop}%
\end{thebibliography}%

\end{document}